\documentclass[12pt,sort&compress]{elsarticle}

\usepackage[utf8]{inputenc}

\usepackage{bm}

\usepackage{mathrsfs}
\usepackage{amsmath}
\usepackage{amsfonts}
\usepackage{amsthm}
\usepackage{color}
\usepackage{sansmath}

\usepackage{array}

\usepackage{caption}
\usepackage{subcaption}
\usepackage{float}
\usepackage{url}
\usepackage{multicol}
\usepackage[top=1in,bottom=1in,left=1in,right=1in]{geometry}
\usepackage[pdfborder={0 0 0},colorlinks,allcolors=blue]{hyperref}
\usepackage{txfonts}
\usepackage{setspace}
\usepackage{arydshln}
\usepackage{enumitem}
\usepackage{lipsum}% for dummy text
\usepackage{siunitx,booktabs}
\usepackage{nth}
\usepackage{accents} % Accents: circle
\usepackage{siunitx}
\theoremstyle{definition}%plain/definition/remark
\newtheorem{remark}{Remark}%[section]
\usepackage{listings}
\usepackage{xcolor}
\usepackage{algorithm}
\usepackage{algorithmicx}
\usepackage{longtable}
\usepackage{multirow}
\usepackage{tikz}
\usetikzlibrary{shapes,arrows,positioning,calc}

\usepackage{listings}
\usepackage{xcolor}
\usepackage{setspace}

\lstdefinestyle{xmlstyle}{
    language=XML,
    basicstyle=\ttfamily\footnotesize,
    keywordstyle=[1]\color{brown},
    keywordstyle=[2]\color{red},
    stringstyle=\color{blue},
    commentstyle=\color{ForestGreen},
    morestring=[b]",
    morecomment=[s]{<!--}{-->},
    morekeywords=[1]{setfrdrawmode,drawcylinder, setmkbound,drawfilestl,drawscale,drawmove,drawrotate,constantsdef,geometry,motion,execution,parameters,xml,case,casedef,mkconfig,definition,pointref,pointmin,pointmax,commands,mainlist,setmkbound,shapeout,objreal,begin,mvnull,special,setshapemode,deformstrucs,bcvel,bcforce,deformstrucbody,density,youngmod,poissratio,constitmodel,avfactor,mapfac,artvisc,notch,p1,p2,p3,p4,fracture,Gc,pflenscale,pflim,restcoef,yieldstress,hardening,kfric,measureplane,nbsrange,timestep,mathexpressions,userexpression,locals,expression,u_mu,u_bulk,gravity,rhop0,hswl,gamma,speedsystem,coefsound,coefh,cflnumber,newvar,drawbox,boxfill,point,size,parameter,posmin,posmax,simulationdomain,setdrawmode,encoding,contcoeff,restrictphi,drawfilevtk},
    morekeywords=[2]{mk,radius,file,objname,autofill,advanced,reverse,x,y,z,angx,angy,angz,app,date,boundcount,fluidcount,dp,comment,units_comment,mkid,ref,mov,id,start,mkbound,value,factor1,factor2,xe,ye,ze,tst,tend,type,auto,mode,solid,key},
    frame=single,
    breaklines=true,
    tabsize=2,
    showstringspaces=false,
    numbers=left,
    numberstyle=\tiny\color{gray},
    numbersep=5pt
}

\lstdefinestyle{cppstyle}{
    language=C++,
    basicstyle=\ttfamily\footnotesize,
    keywordstyle=\color{blue},
    stringstyle=\color{red},
    commentstyle=\color{gray},
    frame=single,
    breaklines=true,
    tabsize=2,
    showstringspaces=false,
    numbers=left,
    numberstyle=\tiny\color{gray},
    numbersep=5pt,
    morekeywords={__global__, __device__, __host__, __shared__, dim3, float3, float4, uint2, tdouble3, tfloat3, typecode}
}

\lstdefinestyle{cudastyle}{
    language=C++,
    basicstyle=\ttfamily\footnotesize,
    keywordstyle=\color{blue},
    stringstyle=\color{red},
    commentstyle=\color{green!60!black},
    frame=single,
    breaklines=true,
    tabsize=2,
    showstringspaces=false,
    numbers=left,
    numberstyle=\tiny\color{gray},
    numbersep=5pt,
    morekeywords={__global__, __device__, __host__, __shared__, __syncthreads, atomicAdd, atomicCAS, dim3, float3, float4, uint2, blockIdx, blockDim, threadIdx, gridDim}
}
\allowdisplaybreaks

\renewcommand{\sectionautorefname}{Section}

\usepackage[dvipsnames]{xcolor}
\newcommand\hl[1]{%
	\bgroup
	\hskip0pt\color{red!80!black}%
	#1%
	\egroup
}

\usepackage{algorithm}
\usepackage{algpseudocode}
\usepackage{algorithmicx}
\definecolor{shadecolor}{cmyk}{0,0,0,0.03}

\newcommand*\subr[1]{\Statex\textbf{#1}}

\newcommand*\qIf[3]{\Statex\quad\textbf{if} $#1 #2 #3$ \textbf{then}}

\newcommand*\EndqIf{\Statex\quad\textbf{end if}}

\usepackage[T1]{fontenc}
\usepackage{amsmath}
\definecolor{icardblue}{HTML}{1F6FEB}
\usepackage{listings}
\usepackage{amsmath} % for \text in math mode

\DeclareRobustCommand{\icard}[1]{%
  \ifmmode
    \text{\lstinline[style=xmlstyle]|<#1>|}%
  \else
    \lstinline[style=xmlstyle]|<#1>|%
  \fi
}
\usepackage{xspace}
\newcommand{\solidname}{SoliDualSPHysics\xspace}

\newcommand{\Eqref}[1]{Eq.~\eqref{#1}}
\newcommand{\Eqsref}[2]{Eqs.~\eqref{#1}--\eqref{#2}}
\journal{}

\usepackage{upgreek}

\newcommand{\etal}{\textit{et al}. }

\begin{document}

\begin{frontmatter}

\title{\solidname: An extension of DualSPHysics for solid mechanics with hyperelasticity, plasticity, and fracture}

\author[synopsys]{Mohammad Naqib Rahimi}
\author[ru]{George~Moutsanidis\corref{cor1}}
\ead{george.moutsanidis@rutgers.edu}
\cortext[cor1]{Corresponding author}

\address[synopsys]{Synopsys Inc., Austin, TX 78746, USA}

\address[ru]{Department of Mechanical and Aerospace Engineering, Rutgers University, Piscataway, NJ 08854, USA}

\begin{abstract}
We introduce \solidname, a novel open-source and GPU-accelerated software that extends DualSPHysics to enable the numerical simulation of hyperelastic, finite-strain plastic, and brittle fracture behavior in deformable solids within a unified smoothed particle hydrodynamics (SPH) software framework. The software implements a total Lagrangian formulation for solid mechanics that allows direct application of external loads and boundary conditions, enabling independent solid mechanics simulations. Brittle fracture is modeled through a phase-field approach coupled with SPH, allowing crack initiation, propagation, and branching under dynamic loading without explicit crack tracking, ad hoc crack-path criteria, or local refinement. The framework also supports user-defined mathematical expressions to prescribe time- and space-dependent quantities, complementing the solid and fracture extensions and enhancing flexibility across existing and future DualSPHysics applications. Leveraging DualSPHysics' native CPU/GPU parallel architecture, the software achieves substantial computational acceleration for large-scale simulations, and the implementation is verified and validated against benchmark numerical problems and experimental data, demonstrating accuracy, robustness, and favorable scaling performance. Comprehensive implementation details and user documentation are provided to ensure reproducibility and to support further development by the community. The framework and source code are freely available through a public GitHub repository.

\noindent \textbf{Program Summary}

\begin{small}
\noindent
{\em Program Title:} SoliDualSPHysics \\
{\em Developer's repository link:} \url{https://github.com/naqibr/SoliDualSPHysics} \\
{\em Licensing provisions:} GNU Lesser General Public License \\
{\em Programming language:} C++ and CUDA \\
{\em Supplementary material:} Available in the repository. The repository includes README documentation, build and execution instructions, XML input examples, post-processing instructions, and several benchmark cases with CPU/GPU run scripts, together with documentation of the expression parser, output files, directory structure, and CPU/GPU implementation components. \\
{\em Nature of problem:} Smoothed particle hydrodynamics (SPH) has been extensively used in fluid dynamics and offers significant advantages for large-deformation solid mechanics due to its meshfree Lagrangian formulation. However, most high-performance SPH frameworks primarily target computational fluid dynamics applications, with limited support for standalone deformable solid mechanics. In particular, the consistent implementation of finite-strain constitutive models, flexible boundary conditions, and physically consistent fracture formulations remains underdeveloped in existing SPH software. Furthermore, integrating advanced solid mechanics and fracture capabilities within a scalable CPU/GPU framework suitable for large-scale simulations introduces additional algorithmic and performance challenges. Addressing these limitations requires a unified SPH formulation for solid and fracture mechanics, together with an efficient parallel implementation within an extensible open-source environment. \\
\noindent {\em Solution method:} SoliDualSPHysics extends the DualSPHysics software to enable standalone deformable solid mechanics and fracture simulations within a unified SPH framework. The solver is based on a total Lagrangian SPH (TLSPH) formulation, and supports hyperelastic constitutive models (St. Venant--Kirchhoff and compressible neo--Hookean) and finite-strain $J_2$ plasticity with isotropic hardening, integrated through an explicit time-stepping scheme. Brittle fracture is incorporated via a phase-field formulation, allowing crack initiation, propagation, branching, and coalescence to emerge naturally without explicit crack tracking or enrichment. Flexible time- and space-dependent velocity and traction boundary conditions are prescribed using user-defined mathematical expressions. Computational efficiency is achieved by leveraging the OpenMP (CPU) and CUDA (GPU) parallel infrastructure of DualSPHysics, allowing scalable simulations involving millions of particles. \\
\noindent {\em Additional comments including Restrictions and Unusual features:} The present release supports isotropic St. Venant--Kirchhoff and compressible neo--Hookean hyperelasticity, isotropic finite-strain $J_2$ plasticity, and brittle phase-field fracture. Plasticity and phase-field fracture are not coupled in the current implementation. Anisotropic constitutive models, composite materials, ductile fracture, and dedicated hourglass-control operators are not included. The software is configured through XML input files and supports user-defined mathematical expressions for time- and space-dependent boundary and loading conditions. GPU execution requires CUDA-compatible hardware.

\end{small}

\end{abstract}


\end{frontmatter}

\section{Introduction}
\label{sec:introduction}

Smoothed particle hydrodynamics (SPH) is a well-established particle-based method with a remarkable application history across many fields. It was originally developed as an interpolation technique for studying astrophysical phenomena \cite{gingold1977smoothed,Lucy1977}, and was later extended to fluid mechanics \cite{MONAGHAN1994399}, solid mechanics \cite{gray2001sph}, and fluid--structure interaction (FSI) \cite{antoci2007numerical}, among others. In SPH, a continuum domain is discretized into a set of particles, commonly referred to as SPH particles, and any functions or their derivatives appearing in the governing conservation equations are approximated through a kernel interpolation (or kernel expansion), which provides a meshfree means of spatial discretization. One of the major advantages of SPH over other existing particle-based and meshfree methods is that the inclusion of new physics is straightforward, and it can thus be easily extended to new applications \cite{Libersky2006}. In the context of solid mechanics, its ability to handle extreme material distortion and significant deformations makes it particularly suitable for simulating large-deformation problems, where conventional Lagrangian mesh-based techniques fail due to
mesh entanglement and the need for frequent mesh updating or remeshing. Many variants of the method have emerged since its initial appearance, and the interested reader is referred to \cite{liu2010smoothed,monaghan2012smoothed,price2012smoothed,violeau2016smoothed,lind2020review,bui2021smoothed,xu2023methodology,gotoh2021entirely,khayyer2022systematic} and the references therein.

In line with the popularity of SPH, several remarkable open-source and high-performance codes have been developed over the years and applied to a wide range of challenging engineering mechanics problems. Many of these frameworks, such as SPHysics \cite{crespo2008application}, GPUSPH \cite{herault2010sph}, AQUAgpusph \cite{cercos2015aquagpusph}, SPlisHSPlasH \cite{abbasi2021numerical}, and openMaelstrom \cite{akhunov2023evaluation}, are primarily focused on computational fluid dynamics (CFD) applications, including free-surface and incompressible flow simulations, and may possess basic FSI functionalities, typically limited to rigid-body motion. A few other codes have included solid mechanics and more extended FSI capabilities. For example, PySPH \cite{ramachandran2021pysph}, in addition to its CFD capabilities, incorporates formulations for elastic dynamics based on the total Lagrangian SPH (TLSPH) approach. SPHinXsys \cite{zhang2021sphinxsys} includes both CFD and FSI features, as well as a simplified approach for modeling structural damage and fracture through a damage factor integrated into the SPH kernel calculations \cite{wu2022modeling}, though its parallel performance is limited compared to GPU-accelerated frameworks. PersianSPH \cite{gholami2018sph} provides similar capabilities but with limited parallelization, while SPHERA \cite{amicarelli2020sphera} is mainly tailored for geophysical and hydraulic problems involving sediment transport and granular materials. Among the available open-source SPH tools, DualSPHysics \cite{crespo2015dualsphysics,dominguez2022dualsphysics} is one of the most widely used frameworks and has been especially influential because of its CPU/GPU implementation and its broad adoption in free-surface-flow simulations. Beyond its core fluid-dynamics functionality, DualSPHysics also includes some fluid--structure interaction capabilities involving deformable bodies, including a total-Lagrangian-SPH-based hydroelastic solver for elastic structures. In addition to the original DualSPHysics framework, recent developments under the name DualSPHysics+ \cite{zhan2025dualsphysics+,zhan2025enhanced1,zhan2025enhanced2} have introduced important enhancements for incompressible free-surface flows, wave-structure interaction, and hydroelastic fluid--structure interaction. These include improved schemes for continuity resolution, pressure-noise mitigation, and energy conservation in the fluid solver, as well as structural advances such as second-order discretization, Riemann-based stabilization, and dynamic hourglass control for elastic structural response.

Taken together, these developments have provided a valuable toolbox for the engineering community and have substantially expanded the scope of SPH-based simulation. At the same time, the current open-source landscape remains weighted toward fluid dynamics, wave-structure interaction, and elastic or hydroelastic structural response. In comparison, open-source SPH support for a broader deformable-solid mechanics framework remains much more limited, particularly when one seeks, within a unified setting, flexible prescription of solid boundary conditions and external loads, constitutive models beyond elasticity, finite-strain plasticity, and physically grounded brittle-fracture modeling. This gap is especially pronounced in the area of brittle fracture, which remains relatively underexplored within the SPH community. Most existing SPH fracture formulations have relied on ad hoc empirical local damage models, “cracking particle” approaches, pseudo-spring and virtual-link methods, and cohesive zone models. However, these approaches come with limitations and drawbacks. For example, local damage models \cite{rabczuk2003simulation} are based on empirical damage laws rather than comprehensive fracture theories and lead to mesh dependency and non-convergent results under refinement. The “cracking particle” approach \cite{Rabczuk2004a} resembles the extended finite element method (XFEM) \cite{moes1999finite,moes2002extended}; therefore, fracture surfaces need to be tracked, and the local enrichment of the approximation space leads to increased computational cost. In the pseudo-spring approach \cite{chakraborty2013pseudo,islam2019total,islam2020pseudo}, damage evolution is based on relatively simple linear damage models, and the softening curve of the damage law may lead to instabilities; previous studies have also shown that these methods are prone to spurious damage patterns. Finally, in cohesive zone models \cite{wang2019new,wang2020simulation,bui2021smoothed}, kinematic enrichment is performed (similar to XFEM), which can lead to increased computational time and more complex implementation.

A more physically consistent alternative has emerged in recent years through phase-field approaches to brittle fracture in SPH \cite{rahimi2022Asmoothed,rahimi2022Modeling,Rahimi2023AnSphbased,rahimi2024IGA}. The phase-field approach to brittle fracture is a variational formulation based on the minimization of an energy functional and originates from Griffith's theory of fracture \cite{Borden2012}. In this framework, cracks are not explicitly introduced into the solid domain; instead, the fracture surface is approximated by a phase-field variable that regularizes the discontinuity over a narrow region. The phase-field variable represents material integrity and provides a smooth transition from the fully intact to the fully damaged state. Its evolution is governed by a partial differential equation, and the fracture problem is therefore reformulated as a system of PDEs that fully determines crack evolution. It is worth noting that there are no conditions or phenomenological rules in order to determine crack nucleation, propagation, branching, and coalescence. Therefore, phase-field models of fracture do not require any numerical tracking of evolving discontinuities.

In this work, we introduce \solidname by extending the open-source software DualSPHysics to enable the simulation of hyperelasticity, plasticity, and brittle fracture in deformable solids within a unified SPH framework. The solid mechanics capabilities of the framework are based on a total Lagrangian SPH formulation and support the direct application of external loads and boundary conditions, making the framework suitable for general deformable-solid simulations. Furthermore, brittle fracture capabilities are introduced through the phase-field approach, which allows the natural capture of crack initiation, propagation, branching, and coalescence without any ad hoc criteria, local refinement, or numerical tracking of evolving discontinuities. The framework includes the following salient capabilities: (1) a user-defined mathematical and logical expression parser, (2) flexible velocity and traction boundary conditions, (3) hyperelastic (neo--Hookean and St. Venant--Kirchhoff) and finite-strain $J_2$ plasticity constitutive models, (4) phase-field modeling of brittle fracture, (5) multi-resolution and independent time-step refinement for solid mechanics, and (6) full CPU and GPU parallelization. Implementation details are provided to facilitate reproducibility and further development by the SPH community. The code is verified and validated against benchmark numerical problems and experimental results, and the GPU implementation is shown to deliver meaningful acceleration and favorable performance for large-scale solid mechanics and fracture simulations. The developed software is released as open source and is freely available through a public GitHub repository, together with example cases and documentation to support reproducibility and further development. We emphasize that the present implementation is restricted to isotropic deformable-solid response, namely isotropic hyperelasticity, isotropic finite-strain $J_2$ plasticity, and phase-field of brittle fracture. Extensions to anisotropic constitutive behavior and composite structures are beyond the present scope. We note, however, that related SPH developments for anisotropic/composite structural modeling have been reported in both Hamiltonian-SPH formulations and in non-Hamiltonian SPH developments \cite{khayyer2021coupled,khayyer20223d,shishova2019tracking,zhang2021integrative}. The present work is instead aimed at providing an open-source, CPU/GPU-accelerated DualSPHysics extension for standalone solid mechanics with unified support for hyperelasticity, finite-strain plasticity, brittle fracture, flexible boundary conditions, and reproducible XML-based workflows.

This paper is organized as follows. In \autoref{sec:formulation}, we review the theoretical formulation underlying the proposed framework. \autoref{sec:dualspharc} describes the extension of DualSPHysics and the implementation details of \solidname. In \autoref{sec:2dcases}, we present a series of numerical examples and a representative performance study that demonstrate the accuracy, robustness, and computational efficiency of the software for problems in hyperelasticity, plasticity, and fracture mechanics. Finally, in \autoref{sec:conclusions} we draw conclusions and outline future research directions.

\section{Formulation}\label{sec:formulation}
In this section, we briefly review the theoretical formulation underlying \solidname. For completeness, and to keep the presentation self-contained, we closely follow the developments presented in our previous work \cite{rahimi2022Asmoothed,rahimi2022Modeling,Kamensky2018,moutsanidis2018hyperbolic}.

\subsection{Momentum Balance}
We consider the problem of nonlinear elastodynamics. Omitting boundary and initial conditions for brevity, the momentum balance in strong form over the undeformed configuration reads
\begin{equation}
    \rho_0 \dot{\mathbf{v}} = \nabla_{0} \cdot \mathbf{P} + \rho_0 \mathbf{f} \; \; \text{in} \; \; \Omega_{0},
\end{equation}
where $\rho_{0}$ is the mass density in the undeformed configuration, $\mathbf{v}$ is the velocity of the material, $\mathbf{P}$ is the first Piola--Kirchhoff stress tensor, $\mathbf{f}$ is a body force per unit mass, $\Omega_{0}$ is the undeformed domain of the solid material, $\nabla_{0}$ denotes derivatives with respect to the undeformed configuration, and the superimposed dot denotes time differentiation. The first Piola--Kirchhoff stress is defined as
\begin{equation}
\mathbf{P} = \mathbf{F} \mathbf{S},
\end{equation}
where $\mathbf{F}$ is the deformation gradient tensor and $\mathbf{S}$ is the second Piola--Kirchhoff stress tensor, defined as
\begin{equation}
\mathbf{F} = \frac{\partial \mathbf{x}}{\partial \mathbf{X}} = \frac{\partial \mathbf{u}}{\partial \mathbf{X}} + \mathbf{I},
\end{equation}
\begin{equation}
\mathbf{S} = \frac{\partial \psi_{e}}{\partial \mathbf{E}},
\end{equation}
respectively. Here, $\mathbf{u}$ is the displacement, $\mathbf{x}$ and $\mathbf{X}$ are the position vectors in the deformed and undeformed configurations, respectively, $\mathbf{I}$ is the identity tensor, $\psi_{e}$ is the elastic strain-energy density defined through an appropriate constitutive model, and $\mathbf{E}$ is the Green--Lagrange strain tensor defined as 
\begin{equation}
\mathbf{E} = \frac{1}{2} \Big(\mathbf{F}^{T} \mathbf{F} - \mathbf{I} \Big).
\end{equation}

\subsection{Phase-Field of Brittle Fracture}

The evolution of brittle fracture is modeled using a phase-field formulation, governed by the following equation
\begin{equation}
\label{eq:phasefield_hyperbolic2}
\frac{2 G_{c} \epsilon_0}{c^2} \ddot{s} + \frac{1}{M} \dot{s} + 2 s \mathcal{H} - G_{c} \bigg ( 2 \epsilon_0 \nabla_0^2 \, s + \frac{1-s}{2 \epsilon_0} \bigg ) = 0 \; \text{in} \; \Omega_{0}.
\end{equation}
Here, $s$ is the phase-field (or damage) variable, which provides a smooth approximation of a fracture surface. It continuously interpolates between the intact material state ($s = 1$) and the fully fractured state ($s = 0$). The parameter $\epsilon_0$ has dimensions of length and controls the width of the regularized crack region. In the limit $\epsilon_0 \to 0$, the phase-field approximation converges to a sharp fracture surface. In practice, $\epsilon_0$ must be chosen sufficiently small so as not to alter the underlying fracture physics, while remaining comparable to or larger than the spatial discretization length to ensure proper numerical regularization. To model the loss of material stiffness, the elastic strain energy density is defined as
\begin{equation}
\label{eq:energy_density}
\psi_{e} = s^2 \psi_{e}^{+} + \psi_{e}^{-}.
\end{equation}
$\psi_{e}^{+}$ and $\psi_{e}^{-}$ are the positive and negative parts of the elastic strain energy density, respectively, which will be defined later on according to the particular constitutive models employed. Evidently, crack propagation is only allowed in tension since the phase-field parameter is applied only to the tensile part of the elastic strain energy. $\mathcal{H}$ is the so-called history functional used to enforce the irreversibility condition (i.e., cracks do not heal), and is defined as
\begin{equation}
\label{eq:irreversibility}
\mathcal{H} (\mathbf{X},t) = \max_{\tau \le t} ( \psi_{e}^{+} (\mathbf{X},\tau)),
\end{equation}
where $t$ is the current time and $\tau$ is a history-time variable. $M$ is a damping parameter controlling the rate at which local damage information diffuses into the bulk material, and in our framework is defined as
\begin{equation} \label{eq:Mparameter1}
M \leq \frac{c}{2 \sqrt{4 G_{c} \epsilon_0 \mathcal{H} + G_{c}^2}},
\end{equation}
so that the phase-field equation does not exhibit a wave-like behavior and evolves monotonically. Finally, $c$ is the characteristic propagation speed in the hyperbolic evolution equation for the phase-field variable $s$, and therefore limits how fast the phase-field disturbance can travel through the undamaged material. Thus, it should not be interpreted as a prescribed crack propagation speed. In the present work, $c$ is set equal to the elastic sound speed of the undamaged material as
\begin{equation}
	c = \sqrt{\frac{\kappa+\frac{4}{3}\mu}{\rho_{0}}},
\end{equation}
where $\kappa = \lambda + \frac{2\mu}{3}$ is the bulk modulus, and $\lambda$ and $\mu$ are the Lam\'e parameters.

\begin{remark}
It is important to note that \Eqref{eq:phasefield_hyperbolic2} is a hyperbolic equation. This contrasts with prevailing phase-field models for brittle fracture, which employ elliptic formulations. As explained in \cite{Kamensky2018,moutsanidis2018hyperbolic,rahimi2022Asmoothed}, elliptic models become problematic when lumped-mass explicit dynamics schemes are employed, such as in SPH, since a global linear system must be solved at every time step to determine the phase-field variable. Although elliptic formulations can in principle be embedded within explicit frameworks, this would require solving a global system at each time step, leading to a substantial increase in computational cost.
\end{remark}

\subsection{Constitutive Modeling}

We consider two hyperelastic constitutive models commonly used in solid mechanics: the St. Venant--Kirchhoff model and a compressible neo--Hookean model. These models are coupled to the phase-field formulation, enabling the simulation of brittle fracture. Additionally, an extension to finite-strain $J_{2}$ plasticity is introduced in a subsequent subsection. At present, this extension is limited to elastoplastic response without fracture, as modeling ductile fracture mechanisms is beyond the scope of the present work.

\subsubsection{St. Venant--Kirchhoff Model}

For a St. Venant--Kirchhoff material model, the elastic strain energy density functional is defined as
\begin{equation}
\psi_{e} = \frac{1}{2} \lambda ( \text{tr}  \mathbf{E})^2 + \mu \text{tr} (\mathbf{E}^2).
\end{equation}
We then define
\begin{equation}\label{eq:elasEnergyP}
    \psi_{e}^{+} = \frac{1}{2} \lambda \{ \text{tr}  \mathbf{E} \}_{+}^2 + \mu \text{tr} (\mathbf{E}^{+} \mathbf{E}^{+}),
\end{equation} 
\begin{equation}\label{eq:elasEnergyN}
    \psi_{e}^{-} = \frac{1}{2} \lambda \{ \text{tr}  \mathbf{E} \}_{-}^2 + \mu \text{tr} (\mathbf{E}^{-} \mathbf{E}^{-}),
\end{equation}
where the following decomposition is employed
\begin{equation}\label{eq:GLstratinP}
\mathbf{E}^{+} = \mathbf{Q} \mathbf{\Lambda}^{+} \mathbf{Q}^{T},
\end{equation}
\begin{equation}\label{eq:GLstratinN}
\mathbf{E}^{-} = \mathbf{Q} \mathbf{\Lambda}^{-} \mathbf{Q}^{T},
\end{equation}
\begin{equation}
\mathbf{E} = \mathbf{Q} \mathbf{\Lambda} \mathbf{Q}^{T}.
\end{equation}
$\Lambda = \text{diag} (\lambda_{1}, \lambda_{2}, \lambda_{3} )$ contains the eigenvalues of $\mathbf{E}$ on its diagonal, $\mathbf{Q}$ has the corresponding eigenvectors as its columns, $\Lambda^{\pm} = \text{diag} (\lambda_{1}^{\pm}, \lambda_{2}^{\pm}, \lambda_{3}^{\pm} )$, and $\{\,.\,\}_{\pm}$ selects the $\pm$ part of its argument, i.e.
\begin{equation}\label{eq:xpm_decomp}
    \{x\}_{\pm} = \left\{\begin{array}{lr}x & x\in\mathbb{R}^\pm\\ 0 &\text{otherwise}\end{array}\right.\text{ .}
\end{equation}
The second Piola--Kirchhoff stress can then be computed by differentiating the strain energy density $\psi_{e}$ with respect to the Green--Lagrange strain tensor $\mathbf{E}$,
\begin{equation}
\label{eq:2PK_SVK1}
\mathbf{S}^{\pm} = \frac{\partial \psi_{e}^{\pm}}{\partial \mathbf{E}} = \lambda~\{ \text{tr}  \mathbf{E} \}_{\pm} \mathbf{I}+2\mu \mathbf{E}^{\pm},
\end{equation}
and
\begin{equation}
\label{eq:2PK_SVK2}
\mathbf{S} = s^2 \mathbf{S}^{+} + \mathbf{S}^{-}.
\end{equation}

\subsubsection{Neo--Hookean Model}
In the neo--Hookean model, the elastic strain energy density functional is expressed as
\begin{equation}
\psi_{e}  = \frac{\mu}{2}(I_1 - 3) - \mu\ln J + \frac{\lambda}{2}(\ln J)^2,
\end{equation}
where $J = \text{det} \mathbf{F}$, $I_1 = \text{tr} (\mathbf{C})$, and $\mathbf{C} = \mathbf{F}^T \mathbf{F}$ is the right Cauchy--Green deformation tensor. Accordingly, the positive and negative parts of the elastic strain energy density are given as
\begin{equation}\label{eq:ksipNH}
\psi_{e}^{+} = \left\{\begin{array}{lr}	U(J) + \overline{\psi}_{e} (\overline{\mathbf{b}}) \, & J \geq 1 \\ \overline{\psi}_{e} (\overline{\mathbf{b}}) \, & J< 1, \end{array}\right.
\end{equation}
\begin{equation}\label{eq:ksinNH}
\psi_{e}^{-} = \left\{\begin{array}{lr}	0 \, & J \geq 1 \\ U(J) \, & J< 1, \end{array}\right.
\end{equation}
where
\begin{equation} \label{eq:UjNeoHok}
U(J) = \frac{1}{2} \kappa~\bigg( \frac{1}{2} (J^2 - 1) - \text{ln} J \bigg),
\end{equation}
\begin{equation} \label{eq:ksiNeoHok}
\overline{\psi}_{e} (\overline{\mathbf{b}}) = \frac{1}{2} \mu \Big( \text{tr} \overline{\mathbf{b}} - 3 \Big),
\end{equation} 
\begin{equation}\label{eq:bmatrix}
\mathbf{b} = \mathbf{F} \mathbf{F}^T,
\end{equation}
\begin{equation}\label{eq:bbarmatrix}
\overline{\mathbf{b}} = J\,^{-2/3} \mathbf{b}.
\end{equation}
The second Piola--Kirchhoff stress can then be computed as
\begin{equation}
\mathbf{S} = 2 \frac{\partial \psi_{e}}{\partial \mathbf{b}},
\end{equation}
which results in
\begin{equation}
\label{eq:2PK_NH1}
\mathbf{S} = 2 \left\{\begin{array}{lr} s^2 \bigg( U'(J) \frac{\partial J}{\partial \mathbf{b}} + \frac{\partial \overline{\psi}_{e} (\overline{\mathbf{b})}}{\partial \mathbf{b}} \bigg) \, & J \geq 1 \\ U'(J) \frac{\partial J}{\partial \mathbf{b}} + s^2 \frac{\partial \overline{\psi}_{e} (\overline{\mathbf{b})}}{\partial \mathbf{b}}  \, & J< 1. 
\end{array}\right.
\end{equation}
The derivatives in the above expression are computed as 
\begin{equation}
U'(J) = \frac{1}{2} \kappa \Big( J - J^{-1} \Big),
\end{equation}
\begin{equation}
\frac{\partial J}{\partial \mathbf{b}} = \frac{\partial \sqrt{\text{det} \mathbf{b}}}{\partial \mathbf{b}} = \frac{1}{2} J \mathbf{b}^{-1},
\end{equation}
\begin{equation}
\frac{\partial \overline{\psi}_{e}}{\partial \mathbf{b}} = \frac{\partial \overline{\psi}_{e}}{\partial \overline{\mathbf{b}}} \frac{\partial \overline{\mathbf{b}}}{\partial \mathbf{b}} = \frac{J^{-2/3}}{2} \mu \bigg( \mathbf{I} - \frac{1}{3} \big( \text{tr} \mathbf{b} \big) \mathbf{b}^{-1} \bigg).
\end{equation}
Substituting the above equations into \Eqref{eq:2PK_NH1} we get
\begin{equation}
\label{eq:2PK_NH2}
    \mathbf{S} = \left\{\begin{array}{lr} s^2 \bigg( \frac{1}{2} \kappa \big( J^2 - 1 \big) \mathbf{b}^{-1} + J^{-2/3} \mu \Big( \mathbf{I} - \frac{1}{3} \big( \text{tr} \mathbf{b} \big) \mathbf{b}^{-1} \Big) \bigg)  \, & J \geq 1 \\ \frac{1}{2} \kappa \big( J^2 - 1 \big) \mathbf{b}^{-1} +s^2 \bigg( J^{-2/3} \mu \Big( \mathbf{I} - \frac{1}{3} \big( \text{tr} \mathbf{b} \big) \mathbf{b}^{-1} \Big) \bigg)  \, & J< 1. 
\end{array}\right.
\end{equation}

\begin{remark}
In the absence of fracture (s = 1), the above constitutive relations reduce to their classical hyperelastic counterparts corresponding to the St. Venant--Kirchhoff and neo--Hookean material models.
\end{remark}

\subsubsection{Extension to Finite-Strain $J_2$ Plasticity}
In addition to the hyperelastic material models described above, we extend the framework to rate-independent $J_2$ elastoplasticity suitable for finite deformations. Unlike small-strain plasticity, finite-strain elastoplasticity requires careful treatment of large rotations and an appropriate elastic--plastic decomposition to ensure frame objectivity and thermodynamic consistency.

We adopt a finite-strain formulation based on the multiplicative decomposition of the deformation gradient \cite{Lee1969,Simo1998}
\begin{equation}
\mathbf{F} = \mathbf{F}_e \mathbf{F}_p,
\end{equation}
where $\mathbf{F}_e$ and $\mathbf{F}_p$ represent the elastic and plastic parts of the deformation gradient, respectively. This decomposition is fundamentally different from the additive strain decomposition $\mathbf{E} = \mathbf{E}^e + \mathbf{E}^p$ and ensures that finite rotations are properly accounted for in the plastic evolution.

In the present implementation the plastic deformation gradient $\mathbf{F}_p$ is stored as the primary internal variable and advanced using an exponential-map return algorithm based on the multiplicative flow rule. The elastic right Cauchy--Green tensor is then computed directly from the multiplicative split as
\begin{equation}
\mathbf{C}_e = \mathbf{F}_e^{T}\mathbf{F}_e
            = \mathbf{F}_p^{-T}\,\mathbf{C}\,\mathbf{F}_p^{-1},
\end{equation}
without recourse to any proxy or simplified relation. The elastic free energy is decomposed into volumetric and isochoric parts,
\begin{equation}
\psi_e(\mathbf{C}_e) = \psi_{\mathrm{vol}}(J_e) + \psi_{\mathrm{iso}}(\bar{\mathbf{C}}_e),
\end{equation}
with
\begin{equation}
J_p = \det\mathbf{F}_p,\qquad
J_e = \det\mathbf{F}_e = J/J_p .
\end{equation}
Plastic flow is enforced to be isochoric so that $J_p = 1$ and consequently $J_e = J$ in the present implementation. The isochoric part of the elastic right Cauchy--Green tensor is
\begin{equation}
\bar{\mathbf{C}}_e = J_e^{-2/3}\,\mathbf{C}_e .
\end{equation}
The volumetric and isochoric contributions are given by \cite{suchocki2025finite} as
\begin{align}
\psi_{\mathrm{vol}}(J) &= \frac{\kappa}{4}\left(J^2 - 1 - 2\ln J\right), \\
\psi_{\mathrm{iso}}(\bar{\mathbf{C}}_e) &= \frac{\mu}{2}\left(\mathrm{tr}(\bar{\mathbf{C}}_e) - 3\right).
\end{align}

The Mandel stress measure employed in the present formulation is defined as the work-conjugate to the plastic velocity gradient \cite{eidel2003elastoplastic,Mandel1974,Weber1990},
\begin{equation}
\mathbf{M} = \mathbf{C}_e \mathbf{S}_e = \mathbf{C}_e \frac{\partial \psi_e}{\partial \mathbf{E}_e},
\end{equation}
where $\mathbf{S}_e$ is the elastic part of the second Piola--Kirchhoff stress and $\mathbf{E}_e = \tfrac{1}{2}(\mathbf{C}_e - \mathbf{I})$ is the elastic Green--Lagrange strain. For the hyperelastic model above, the deviatoric Mandel stress is
\begin{equation}
\mathbf{M}_{\mathrm{dev}} = \mu\,\mathrm{dev}(\bar{\mathbf{C}}_e),
\quad\text{where}\quad
\mathrm{dev}(\bar{\mathbf{C}}_e) = \bar{\mathbf{C}}_e - \tfrac{1}{3}\mathrm{tr}(\bar{\mathbf{C}}_e)\,\mathbf{I}.
\end{equation}
Plastic yielding is governed by a von Mises ($J_2$) yield criterion expressed in terms of the Mandel stress as
\begin{equation}
f(\mathbf{M},\bar{\varepsilon}^p) = \sigma_{\mathrm{eq}} - \sigma_y(\bar{\varepsilon}^p) \le 0,
\end{equation}
where the equivalent stress is
\begin{equation}
\sigma_{\mathrm{eq}} = \sqrt{\tfrac{3}{2}\,\mathbf{M}_{\mathrm{dev}}:\mathbf{M}_{\mathrm{dev}}},
\end{equation}
and $\bar{\varepsilon}^p$ denotes the accumulated equivalent plastic strain. Linear isotropic hardening is adopted, $\sigma_y(\bar{\varepsilon}^p) = \sigma_{y0} + H\,\bar{\varepsilon}^p$, with initial yield stress $\sigma_{y0}$ and hardening modulus $H$.

An associative flow rule is postulated in the intermediate (plastic) configuration. Adopting the flow direction
\begin{equation}
\mathbf{N} \;=\; \frac{\partial f}{\partial \mathbf{M}}
       \;=\; \frac{3}{2\sigma_{\mathrm{eq}}}\,\mathbf{M}_{\mathrm{dev}},
\end{equation}
which is symmetric, deviatoric, and satisfies $\|\mathbf{N}\|=\sqrt{3/2}$, the plastic deformation gradient evolves multiplicatively as
\begin{equation}
\dot{\mathbf{F}}_p\,\mathbf{F}_p^{-1} \;=\; \dot{\gamma}\,\mathbf{N},
\end{equation}
and the equivalent plastic strain accumulates as
\begin{equation}
\dot{\bar{\varepsilon}}^p \;=\; \sqrt{\tfrac{2}{3}}\,\|\dot{\gamma}\,\mathbf{N}\|
                            \;=\; \dot{\gamma},
\end{equation}
i.e.\ with the present normalization of $\mathbf{N}$ the plastic multiplier coincides with the equivalent plastic strain rate. The loading-unloading behavior is governed by the standard Kuhn--Tucker conditions, which enforce plastic admissibility, non-negative plastic flow, and consistency during active plastic loading,
\begin{equation}
\dot{\gamma}\ge0,\qquad f\le 0,\qquad \dot{\gamma}\,f = 0.
\end{equation}

The constitutive update is performed by an exponential-map predictor--corrector scheme operating directly on $\mathbf{F}_p$. At each material point and time step $t_n\to t_{n+1}$, the trial elastic state is obtained by freezing the plastic deformation gradient,
\begin{equation}
\mathbf{F}_p^{\mathrm{tr}} = \mathbf{F}_p^{n},\qquad
\mathbf{C}_e^{\mathrm{tr}} = (\mathbf{F}_p^{n})^{-T}\,\mathbf{C}^{n+1}\,(\mathbf{F}_p^{n})^{-1},
\end{equation}
from which $\bar{\mathbf{C}}_e^{\mathrm{tr}}$, the trial deviatoric Mandel stress $\mathbf{M}_{\mathrm{dev}}^{\mathrm{tr}}$, and the trial equivalent stress $\sigma_{\mathrm{eq}}^{\mathrm{tr}}$ are evaluated as above. The trial flow direction
\begin{equation}
\mathbf{N}^{\mathrm{tr}} = \frac{3}{2\sigma_{\mathrm{eq}}^{\mathrm{tr}}}\,\mathbf{M}_{\mathrm{dev}}^{\mathrm{tr}},
\end{equation}
is computed once at the trial state and held fixed throughout the corrector, in the spirit of the standard return-mapping for $J_2$ plasticity. If $f(\mathbf{M}^{\mathrm{tr}},\bar{\varepsilon}^{p,n}) \le 0$, the step is elastic and $\mathbf{F}_p^{n+1} = \mathbf{F}_p^{n}$. Otherwise, the plastic increment $\Delta\gamma>0$ is determined from the discrete consistency condition
\begin{equation}
r(\Delta\gamma) \;\equiv\; \sigma_{\mathrm{eq}} \bigl(\mathbf{F}_p(\Delta\gamma)\bigr) \;-\; \sigma_y \bigl(\bar{\varepsilon}^{p,n} + \Delta\gamma\bigr) \;=\; 0,
\label{eq:J2_residual}
\end{equation}
where the candidate plastic deformation gradient is updated by the exponential map of the frozen flow direction,
\begin{equation}
\mathbf{F}_p(\Delta\gamma) \;=\; \exp \bigl(\Delta\gamma\,\mathbf{N}^{\mathrm{tr}}\bigr)\,\mathbf{F}_p^{n},
\label{eq:expmap}
\end{equation}
followed by the volume-preserving projection that enforces plastic incompressibility ($J_p=1$),
\begin{equation}
\mathbf{F}_p(\Delta\gamma) \;\leftarrow\; \bigl(\det\mathbf{F}_p(\Delta\gamma)\bigr)^{-1/3}\,\mathbf{F}_p(\Delta\gamma).
\label{eq:Fp_iso_proj}
\end{equation}
For each candidate $\mathbf{F}_p(\Delta\gamma)$, the elastic right Cauchy--Green tensor $\mathbf{C}_e(\Delta\gamma) = \mathbf{F}_p^{-T}(\Delta\gamma)\,\mathbf{C}^{n+1}\,\mathbf{F}_p^{-1}(\Delta\gamma)$ and the corresponding $\sigma_{\mathrm{eq}}(\Delta\gamma)$ are recomputed, so that \Eqref{eq:J2_residual} is a genuinely nonlinear scalar equation in $\Delta\gamma$ rather than the linearized closed-form expression typical of small-strain $J_2$ algorithms.

\Eqref{eq:J2_residual} is solved by an Illinois-modified regula--falsi \cite{dowell1971modified} (false-position) bracketing iteration. The iteration is initialized with the bracket
\begin{equation}
\Delta\gamma_L = 0,\qquad
\Delta\gamma_R^{(0)} = \max\!\left(\frac{\sigma_{\mathrm{eq}}^{\mathrm{tr}} - \sigma_y(\bar{\varepsilon}^{p,n})}{3\mu + H},\;\epsilon\right),
\end{equation}
which corresponds to a small-strain return-mapping estimate; the upper endpoint is doubled until $r(\Delta\gamma_R)\le 0$ so that the root is bracketed. The Illinois update then alternates the false-position step with a halving of the stale endpoint residual, ensuring monotonic bracket contraction and avoiding the stagnation typical of plain regula--falsi:
\begin{equation}
\Delta\gamma^{(k+1)} \;=\; \Delta\gamma_R^{(k)} \;-\; r_R^{(k)}\,
\frac{\Delta\gamma_R^{(k)} - \Delta\gamma_L^{(k)}}{r_R^{(k)} - r_L^{(k)}},
\end{equation}
with the half-residual safeguard applied to the endpoint that has not been replaced. The iteration is terminated when the absolute residual or the bracket width falls below pre-set tolerances. At convergence the plastic deformation gradient is updated through \Eqsref{eq:expmap}{eq:Fp_iso_proj} as
\begin{equation}
\mathbf{F}_p^{n+1} \;=\; \bigl(\det\mathbf{F}_p^{*}\bigr)^{-1/3}\,\mathbf{F}_p^{*},
\qquad
\mathbf{F}_p^{*} \;=\; \exp \bigl(\Delta\gamma\,\mathbf{N}^{\mathrm{tr}}\bigr)\,\mathbf{F}_p^{n},
\end{equation}
and the equivalent plastic strain by
\begin{equation}
\bar{\varepsilon}^{p,n+1} \;=\; \bar{\varepsilon}^{p,n} + \Delta\gamma .
\end{equation}
The matrix exponential in \Eqref{eq:expmap} is evaluated by a scaling-and-squaring Taylor scheme, which is well suited to GPU evaluation since the argument $\Delta\gamma\,\mathbf{N}^{\mathrm{tr}}$ is typically small over a single time step.

Once $\mathbf{F}_p^{n+1}$ is known, the updated elastic state is recomputed from
\begin{equation}
\mathbf{C}_e^{\,n+1} = (\mathbf{F}_p^{n+1})^{-T}\,\mathbf{C}^{n+1}\,(\mathbf{F}_p^{n+1})^{-1},
\end{equation}
and the deviatoric Mandel stress $\mathbf{M}_{\mathrm{dev}}^{\,n+1} = \mu\,\mathrm{dev}\!\left(J^{-2/3}\mathbf{C}_e^{\,n+1}\right)$ is evaluated. The elastic second Piola--Kirchhoff stress in the intermediate configuration is then obtained from the Mandel definition,
\begin{equation}
\mathbf{S}_e^{\,n+1}
\;=\; (\mathbf{C}_e^{\,n+1})^{-1}\,\mathbf{M}_{\mathrm{dev}}^{\,n+1},
\end{equation}
and pulled back to the reference configuration through the standard transformation
\begin{equation}
\mathbf{S}_{\mathrm{iso}}^{\,n+1}
\;=\; (\mathbf{F}_p^{n+1})^{-1}\,\mathbf{S}_e^{\,n+1}\,(\mathbf{F}_p^{n+1})^{-T}.
\end{equation}
Adding the volumetric contribution that follows from $\psi_{\mathrm{vol}}(J)$ and the assumption of plastic incompressibility,
\begin{equation}
\mathbf{S}_{\mathrm{vol}}^{\,n+1} \;=\; \frac{\kappa}{2}\bigl(J^{2}-1\bigr)\,\mathbf{C}^{-1},
\end{equation}
the total second Piola--Kirchhoff stress used to evaluate the SPH internal forces is
\begin{equation}
\mathbf{S}^{n+1}
\;=\; (\mathbf{F}_p^{n+1})^{-1}\bigl[(\mathbf{C}_e^{\,n+1})^{-1}\,\mathbf{M}_{\mathrm{dev}}^{\,n+1}\bigr](\mathbf{F}_p^{n+1})^{-T}
   \;+\; \tfrac{\kappa}{2}\bigl(J^{2}-1\bigr)\,\mathbf{C}^{-1}.
\end{equation}
The plastic deformation gradient $\mathbf{F}_p^{n+1}$ and the equivalent plastic strain $\bar{\varepsilon}^{p,n+1}$ are stored as internal variables for the next time step.

In the two-dimensional plane-strain regime the formulation is implemented via a three-dimensional embedding: the in-plane components of $\mathbf{F}$ and $\mathbf{C}$ are assembled into $3\times3$ tensors with $F_{22}=1$, $F_{12}=F_{21}=F_{23}=F_{32}=0$, and $\mathbf{F}_p$ is constrained to share the same block-diagonal structure. All tensor operations of the return-mapping algorithm are then performed in three dimensions, which guarantees that $\mathbf{C}_e$, $\mathbf{M}_{\mathrm{dev}}$, and $\mathbf{N}^{\mathrm{tr}}$ remain well-defined and that the isochoric projection of $\mathbf{F}_p$ acts consistently in both 2D and 3D simulations.

The present formulation is motivated by the multiplicative decomposition $\mathbf{F}=\mathbf{F}_e\mathbf{F}_p$, introduced by Lee \cite{Lee1969} and developed extensively in finite-strain plasticity by Simo et al.\ \cite{Simo1992,Simo1998}. By taking $\mathbf{F}_p$ itself as the internal variable and using an exponential-map update for the plastic flow, plastic incompressibility is enforced at the algorithmic level, up to the tolerance of the matrix-exponential approximation and floating-point round-off, with the projection in \Eqref{eq:Fp_iso_proj} serving as a numerical safeguard. The formulation is frame-indifferent under superposed rigid-body motions, and the Mandel stress provides the natural work-conjugate driving force for the inelastic flow. The resulting return-mapping procedure is consistent with the standard thermodynamic structure of finite-strain plasticity and is implemented within the GPU-oriented computational framework adopted in this work.

\begin{remark}
Plastic deformation leads to irreversible energy dissipation, while the elastic strain-energy density governs only the recoverable response. For this reason, and because modeling ductile fracture mechanisms lies beyond the scope of the present work, the $J_2$ elastoplastic model is employed without coupling to the phase-field formulation. When plasticity is activated, the framework therefore describes elastoplastic solid mechanics without fracture.
\end{remark}

\begin{remark}
\label{rem:additive_strain_issues}
While additive decomposition $\mathbf{E}=\mathbf{E}^e+\mathbf{E}^p$ of the Green--Lagrange strain is often used in finite element implementations for moderate strains, its extension to large-deformation total Lagrangian frameworks requires additional assumptions and may exhibit important theoretical and algorithmic limitations, as discussed extensively in the computational plasticity literature \cite{Simo1998,Belytschko2000}:

\begin{enumerate}[label=(\roman*)]
\item The additive decomposition is rigorously valid only for infinitesimal strains where the distinction between reference and current configurations is negligible \cite{Lubliner1990}. When extended to finite strains, the use of Green--Lagrange strain $\mathbf{E}=\tfrac{1}{2}(\mathbf{C}-\mathbf{I})$ as a primary strain measure may lead to nonphysical response under large rotations and volumetric deformations for certain constitutive choices (e.g., St.\ Venant--Kirchhoff) \cite{Bonet2008}.

\item Classical continuum mechanics requires that constitutive equations be invariant under superposed rigid body motions (frame-indifference or objectivity) \cite{Truesdell1960,Marsden1983}. For finite deformations, this requirement is more naturally accommodated in formulations based on multiplicative decomposition of the deformation gradient, whereas additive strain decompositions generally require additional rotational treatments or objective stress/strain rates to remain consistent under large rotations \cite{Naghdi1990,Xiao1998}.

\item The dissipation inequality for rate-independent plasticity requires careful definition of driving forces and flow rules. In multiplicative finite-strain plasticity, the Mandel stress provides the natural work-conjugate driving force associated with the inelastic flow and is therefore widely used in constitutive updates \cite{Simo1988,Miehe1995}. Additive decompositions, when extended beyond small strains, generally require additional assumptions to ensure thermodynamic consistency \cite{Lubarda2004}.

\item $J_2$ plasticity is predicated on volume-preserving plastic flow. In the present multiplicative formulation, this constraint is enforced at the algorithmic level, up to numerical tolerance, by combining a deviatoric exponential-map update with the isochoric projection \Eqref{eq:Fp_iso_proj} applied to $\mathbf{F}_p$ \cite{Lee1969,Simo1992,Simo1998}. In contrast, enforcing $\mathrm{tr}(\dot{\mathbf{E}}^p)=0$ in an additive framework does not, in general, guarantee plastic incompressibility at finite strains, particularly under complex loading paths involving large rotations \cite{Simo1998}.

\end{enumerate}
\end{remark}

\subsection{Total Lagrangian SPH Discretization}

In this subsection, we present how the governing equations (i.e., momentum balance and phase-field of fracture) are discretized employing a total Lagrangian SPH approach. Throughout this section, subscripts $i$ and $j$ denote particle labels, while superscripts $k$ and $q$ denote Cartesian component indices. We begin by discretizing the deformation gradient for particle $i$ based on the standard TLSPH difference-form derivative operator as
\begin{equation}
	\label{eq:deformGradTLSPH}
	\mathbf{F}_i = \mathbf{I}+ \frac{1}{\rho_{0i}}
	\sum_{j=1}^{N_{i}} m_{0j}~ (\mathbf{u}_{j}-\mathbf{u}_{i})\otimes \nabla_0 W_{0ij},
\end{equation}
or in index notation as
\begin{equation}
	\label{eq:deformGradTLSPHindex}
	F^{kq}_i = \delta^{kq} + \frac{1}{\rho_{0i}}
	\sum_{j=1}^{N_{i}} m_{0j}~ u^{k}_{ji}~\frac{\partial W_{0ij}}{\partial X_{j}^q}.
\end{equation}
Here, $N_{i}$ is the total number of particles located within the interpolation space, also known as the neighborhood, influence, or support domain, of particle $i$, and $\rho_{0i}$ is the density of particle $i$ in the undeformed configuration. $j$ is a neighbor particle to $i$, $m_{0j}$ is its mass in the undeformed state, $W_{0ij}$ is the kernel function relating particles $i$ and $j$, $X_{j}^q$ is the $q$ component of the initial coordinate of particle $j$, $u_{ji}^k$ is the $k$ component of the displacement difference vector $\mathbf{u}_{ji}=\mathbf{u}_{j}-\mathbf{u}_{i}$, and $\delta^{kq}$ is the Kronecker delta. $\partial W_{0ij}/ \partial X_{j}^q$ denotes the corrected derivative of the Lagrangian kernel. In the present implementation, a first-order kernel-gradient correction tensor is employed in the reference configuration, such that
\begin{equation}
\widetilde{\nabla}_0 W_{0ij} = \mathbf{L}_i \cdot \nabla_0 W_{0ij},
\end{equation}
where the correction tensor $\mathbf{L}_i$ is evaluated as
\begin{equation}
\mathbf{L}_i =
-\left(
\sum_{j=1}^{N_i}
\left(
\nabla_0 W_{0ij} \otimes \mathbf{r}_{0ij}
\right)
V_{0j}
\right)^{-1},
\end{equation}
with $\mathbf{r}_{0ij}=\mathbf{X}_{i}-\mathbf{X}_{j}$ and $V_{0j}$ denoting the volume of particle $j$ in the reference configuration. This correction restores first-order consistency of the SPH gradient approximation. For brevity, the tilde notation is omitted in the remainder of this work, and $\nabla_0 W_{0ij}$ or $\partial W_{0ij}/\partial X_j^q$ is henceforth understood to refer to the corrected kernel gradient. Full details of this correction procedure can be found in our earlier work \cite{rahimi2022Asmoothed}.

Similarly, the momentum balance equation for particle $i$ can be written as 
\begin{equation}
	\label{eq:MomentumEqTLSPH}
	\frac{d \mathbf{v}_i}{dt} = \sum_{j=1}^{N_{i}} 
	m_{0j}
	\left( 
	\frac{\mathbf{P}_i}{\rho^2_{0i}} +
	\frac{\mathbf{P}_j}{\rho^2_{0j}} +
	\mathbf{P}_{vij}
	\right)
	\cdot \nabla_0 W_{0ij} + \mathbf{f}_{0i},
\end{equation}
or in index notation as
\begin{equation}
	\label{eq:MomentumEqTLSPHindex}
	\frac{d v^k_i}{dt} = \sum_{j=1}^{N_{i}} 
	m_{0j}
	\left( 
	\frac{P^{kq}_i}{\rho^2_{0i}} +
	\frac{P^{kq}_j}{\rho^2_{0j}} +
	P^{kq}_{vij}
	\right)
	~\frac{\partial W_{0ij}}{\partial X_{j}^q} + f_{0i}^k,
\end{equation}
where Einstein's summation rule is employed for the repeated index $q$. $\mathbf{P}_i$ and $\mathbf{P}_j$ are the first Piola--Kirchhoff stress tensors for particles $i$ and $j$, respectively.

\begin{remark}
\Eqsref{eq:deformGradTLSPH}{eq:deformGradTLSPHindex} define the TLSPH approximation of the deformation gradient used in the constitutive update, whereas \Eqsref{eq:MomentumEqTLSPH}{eq:MomentumEqTLSPHindex} define the corresponding strong-form TLSPH discretization of the momentum balance. In the present work, the internal force in \Eqref{eq:MomentumEqTLSPH} is adopted in a symmetrized TLSPH form and is not derived from the first variation of a discrete particle potential associated with \Eqref{eq:deformGradTLSPH}. Therefore, exact variational consistency is not claimed. This type of strong-form TLSPH discretization is, however, common in particle-based solid mechanics and has been successfully employed in a variety of TLSPH formulations \cite{islam2019total,o2021fluid}. The results reported in those works, as well as in the present study, indicate that the approach remains accurate and robust for practical computations.
\end{remark}

Furthermore, an artificial viscosity term is included to mitigate numerical instabilities (e.g., spurious modes and shock-like jumps). Following \cite{monaghan1983374}, we define
\begin{equation}
\mathbf{P}_{vij} = \det(\mathbf{F}_{i})\,\pi_{ij}\,\mathbf{F}_{i}^{-1},
\end{equation}
with
\begin{equation}
    \label{eq:artViscCoef}
	\pi_{ij}=\frac{1}{\rho_{0i}}\left(\beta_2 G_{ij}^2-\beta_1\,c_{0i}G_{ij}\right),
\end{equation}
where
\begin{equation}
    \label{eq:speedofsound}
	c_{0i} = \sqrt{\frac{\kappa+\frac{4}{3}\mu}{\rho_{0i}}}
\end{equation}
is the (reference) sound speed, $\beta_1$ and $\beta_2$ are scalar coefficients, and
\begin{equation}
	G_{ij}=\frac{h\,(\mathbf{v}_{i}-\mathbf{v}_{j}) \cdot (\mathbf{X}_{i}-\mathbf{X}_{j})}{r^2_{0ij}+0.001\,h^2}.
\end{equation}
In the numerical implementation, the artificial-viscosity contribution is activated only for particle pairs in relative compression. More specifically, the quantity $G_{ij}$ is evaluated only when
\begin{equation}
(\mathbf{v}_i-\mathbf{v}_j)\cdot(\mathbf{x}_i-\mathbf{x}_j) < 0,
\end{equation}
and otherwise the artificial-viscosity term is set to zero. This restriction prevents nonphysical viscous action during particle separation. In the present work, artificial viscosity is used as a compact numerical regularization of compressive particle interactions within the TLSPH discretization. Since the artificial-viscosity coefficients employed are intentionally small, its effect on global conservation is weak in practice, while still providing useful suppression of spurious oscillations and some regularization against spurious zero-energy modes, thus improving overall robustness.

Finally, the phase-field equation for particle $i$ is discretized as
\begin{equation} \label{eq:phasefieldddot}
    \ddot{s}_i =\frac{c^2}{2 \epsilon_0}  \left ( 2 \epsilon_0 \nabla_0^2 s_i + \frac{1-s_i}{2 \epsilon_0} - \frac{2\sqrt{4 \epsilon_0 \mathcal{H}_i/G_{ci}+1}}{c} \dot{s}_i - 2 s_i \mathcal{H}_i/G_{ci} \right ) \text~{,}
\end{equation}
in which we have substituted $M_i = c/(2 \sqrt{4 G_{ci} \epsilon_0 \mathcal{H}_i + G_{ci}^2})$. The Laplacian of the phase field above is calculated using the SPH Laplacian operator as
\begin{equation}
	\label{eq:phasefieldLapLacian}
	\nabla_0^2 s_{i} = 2
	\sum_{j=1}^{N_i} (s_{i}-s_{j})~V_{0j}~ \frac{r^{q}_{0ij}}{|\mathbf{r}_{0ij}|^2}~\frac{\partial W_{0ij}}{\partial X_{j}^q},
\end{equation}
where $\mathbf{r}_{0ij} = \mathbf{X}_{i} - \mathbf{X}_{j}$ is the initial distance vector between particles $i$ and $j$. It should be further pointed out that when phase-field of fracture is employed in this work, the deformation gradient is computed as
\begin{equation} \label{eq:defGrad_phase}
    F_{i}^{kq} = \left\{\begin{array}{lr}
    \frac{1}{\rho_{0i}}
	\sum_{j=1}^{N_i} m_{0j}~ u^{k}_{ji}~\frac{\partial W_{0ij}}{\partial X_{j}^q}
    +\delta^{kq} & s_{i}>s_l\\ \delta^{kq} &\text{otherwise}\end{array}\right.\text~{,}
\end{equation}
where $s_l=0.1$ is the phase-field limit introduced to improve the numerical stability. Specifically, particles with $s_i \leq s_l$ are treated as highly damaged (``soft'') particles, for which the stress is already negligible. In this regime, replacing the deformation gradient by the identity stabilizes the computation without materially affecting the fracture response, since these particles contribute primarily through inertia rather than internal forces. As in our previous work \cite{rahimi2022Asmoothed}, $s_l$ is chosen as a small numerical threshold and should be kept as small as possible so as not to interfere with the fracture physics.

\subsection{Time Integration}
After spatial discretization, the semi-discrete system is advanced in time using explicit time integration schemes. Two alternatives are available: a Verlet scheme and a symplectic (leapfrog-type) scheme, presented in Algorithm~1 and Algorithm~2, respectively. The stable time step is selected to satisfy stability constraints accounting for
\begin{itemize}
    \item a CFL-type constraint based on wave propagation and particle velocity,
    \item an acceleration-based constraint associated with explicit time integration.
\end{itemize}
Accordingly, the time-step $\Delta t$ is chosen to satisfy
\begin{equation}
\Delta t \le C_{\mathrm{CFL}}\,
\min\!\left(
\Delta t_{v},\;
\Delta t_{a}
\right),
\label{eq:dt_criterion}
\end{equation}
where $C_{\mathrm{CFL}}$ is a user-defined factor. The individual bounds are defined as
\begin{align}
\Delta t_{v} &= \frac{h}{c + \lVert \mathbf{v}\rVert_{\max}},
\label{eq:dt_vel}\\[4pt]
\Delta t_{a} &= \sqrt{\frac{h}{\lVert \mathbf{a}\rVert_{\max}}},
\label{eq:dt_acc}
\end{align}
where $h$ denotes the SPH smoothing length (i.e., the kernel interaction radius), $\mathbf{v}$ is the particle velocity vector, and $\mathbf{a}$ is the particle acceleration vector. The quantities $\lVert \mathbf{v}\rVert_{\max}$ and $\lVert \mathbf{a}\rVert_{\max}$ denote the maximum magnitudes of particle velocity and acceleration, respectively, over the computational domain. As previously mentioned, $c$ denotes a reference material wave speed computed from the undeformed elastic constants and density. It is employed in the CFL restriction as a practical and conservative characteristic speed for explicit time integration, rather than as the exact instantaneous wave speed of the nonlinear constitutive response. For the hyperelastic models, it reflects the initial elastic stiffness, whereas for the finite-strain $J_2$ model it remains conservative since yielding reduces the effective tangent stiffness.

\begin{remark}
It should be noted that the present implementation does not employ a dedicated hourglass-control scheme. Numerical stabilization is achieved through the corrected TLSPH discretization, artificial viscosity, and a conservative explicit time-step restriction. Based on our previous experience, artificial viscosity provides a certain degree of numerical regularization against spurious zero-energy modes, although it is not as explicit or systematic as a dedicated hourglass-control scheme. For the benchmark cases considered in this work, the adopted strategy proved sufficient in practice. We nevertheless note that alternative stabilization approaches exist in the recent SPH literature, including dedicated hourglass-control techniques and Riemann-based structural stabilization procedures \cite{ganzenmuller2015hourglass,zhan2019stabilized,shimizu2022implicit,khayyer2024improved}.
\end{remark}

\begin{algorithm}[H]
\caption{Verlet Time Integration (single step for a specific particle)}
\begin{algorithmic}
\State \textbf{Given:} $\mathbf{u}_n$, $\mathbf{v}_n$, $\mathbf{a}_n$, $s_n$, $\dot{s}_n$, $\ddot{s}_n$

\subr{Step 1: Contact Forces}
\qIf{N_{\mathrm{bodies}}}{>}{1}
    \State \hspace{2em} Compute contact forces between bodies
\EndqIf
\Statex

\subr{Step 2: Internal Forces}
\State Compute deformation gradient
$\mathbf{F}$
\State Compute strain
$\mathbf{E} = \frac{1}{2}\left(\mathbf{F}^T \mathbf{F} - \mathbf{I}\right)$
\State Compute stress $\mathbf{S}$ through an appropriate constitutive model
\qIf{\texttt{fracture}}{=}{\texttt{true}}
    \State \hspace{2em} Compute phase-field second derivative $\ddot{s}_{n+1}$
    \State \hspace{2em} Degrade stress $\mathbf{S} \gets s_n^2 \mathbf{S}^{+} + \mathbf{S}^{-}$
\EndqIf
\State Compute first Piola--Kirchhoff stress $\mathbf{P} = \mathbf{F} \mathbf{S}$
\State Compute acceleration
$\mathbf{a}$
\State Add artificial viscosity $\mathbf{a} \gets \mathbf{a} + \mathbf{a}^{\mathrm{visc}}$
\State Add external forces $\mathbf{a} \gets \mathbf{a} + \mathbf{f}^{\mathrm{ext}}$
\Statex

\subr{Step 3: Apply Boundary Conditions}
\State Apply velocity BCs (modify $\mathbf{v}$ directly)
\State Apply force BCs (add to $\mathbf{a}$)
\Statex

\subr{Step 4: Update}
\State $\mathbf{v}_{n+1} = \mathbf{v}_n + \Delta t\,\mathbf{a}_n$
\State $\mathbf{u}_{n+1} = \mathbf{u}_n + \Delta t\,\mathbf{v}_{n+1}$
\State $\mathbf{x}_{n+1} = \mathbf{X} + \mathbf{u}_{n+1}$
\qIf{\texttt{fracture}}{=}{\texttt{true}}
    \State \hspace{2em} $\dot{s}_{n+1} = \dot{s}_n + \Delta t \ddot{s}_{n+1}$
    \State \hspace{2em} $s_{n+1} = s_n + \Delta t \dot{s}_{n+1}$
\EndqIf
\Statex

\subr{Step 5: Update Global Arrays}
\State Copy updated positions/velocities to main particle arrays
\end{algorithmic}
\end{algorithm}

\begin{algorithm}[H]
\caption{Symplectic (Leapfrog) Time Integration (single step for a specific particle)}
\begin{algorithmic}
\State \textbf{Given:} $\mathbf{u}_n$, $\mathbf{v}_{n}$, $\mathbf{a}_{n-\frac{1}{2}}$, $s_n$, $\dot{s}_n$, $\ddot{s}_{n-\frac{1}{2}}$

\subr{Step 1: Predictor}
\State $\mathbf{v}_{n+\frac{1}{2}} = \mathbf{v}_{n} + \frac{\Delta t}{2}\mathbf{a}_{n-\frac{1}{2}}$
\State $\mathbf{u}_{n+\frac{1}{2}} = \mathbf{u}_n + \frac{\Delta t}{2}\mathbf{v}_{n}$
\State $\mathbf{x}_{n+\frac{1}{2}} = \mathbf{X} + \mathbf{u}_{n+\frac{1}{2}}$
\qIf{\texttt{fracture}}{=}{\texttt{true}}
    \State \hspace{2em} $\dot{s}_{n+\frac{1}{2}} = \dot{s}_n + \frac{\Delta t}{2} \ddot{s}_{n-\frac{1}{2}}$
    \State \hspace{2em} $s_{n+\frac{1}{2}} = s_n + \frac{\Delta t}{2} \dot{s}_{n}$
\EndqIf
\Statex

\subr{Step 2: Contact Forces}
\qIf{N_{\mathrm{bodies}}}{>}{1}
    \State \hspace{2em} Compute contact forces between bodies
\EndqIf
\Statex

\subr{Step 3: Internal Forces} (at the $n+\frac{1}{2}$ level)
\State Compute deformation gradient
$\mathbf{F}$
\State Compute strain
$\mathbf{E} = \frac{1}{2}\left(\mathbf{F}^T \mathbf{F} - \mathbf{I}\right)$
\State Compute stress $\mathbf{S}$ through an appropriate constitutive model
\qIf{\texttt{fracture}}{=}{\texttt{true}}
    \State \hspace{2em} Compute phase-field second time derivative $\ddot{s}_{n+\frac{1}{2}}$
    \State \hspace{2em} Degrade stress $\mathbf{S} \gets s_{n+\frac{1}{2}}^2 \mathbf{S}^{+} + \mathbf{S}^{-}$
\EndqIf
\State Compute first Piola--Kirchhoff stress $\mathbf{P} = \mathbf{F} \mathbf{S}$
\State Compute acceleration
$\mathbf{a}_{n+\frac{1}{2}}$
\State Add artificial viscosity
$\mathbf{a}_{n+\frac{1}{2}} \gets \mathbf{a}_{n+\frac{1}{2}} + \mathbf{a}_{\mathrm{visc}}$
\State Add external forces
$\mathbf{a}_{n+\frac{1}{2}} \gets \mathbf{a}_{n+\frac{1}{2}} + \mathbf{f}_{\mathrm{ext}}$
\Statex

\subr{Step 4: Apply Boundary Conditions}
\State Apply velocity BCs (modify $\mathbf{v}$ directly)
\State Apply force BCs (add to $\mathbf{a}$)
\Statex

\subr{Step 5: Corrector}
\State $\mathbf{v}_{n+1} = \mathbf{v}_{n} + \Delta t \, \mathbf{a}_{n+\frac{1}{2}}$
\State $\mathbf{u}_{n+1} = \mathbf{u}_{n+\frac{1}{2}} + \frac{\Delta t}{2}\mathbf{v}_{n+1}$
\State $\mathbf{x}_{n+1} = \mathbf{X} + \mathbf{u}_{n+1}$
\qIf{\texttt{fracture}}{=}{\texttt{true}}
    \State \hspace{2em} $\dot{s}_{n+1} = \dot{s}_{n} + \Delta t \ddot{s}_{n+\frac{1}{2}}$
    \State \hspace{2em} $s_{n+1} = s_{n+\frac{1}{2}} + \frac{\Delta t}{2} \dot{s}_{n+1}$
\EndqIf
\Statex

\subr{Step 6: Update Global Arrays}
\State Copy updated positions/velocities to main particle arrays
\end{algorithmic}
\end{algorithm}

\section{DualSPHysics Architecture and Extension to SoliDualSPHysics} \label{sec:dualspharc}

DualSPHysics is a modular, GPU-accelerated open-source SPH framework designed for high-performance simulations of fluid dynamics and multi-physics problems \cite{dominguez2022dualsphysics,crespo2015dualsphysics,o2021fluid}. Its architecture supports solvers for single- and multi-phase flows, fluid--structure interaction (FSI), and coupling with external libraries such as Project Chrono and MoorDyn. Limited support for deformable structures has also been introduced \cite{o2021fluid}; however, these features are primarily restricted to small-strain elasticity within FSI settings. The present work extends DualSPHysics to enable standalone simulations of hyperelasticity, finite-strain plasticity, and brittle fracture in deformable solids, together with complex boundary conditions and user-defined expressions. In what follows, we first outline the structure of the original framework and then describe how the governing equations and total Lagrangian SPH discretization presented in \autoref{sec:formulation} are incorporated through the newly developed solid mechanics and phase-field fracture solvers.

\subsection{DualSPHysics Framework}
\begin{figure}[!htbp]
    \centering
    \includegraphics[width=1.0\linewidth]{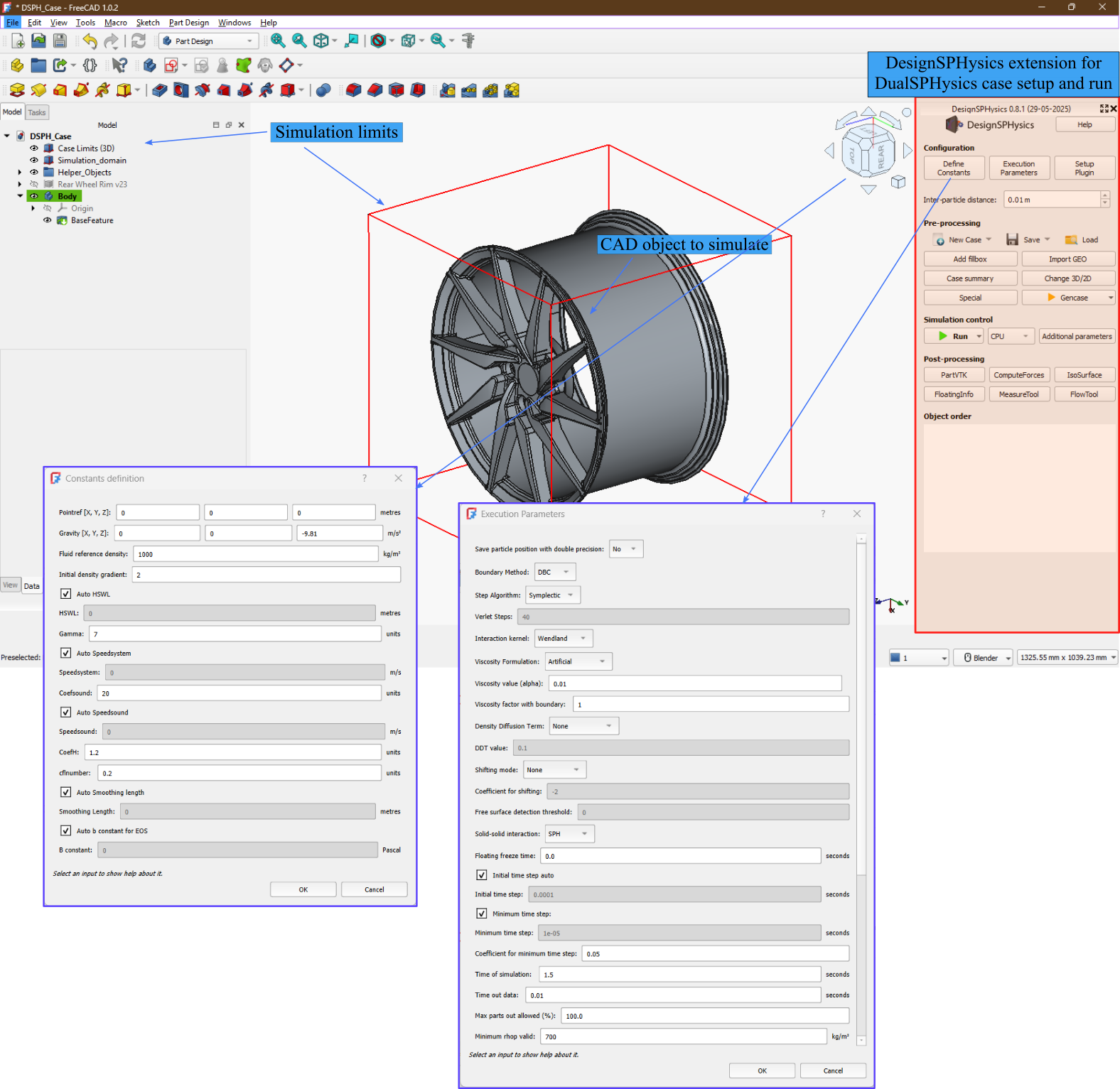}
    \caption{DesignPhysics FreeCAD plugin for DualSPHysics. The GUI enables interactive setup of fluid simulation cases; support for deformable solid mechanics is not included in the standard release.}
    \label{fig:FreeCadGUI}
\end{figure}

The base DualSPHysics code includes the following solvers and features:
\begin{itemize}
    \item \textbf{Single-phase free-surface flows}: Capability to simulate dam breaks, wave propagation, and open-channel flows.
    \item \textbf{Multi-phase systems}: Support for Newtonian--Newtonian, gas--liquid, and liquid--granular interactions.
    \item \textbf{Coupled physics}: Integration with the Discrete Element Method (DEM), Project Chrono (rigid-body dynamics), and MoorDyn (mooring systems).
    \item \textbf{Fluid boundary conditions}: Support for walls, periodic domains, inflow/outflow boundaries, and advanced viscosity models (laminar and LES turbulence).
    \item \textbf{Pre- and post-processing tools}: GenCase (particle generation), the DesignPhysics FreeCAD GUI, and PartVTK/IsoSurface utilities for visualization in ParaView.
\end{itemize}
The DesignPhysics FreeCAD GUI provides a user-friendly interface and supports various input options but does not currently include support for deformable solid mechanics, as shown in \autoref{fig:FreeCadGUI}. The framework employs a hybrid CPU--GPU architecture, using CUDA kernels for core SPH operations (e.g., neighbor search and force calculations) and OpenMP for CPU parallelization. Its modular design facilitates extensions while maintaining compatibility with existing solvers.

\newpage
\subsection{Base DualSPHysics XML Input}
DualSPHysics simulations are configured via XML files (\autoref{fig:htmlinputdualsph}), structured as:
\begin{lstlisting}[style=xmlstyle]
<constantsdef> ... </constantsdef>
<geometry> ... </geometry>
<motion> ... </motion>
<execution> ... </execution>
<parameters> ... </parameters>
\end{lstlisting}

\begin{figure}[!htbp]
    \centering
    \begin{lstlisting}[style=xmlstyle]
<?xml version="1.0" encoding="UTF-8" ?>
<case app="DesignSPHysics">
  <casedef>
    <constantsdef>
      ...
    </constantsdef>
    </mkconfig>
    <geometry>
      <definition dp="0.004">
        ...
      </definition>
      <commands>
        <mainlist>
        <setmkbound mk="0"/>
        <drawfilestl file="Wheel.stl" />
        </mainlist>
      </commands>
    </geometry>
  </casedef>
  <execution>
    <special>
    ...
    </special>
    <parameters>
    ...  
    </parameters>
  </execution>
</case>
\end{lstlisting}
    \caption{Structure of a DualSPHysics XML input file. In the standard framework, solid geometries are treated as rigid boundaries; modeling deformable structures requires the extensions introduced in this work.}
    \label{fig:htmlinputdualsph}
\end{figure}
Case constants such as gravity, reference fluid density, smoothing length, and CFL coefficient are defined in the \icard{constantsdef} section. The \icard{geometry} section defines the computational domain using CAD files or built-in commands for simple geometries, as documented in the official DualSPHysics documentation. By default, solid geometries are treated as rigid boundaries within the DualSPHysics FSI solver. Therefore, any moving or deformable structure must be defined as a boundary with prescribed motion in the \icard{motion} section. Definitions related to special options, such as inlet/outlet conditions, Project Chrono and MoorDyn coupling, and piston motion, are specified in the \icard{special} section. The \icard{parameters} section includes numerical settings such as the kernel type and time-integration constants.

\subsection{Extensions for Solid and Fracture Mechanics in \solidname}
The development of a graphical user interface to support user-defined expressions, solid dynamics, plasticity, and fracture mechanics is beyond the scope of the present work and is left for future development. At this stage, all functionality is configured through the XML input files. To enable user-defined expressions, deformable solid mechanics, plasticity, and fracture mechanics, new input blocks are introduced under \icard{special}\icard{mathexpressions} and \icard{special}\icard{deformstrucs}. The overall workflow is outlined below.

\begin{figure}[!htbp]
  \centering
   \begin{subfigure}[b]{0.8\linewidth}
     \begin{lstlisting}[style=xmlstyle]
<geometry>
  ...
  <commands>
    <mainlist>
    <setshapemode>actual | dp | bound</setshapemode>
    <setmkbound mk="0"/>
    <drawfilestl file="Wheel.stl" />
    <setmkbound mk="1"/>
    <drawfilestl file="Tube.stl"/>
    </mainlist>
  </commands>
</geometry>
<motion>
  <objreal ref="0">
    <begin mov="1" start="0" />
    <mvnull id="1" />
  </objreal>
</motion>
...
<special>
  <deformstrucs>
    <deformstrucbody mkbound="0">
      <bcforce type="2" mkid="1" y="5.0e6" comment="Traction boundary condition applied as 5 MPa in y direction" />
      ...
    </deformstrucbody>
  </deformstrucs>
</special>
\end{lstlisting}
     \caption{}
   \end{subfigure}
   \begin{subfigure}[b]{0.7\linewidth}
         \includegraphics[width=\linewidth]{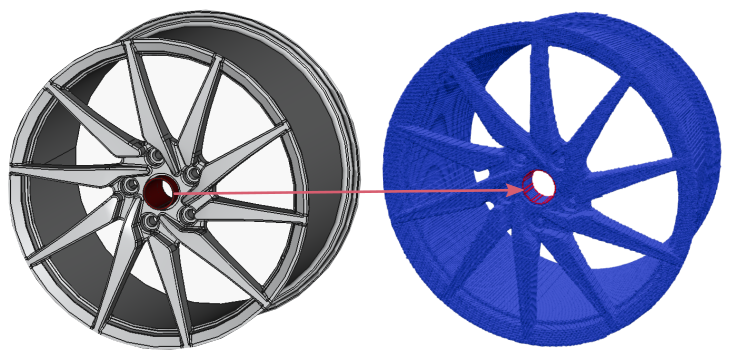}
         \caption{}
     \end{subfigure}
    \caption{(a) Geometry and boundary condition definitions in XML input file in \solidname. (b) Visual representation of geometry and boundary conditions in \solidname.}
    \label{fig:htmlinputdualsphfracture}
\end{figure}

First, the geometry of interest is defined in the \icard{geometry} section (e.g., via CAD files) using distinct \texttt{mk} identifiers. Deformable bodies are then flagged as moving objects in the \icard{motion} section. Once defined, boundary conditions can be applied either directly through user-defined expressions or indirectly through proximity-based auxiliary geometries. For instance, consider a car wheel deforming under pressure from an axle. The axle may be introduced as an auxiliary rigid geometry positioned within a distance \texttt{dp} of the intended contact region, as illustrated in \autoref{fig:htmlinputdualsphfracture}. Boundary conditions for each \icard{deformstrucbody} can be prescribed either as velocity constraints or as applied surface tractions.

For force-type loading, \solidname\ provides two equivalent specification routes:
\begin{enumerate}
    \item direct prescription of surface traction (e.g., via a user-defined expression), or
    \item definition of an auxiliary rigid body whose proximity/contact with the deformable structure generates an equivalent traction-like load.
\end{enumerate}
The second option is intended solely as a mechanism to impose an external load and should not be confused with general contact interactions between deformable bodies. The material properties and main structural parameters are specified within the \icard{execution} section under the \icard{special}\icard{deformstrucs} subsection for each \icard{deformstrucbody mkbound="mkid"}, as shown in \autoref{fig:htmlinputdeformstruc}.

\begin{figure}[!htbp]
    \centering
         \begin{lstlisting}[style=xmlstyle]
<deformstrucs>
<deformstrucbody mkbound="0">
  <density value="7800.0" comment="Mass density" units_comment="kg/m^3" />
  <youngmod value="210.0e9" comment="Young's Modulus" units_comment="Pa" />
  <poissratio value="0.3" comment="Poisson's ratio" />
  <constitmodel value="1" comment="Constitutive model 1:SVK" />
  <artvisc factor1="0.2" factor2="0.0" comment=" Art. visc. factors" />
  <mapfac value="4" comment="Map factor: x4 refinement" />
  ...
</deformstrucbody>
</deformstrucs>
\end{lstlisting}
    \caption{Properties of deformable structures in \solidname.}
    \label{fig:htmlinputdeformstruc}
\end{figure}

\subsubsection{Modeling Fracture}
To enable fracture modeling via the phase-field approach, the input cards shown in \autoref{fig:htmlinputdeformstrucfrac} must be specified. It should be noted that fracture modeling is not compatible with $J_2$ plasticity. If $J_2$ plasticity is activated (\icard{constitmodel value="3"}), the fracture option is automatically disabled.
\begin{figure}[!htbp]
    \centering
\begin{lstlisting}[style=xmlstyle]
<deformstrucbody mkbound="1">
  ...
  <!--Enables fracture. Default: false. Not available with J2 plasticity.-->
  <fracture value="true" comment="Enables fracture" />
  <!--Critical energy release rate. Required if fracture is enabled.-->
  <Gc value="2700.0" comment="Fracture toughness  [J/m^2]" />
  <!-- Phase-field length scale. Required if fracture is enabled.-->
  <pflenscale value="0.002" comment="Length scale [m]" />
  <!-- Phase-field limit for soft particles. Default: 0.1.-->
  <pflim value="0.1" comment="Phase-field threshold for soft particles" />
  <!-- Pre-existing notch plane defined by 4 points in 3D.-->
  <!-- Up to 512 notches can be applied to a body.-->
  <notch>  
    <p1 x="-2.0e-3" y="-5.0e-3" z="0.02" />
    <p2 x="50.0e-3" y="-5.0e-3" z="0.02" />
    <p3 x="50.0e-3" y="25.0e-3" z="0.02" />
    <p4 x="-2.0e-3" y="25.0e-3" z="0.02" />
  </notch>
</deformstrucbody>
\end{lstlisting}
\caption{Fracture configuration for deformable solids.}
    \label{fig:htmlinputdeformstrucfrac}
\end{figure}

\subsubsection{Application of Boundary Conditions}
Velocity and force boundary conditions can be applied using the
\icard{bcvel} and \icard{bcforce} input cards, respectively, as shown in \autoref{fig:bcvelbcforinput}.
\begin{figure}[!htbp]
    \centering
\begin{lstlisting}[style=xmlstyle]
<deformstrucbody mkbound="1">
  ...
  <!--Force boundary condition type 1: point force [N], 2: surface distributed [N/m^2], 3: acceleration.-->
  <bcforce type="..." x="..." ye="2" ze="1" tst="..." tend="..." comment="Force boundary condition, x is constant value, y and z are expressions id=2 and expression id=1, respectively. Applied between times tst and tend" />
  <bcforce mkid="4" xe="2" comment="Applies force type=2 on surface near body with mk_bound=4 in x direction according to expression id=2" />

  <!--Velocity boundary condition.-->
  <bcvel xe="2" ye="2" z="0.0" tst="..." tend="..." comment="Velocity boundary condition, x and y are expression id=2, z is constant value. Applied between times tst and tend to the entire body" />
  <bcvel mkid="4" xe="2" ye="2" z="0.0" tst="..." tend="..." comment="Velocity boundary condition, x and y are expression id=2, z is constant value. Applied between times tst and tend to particles on surface near body with mk_bound=4" />
</deformstrucbody>
\end{lstlisting}
\caption{Input format for velocity and force boundary conditions.}
    \label{fig:bcvelbcforinput}
\end{figure}
Each of these sections can have the following attributes:
\begin{itemize}
    \item \textbf{mkid}: The \texttt{mkbound} identifier of the auxiliary geometry from which the boundary condition is applied. If not specified, the condition is applied to the entire deformable structure body.
    \item \textbf{x}: If provided, applies the given value to the $x$-component of velocity or force.
    \item \textbf{y}: If provided, applies the given value to the $y$-component of velocity or force.
    \item \textbf{z}: If provided, applies the given value to the $z$-component of velocity or force.
    \item \textbf{xe}: If provided, applies a user-defined expression to the $x$-component of velocity or force.
    \item \textbf{ye}: If provided, applies a user-defined expression to the $y$-component of velocity or force.
    \item \textbf{ze}: If provided, applies a user-defined expression to the $z$-component of velocity or force.
    \item \textbf{tst}: Starting time of the boundary condition. If not specified, the condition is applied from the beginning of the simulation.
    \item \textbf{tend}: Ending time of the boundary condition. If not specified, the condition remains active for the entire duration of the simulation.
    \item \textbf{type}: (Force input card only.) Specifies the type of force boundary condition: (1: point force [N], 2: surface-distributed force $[\text{N}/\text{m}^2]$, 3: acceleration $[\text{m}/\text{s}^2]$).
\end{itemize}
\subsubsection{Measuring Quantities}
The input card shown in \autoref{fig:htmlinputmeasplane} can be used to measure quantities such as average displacement and total force on particles located on or near a surface defined by four points.
\begin{figure}[!htbp]
    \centering
    \begin{minipage}{0.49\linewidth}
\begin{lstlisting}[style=xmlstyle] 
<measureplane>  
  <p1 x="..." y="..." z="..." />
  <p2 x="..." y="..." z="..." />
  <p3 x="..." y="..." z="..." />
  <p4 x="..." y="..." z="..." />
</measureplane>
\end{lstlisting}
\end{minipage}
\caption{Input card for measurements on a specific surface.}
    \label{fig:htmlinputmeasplane}
\end{figure}
At present, each \icard{measureplane} supports output of average displacement and total force only. The implementation can be extended in the future to include additional quantities, such as energy or stress.

\subsubsection{Full List of Keywords in deformstrucs Class}
The complete set of supported keywords and their functionalities is summarized in \autoref{fig:htmlinputdeformstrucfulllist}.

\begin{lstlisting}[style=xmlstyle, caption={Full list of input cards for modeling deformable structures in \solidname.}, label={fig:htmlinputdeformstrucfulllist}]
<deformstrucs>
 <!-- If given, will overwrite the adaptive timestep for deformstrucbodies -->
 <timestep value="..." comment="User-defined time step value" />
 <!-- If given, will multiply the contact potential by the given factor -->
 <contcoeff value="..." comment="Contact potential multiplier" />
 <deformstrucbody mkbound="1">
  <bcforce ... comment="Force boundary condition" />
  <bcvel ... comment="Velocity boundary condition" />
  <mapfac value="4" comment="x4 refinement, default: 1" />
  <nbsrange value="1" comment="If provided, restricts neighbor search to 1 particle in each direction. Default is original support length" />
  <density value="..." comment="Mass density [kg/m^3]" />
  <artvisc factor1="..." factor2="..." comment="Artificial viscosity factors, default: factor1=0.2 factor2=0" />
  <!-- Elastic constants - Option 1: Young's modulus and Poisson's ratio -->
  <youngmod value="..." comment="Young's Modulus [Pa]" />
  <poissratio value="..." comment="Poisson's ratio [-]" />
  
  <!-- Elastic constants - Option 2: Lame parameters directly -->
  <!-- <u_lambda value="..." comment="Lame lambda [Pa]" /> -->
  <!-- <u_mu value="..." comment="Lame mu (shear modulus) [Pa]" /> -->
  
  <!-- Elastic constants - Option 3: Bulk and shear modulus -->
  <!-- <u_bulk value="..." comment="Bulk modulus [Pa]" /> -->
  <!-- <u_mu value="..." comment="Shear modulus [Pa]" /> -->
  
  <!-- Constitutive model: 1=SVK (default), 2=neo--Hookean, 3=J2 -->
  <constitmodel value="2" comment="1:SVK, 2:neo--Hookean, 3:J2 plasticity" />
  
  <!-- For J2 plasticity only (required if constitmodel=3) -->
  <yieldstress value="..." comment="Initial yield stress [Pa]" />
  <hardening value="..." comment="Hardening modulus [Pa]" />

  <!-- Used for multi-body contact between deformstrucs -->
  <restcoef value="..." comment="Restitution coefficient, default=1.0" />
  <kfric value="..." comment="Friction coefficient, default=0.0" />
  
  <!--Enables fracture. Default: false. Not available with J2 plasticity.-->
  <fracture value="true" comment="Enables fracture" />
  <!-- If given, will restrict the damage of particles satisfying the expression to the value of the expression (can only be between 0 and 1)-->
  <restrictphi value="expressionID" comment="Restricts phase-field propagation according to expression ID"/>
  <!--Critical energy release rate. Required if fracture is enabled.-->
  <Gc value="..." comment="Fracture toughness  [J/m^2]" />
  <!-- Phase-field length scale. Required if fracture is enabled.-->
  <pflenscale value="..." comment="Length scale [m]" />
  <!-- Phase-field limit for soft particles. Default: 0.1.-->
  <pflim value="..." comment="Phase-field threshold for soft particles" />
  <!-- Pre-existing notch surfaces defined by 4 points in 3D.-->
  <!-- Up to 512 notches can be applied to a body.-->
  <notch>  
  ...
  </notch>
  <notch>  
  ...
  </notch>

  <!-- Measuring surfaces defined by 4 points in 3D.-->
  <!-- Up to 512 measureplanes can be applied to a body.-->
  <measureplane>  
  ...
  </measureplane>
  <measureplane>  
  ...
  </measureplane>
 </deformstrucbody>

 <deformstrucbody mkbound="2">
  ...
 </deformstrucbody>

 <deformstrucbody mkbound="3">
  ...
 </deformstrucbody>
 ...
</deformstrucs>
\end{lstlisting}

\subsection{User-Defined Expressions}
\solidname\ supports user-defined mathematical and logical expressions within the \icard{special}\icard{mathexpressions} section. The input format is shown in \autoref{fig:userdefinexpput}.
\begin{figure}[!htbp]
  \centering
  \begin{lstlisting}[style=xmlstyle]
<execution>
  <special>
  <mathexpressions>
    <userexpression id="1" comment="Math expression">
    <locals value="L0=0.2; kw=9.375; cs=57.0"/>
    <expression value="if(x0<=0.0,0.0,if(t<=0.0,0.01 * cs * ((cos(kw*L0)+cosh(kw*L0))*(cosh(kw*x0)-cos(kw*x0)) + (sin(kw*L0)-sinh(kw*L0))*(sinh(kw*x0)-sin(kw*x0)))/ ((cos(kw*L0)+cosh(kw*L0))*(cosh(kw*L0)-cos(kw*L0)) + (sin(kw*L0)-sinh(kw*L0))*(sinh(kw*L0)-sin(kw*L0))),skip))"/>
    </userexpression>
    <userexpression id="2" comment="Math expression">
    <expression value="if(x0<0.0,0.0,skip)"/>
    </userexpression>
  </mathexpressions>
  </special>
</execution>
    \end{lstlisting}
    \caption{Input format for user-defined space- and time-dependent mathematical and logical expressions.}
    \label{fig:userdefinexpput}
\end{figure}
The following features are supported:
\begin{itemize}
    \item Built-in variables: initial position (\texttt{x0}, \texttt{y0}, \texttt{z0}), current position (\texttt{x}, \texttt{y}, \texttt{z}), displacement (\texttt{ux}, \texttt{uy}, \texttt{uz}), time (\texttt{t}), timestep (\texttt{dt}), and particle spacing (\texttt{dx}).
    
    \item Arithmetic operators: $+,~-,~*,~/$.
    
    \item Mathematical functions: \texttt{log}, \texttt{ln}, \texttt{pow}, \texttt{sqrt}, \texttt{abs}.
    
    \item Trigonometric and hyperbolic functions: \texttt{sin}, \texttt{cos}, \texttt{tan}, \texttt{cot}, \texttt{sinh}, \texttt{cosh}, \texttt{tanh}, \texttt{coth}.
    
    \item Logical operators: $<,~>,~<=,~>=,~==,~!=$, \texttt{and}, \texttt{or}.
    
    \item Nested \texttt{if} clauses.
    
    \item Local variable definitions via the \texttt{<locals>} tag.
    
    \item The keyword \texttt{Skip} or \texttt{skip}, which causes a branch of an \texttt{if} clause to perform no operation. This is particularly useful when applying expressions to only a subset of particles.
\end{itemize}

The \texttt{<userexpression>} block with $\text{id}=2$ shown in \autoref{fig:userdefinexpput} assigns a constant zero value to particles satisfying $X_1 < 0$, while the block with $\text{id}=1$ corresponds to
\begin{equation} 
    F(X_1,t)=\left\{\begin{array}{lr} 0.0 & X_1 \le 0.0\\
    0.01 c_s \frac{f(X_{1})}{f(L_0)}&X_1>0.0~and~t \le 0.0\\
    \text{do~nothing} & \text{otherwise,}
    \end{array}\right.
\end{equation}
where the function $f(x)$ is
\begin{equation}
\begin{aligned}
f(x) = & \,[\cos(k_w L_0) + \cosh(k_w L_0)] \,[\cosh(k_w x) - \cos(k_w x)] \\
& + [\sin(k_w L_0) - \sinh(k_w L_0)] \,[\sinh(k_w x) - \sin(k_w x)] .
\end{aligned}
\end{equation}
This example corresponds to the boundary conditions of a freely oscillating cantilever beam, where an initial velocity field is prescribed while the left end remains constrained throughout the simulation. \autoref{tab:expr_variables_detailed}--\autoref{tab:expr_operators_detailed} summarize the supported variables, functions, and operators.

\begin{table}[H]
\centering
\caption{Built-in expression variables}
\label{tab:expr_variables_detailed}
\begin{tabular}{lll}
\toprule
\textbf{Variable} & \textbf{Description} & \textbf{Units} \\
\midrule
\texttt{x0} & Initial $X$-position & m\\
\texttt{y0} & Initial $Y$-position & m\\
\texttt{z0} & Initial $Z$-position & m\\
\texttt{x} & Current $X$-position & m\\
\texttt{y} & Current $Y$-position & m\\
\texttt{z} & Current $Z$-position & m\\
\texttt{ux} & $X$-displacement & m\\
\texttt{uy} & $Y$-displacement & m\\
\texttt{uz} & $Z$-displacement & m\\
\texttt{t} & Current time & s\\
\texttt{dt} & Current time step & s\\
\texttt{dx} & Particle spacing & m\\
\bottomrule
\end{tabular}
\end{table}

\begin{table}[H]
\centering
\caption{Mathematical functions}
\label{tab:expr_functions_detailed}
\begin{tabular}{llll}
\toprule
\textbf{Function} & \textbf{Arguments} & \textbf{Description} & \textbf{Example} \\
\midrule
\texttt{sin(x)} & 1 & Sine & \texttt{sin(2*3.14159*t)} \\
\texttt{cos(x)} & 1 & Cosine & \texttt{cos(omega*t)} \\
\texttt{tan(x)} & 1 & Tangent & \texttt{tan(angle)} \\
\texttt{sinh(x)} & 1 & Hyperbolic sine & \texttt{sinh(x0)} \\
\texttt{cosh(x)} & 1 & Hyperbolic cosine & \texttt{cosh(x0)} \\
\texttt{tanh(x)} & 1 & Hyperbolic tangent & \texttt{tanh(x0/L)} \\
\texttt{cot(x)} & 1 & Cotangent & \texttt{cot(angle)} \\
\texttt{coth(x)} & 1 & Hyperbolic cotangent & \texttt{coth(x0)} \\
\texttt{sqrt(x)} & 1 & Square root & \texttt{sqrt(x0*x0+y0*y0)} \\
\texttt{log(x)} & 1 & Base-10 logarithm & \texttt{log(10)} \\
\texttt{ln(x)} & 1 & Natural logarithm & \texttt{ln(J)} \\
\texttt{pow(x,y)} & 2 & Power: $x^y$ & \texttt{pow(r,1.5)} \\
\texttt{abs(x)} & 1 & Absolute value & \texttt{abs(ux)} \\
\texttt{if(c,t,f)} & 3 & Conditional & \texttt{if(x0<0.5,1,0)} \\
\bottomrule
\end{tabular}
\end{table}

\begin{table}[H]
\centering
\caption{Expression operators with precedence}
\label{tab:expr_operators_detailed}
\begin{tabular}{llll}
\toprule
\textbf{Operator} & \textbf{Precedence} & \textbf{Associativity} & \textbf{Description} \\
\midrule
\texttt{or} & 1 & Left & Logical OR \\
\texttt{and} & 2 & Left & Logical AND \\
\texttt{<, >, <=, >=, ==, !=} & 3 & Left & Comparison \\
\texttt{+, -} & 4 & Left & Addition, Subtraction \\
\texttt{*, /} & 5 & Left & Multiplication, Division \\
\texttt{\^{}} & 6 & Right & Exponentiation \\
\bottomrule
\end{tabular}
\end{table}

\subsection{Deformable-Structure Output Files}
\solidname\ generates the following output files for deformable-structure simulations:

\begin{itemize}
    \item \textbf{Domain information}: Information such as TLSPH cell division, number of neighbors, and initial particle velocities is written to the \texttt{TLSPH\_INFO} directory.
    
    \item \textbf{Energy tracking}: The file \texttt{DeformStruc\_Energies.csv} records the strain energy, kinetic energy, and fracture or plastic energy of each deformable structure over time.
    
    \item \textbf{Measurements}: The files \texttt{MeasuringPlData*.csv} contain measurement data associated with each \icard{measureplane}.
    
    \item \textbf{VTK files}: The directory \texttt{DeformStruc/} contains VTK files with particle displacements, Cauchy stress components, and phase-field or plastic strain variables.
\end{itemize}

\subsection{SoliDualSPHysics CPU/GPU Implementation}
The core algorithms and C++ functions for deformable structures are implemented in the \solidname\ codebase within the files \texttt{JSphCpuSingle.h/.cpp} and \texttt{JSphGpuSingle.h/.cpp}, corresponding to OpenMP-based CPU and GPU parallel execution, respectively. The CUDA kernels are implemented in \texttt{JSphGpu\_DefStruc\_ker.h/.cu}, while additional CUDA-optimized macros and inline functions are defined in \texttt{TypesDef\_GPU.h}. The files \texttt{JSphCpu\_ExpressionParser.h/.cpp} and \texttt{JSphGpu\_ExpressionParser.h/.cu} contain the implementation and compilation logic for user-defined mathematical expressions on CPU and GPU architectures, respectively. The CPU implementation follows the same algorithmic structure as the formulation and time-integration procedures described in \autoref{sec:formulation}. In each time step, the solver updates the deformable-structure state by evaluating the deformation gradient, computing the constitutive response, assembling internal and artificial-viscosity forces, applying prescribed velocity and force boundary conditions, advancing the kinematic variables, and writing the requested output quantities. The GPU implementation follows the same logical workflow and naming convention, with computationally intensive operations parallelized across CUDA thread blocks.

\subsection{Features and Capabilities of \solidname}

The main features and capabilities of \solidname\ are summarized below:

\begin{enumerate}
    \item Support for user-defined time- and space-dependent mathematical and logical expressions.
    
    \item Flexible velocity and traction boundary conditions, applicable globally or to selected structural groups.
    
    \item Hybrid CPU--GPU parallel implementation based on OpenMP and CUDA.
    
    \item Multi-resolution capability allowing independent particle-spacing refinement for each deformable structure body.
    
    \item Independent time-step refinement for solid mechanics, enabling optimal stability constraints without interfering with other time-integration parameters.
    
    \item Support for hyperelastic constitutive models (St. Venant--Kirchhoff and neo--Hookean) and finite-strain $J_2$ plasticity with isotropic hardening.
    
    \item Phase-field modeling of brittle fracture based on a hyperbolic formulation.
    
    \item Tension-driven crack propagation enforced through  decomposition of the strain tensor.
    
    \item Definition of pre-existing cracks using geometric notch specifications.
    
    \item Local measurement of displacements and forces on user-defined surface regions.
    
    \item DEM-based contact interaction between deformable structures, leveraging the existing DEM contact implementation available in DualSPHysics.
\end{enumerate}

\section{Numerical Examples}
\label{sec:2dcases}

In this section, we demonstrate the capabilities of \solidname through a set of numerical examples. Simulations were executed on two machines, each in CPU-only and GPU-accelerated modes (four compute configurations total). Machine~1 is a personal Windows~11 Pro laptop with an Intel Core Ultra~9 185H CPU (16 cores / 22 logical processors) and 63.5\,GB RAM, equipped with a CUDA-capable NVIDIA GeForce RTX~4070 Laptop GPU (8\,GB VRAM) (NVIDIA-SMI 560.94; CUDA 12.6). Machine~2 is a Rutgers University workstation running Ubuntu~24.04.3 with an Intel Xeon w9-3595X CPU (60 physical cores / 120 hardware threads) and 93\,GB RAM, equipped with an NVIDIA RTX~4000 Ada Generation GPU (20\,GB VRAM) (NVIDIA-SMI/driver 580.95.05; CUDA driver 13.0; CUDA compilation tools 12.0). Unless stated otherwise, CPU runs use 22 threads on Machine~1 and 96 threads on Machine~2, while GPU runs use a single GPU. Performance is evaluated separately for a representative case in the final subsection. The input files for all cases are provided in the Appendices.

\subsection{Elasto-Dynamic Cases}
\subsubsection{Free Oscillation of a Cantilever Beam}\label{sec:freeoscibeam}
\begin{figure}[!htbp]
    \centering
    \includegraphics[width=0.5\linewidth]{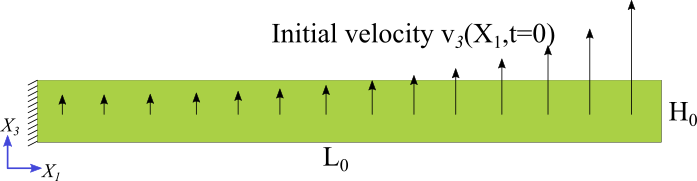}
    \caption{Free oscillation of a cantilever beam. Problem setup.}
    \label{fig:2dcatilevergeo}
\end{figure}

We consider the free oscillation of a cantilever beam, a classical benchmark for assessing dispersion and numerical dissipation in total Lagrangian SPH formulations \cite{landau1986theory,gray2001sph,Rahimi2023AnSphbased}. The beam has length $L_0=0.2$~m and height $H_0=0.02$~m and is clamped at $X_1=0$, as illustrated in \autoref{fig:2dcatilevergeo}. The material is modeled using the St.~Venant--Kirchhoff constitutive law with $\mu=0.715$~MPa, $\kappa=3.25$~MPa, and density $\rho_0=1000~\mathrm{kg/m^3}$, resulting in a sound speed $c_s=57~\mathrm{m/s}$ \cite{landau1986theory,gray2001sph,Rahimi2023AnSphbased}. A two-dimensional plane-strain setting is adopted with unit out-of-plane thickness $W_0=1$~m.
The base particle spacing is $\mathrm{dp}=1$~mm. Mapping factors $\mathrm{mapfac}=2,4,8$ are considered, corresponding to spatial resolutions $\Delta x = 0.5$~mm, $0.25$~mm, and $0.125$~mm in the deformable body. The beam is initialized with a spatially varying velocity field in the $X_3$-direction,
\begin{equation} \label{eq:freeoscivel}
    v_3(X_1,t=0)=0.01 c_s \frac{f(X_1)}{f(L_0)},
\end{equation}
where
\begin{equation}
\begin{split}
f(\bar{x})=&[\cos(k_w L_0)+\cosh(k_w L_0)] [ \cosh(k_w \bar{x}) - \cos(k_w \bar{x})] \\ 
+&[\sin(k_w L_0)-\sinh(k_w L_0)] [ \sinh(k_w \bar{x}) - \sin(k_w \bar{x})] ,
\end{split}
\end{equation}
and $k_w L_0 = 1.875$ corresponds to the fundamental vibration mode. This velocity profile is implemented via a user-defined expression (\autoref{fig:userdefinexpput}) and applied through the XML input file, as shown in \autoref{fig:defstrucdefcantbeam}.
\begin{figure}[!htbp]
    \centering
    \begin{lstlisting}[style=xmlstyle]
<deformstrucbody mkbound="1">
  <bcvel ze="1" xe="2" ye="2" comment="Velocity BC" />
  <density value="1000.0" comment="Mass density" units_comment="kg/m^3" />
  <u_mu value="0.715e6" comment="Shear modulus" units_comment="Pa" />
  <u_bulk value="3.25e6" comment="Bulk modulus" units_comment="Pa" />
  <constitmodel value="1" comment="Constitutive model 1:SVK" />
  <artvisc factor1="0.015" factor2="0.01" comment=" Art. Visc." />
  <mapfac value="4" comment="x4 refinement" />
  <measureplane comment="Measure tip disp.">
    <p1 x="199.999e-3" y="#Lys" z="#LzMn" />
    <p2 x="199.999e-3" y="#Lyf + 0.5e-3" z="#LzMn" />
    <p3 x="199.999e-3" y="#Lyf + 0.5e-3" z="#LzMp" />
    <p4 x="199.999e-3" y="#Lys" z="#LzMp" />
  </measureplane>
</deformstrucbody>
    \end{lstlisting}
    \caption{Free oscillation of a cantilever beam. XML definition of the deformable structure body and measurement plane used to extract the free-end displacement.}
    \label{fig:defstrucdefcantbeam}
\end{figure}

Under linear Euler--Bernoulli beam theory, the analytical solution for the free-end deflection, strain energy, and kinetic energy is
\begin{equation}
    u_{3}(L_0,t) = \frac{0.01 c_s}{\omega_1}\sin(\omega_1 t),
\end{equation}
\begin{equation}
    E_{e} = \frac{\rho_0 L_0 H_0 W_0}{8}(0.01 c_s)^2 \sin^2(\omega_1 t),
\end{equation}
\begin{equation}
    E_{k} = \frac{\rho_0 L_0 H_0 W_0}{8}(0.01 c_s)^2 \cos^2(\omega_1 t),
\end{equation}
where
\begin{equation}
    \omega_1 = k_w^2 \sqrt{\frac{EI}{\rho_0 H_0 W_0}}, 
    \qquad
    I = \frac{W_0 H_0^3}{12}.
\end{equation}
Since the present formulation is finite-strain hyperelastic, small deviations from the linear analytical solution are expected due to geometric nonlinear effects, particularly at larger oscillation amplitudes.

\begin{figure}[!htbp]
    \centering
     \begin{subfigure}[b]{0.49\linewidth}
         \includegraphics[width=\linewidth]{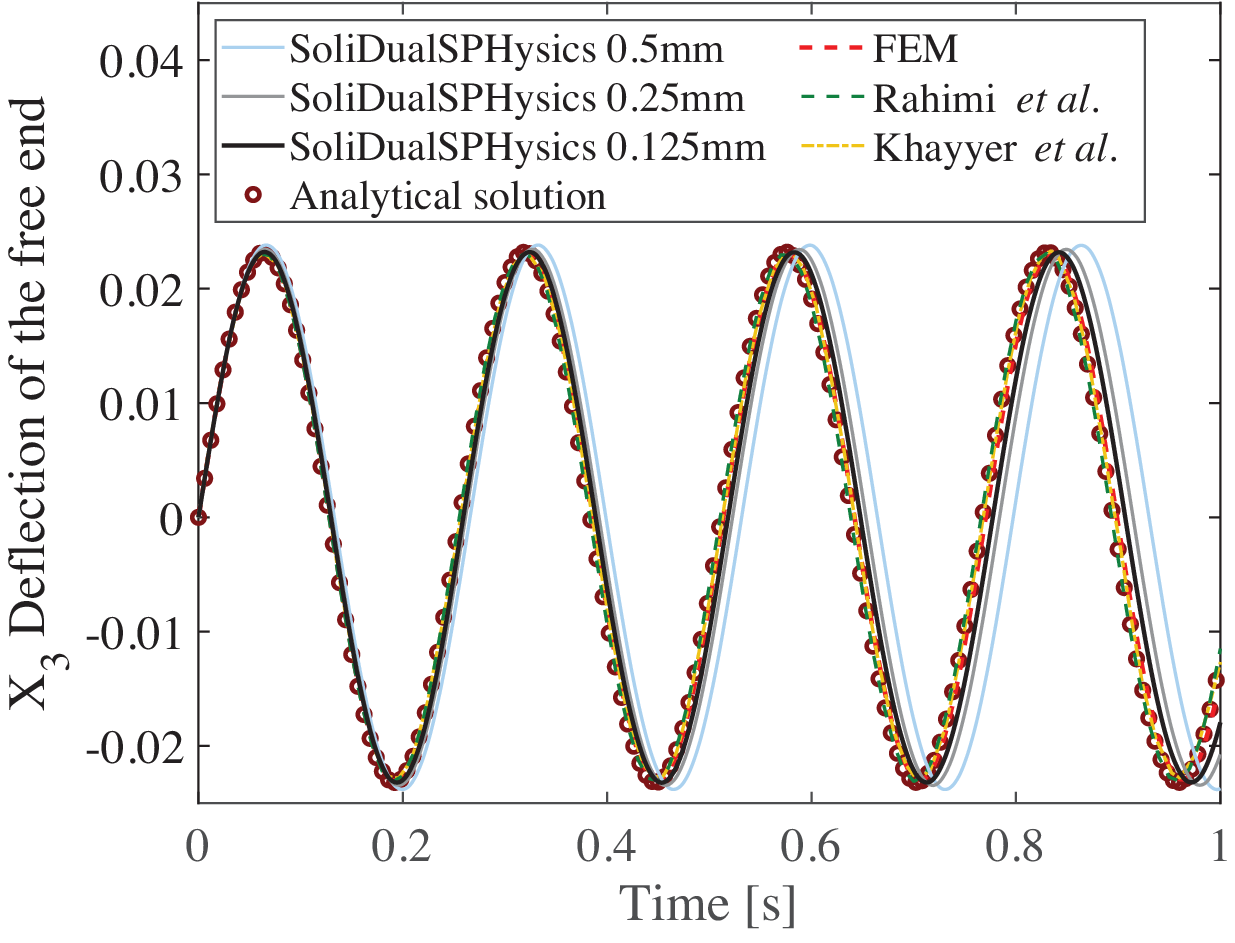}
         \caption{}
     \end{subfigure}
     \begin{subfigure}[b]{0.49\linewidth}
         \includegraphics[width=\linewidth]{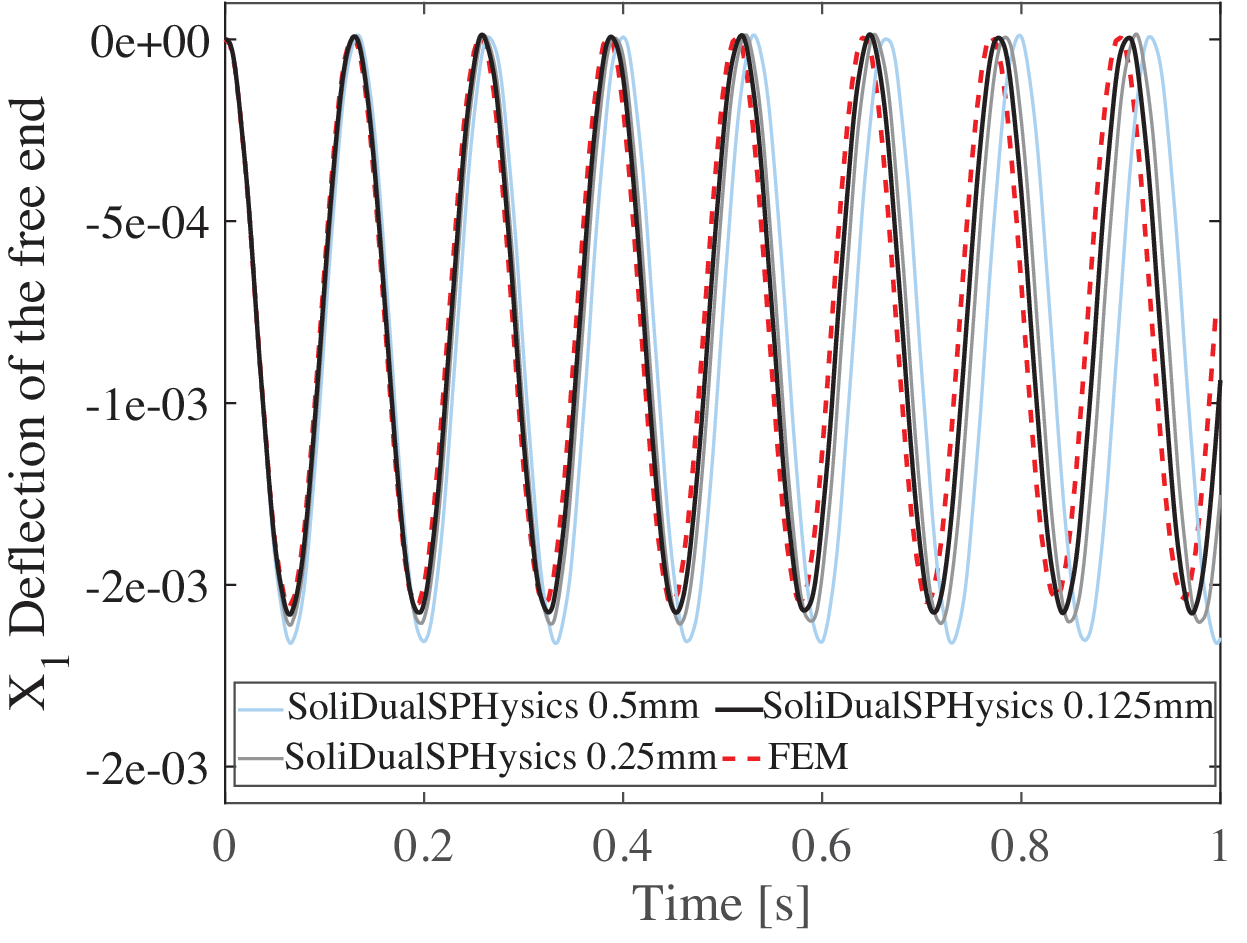}
         \caption{}
     \end{subfigure}
     \begin{subfigure}[b]{0.49\linewidth}
         \includegraphics[width=\linewidth]{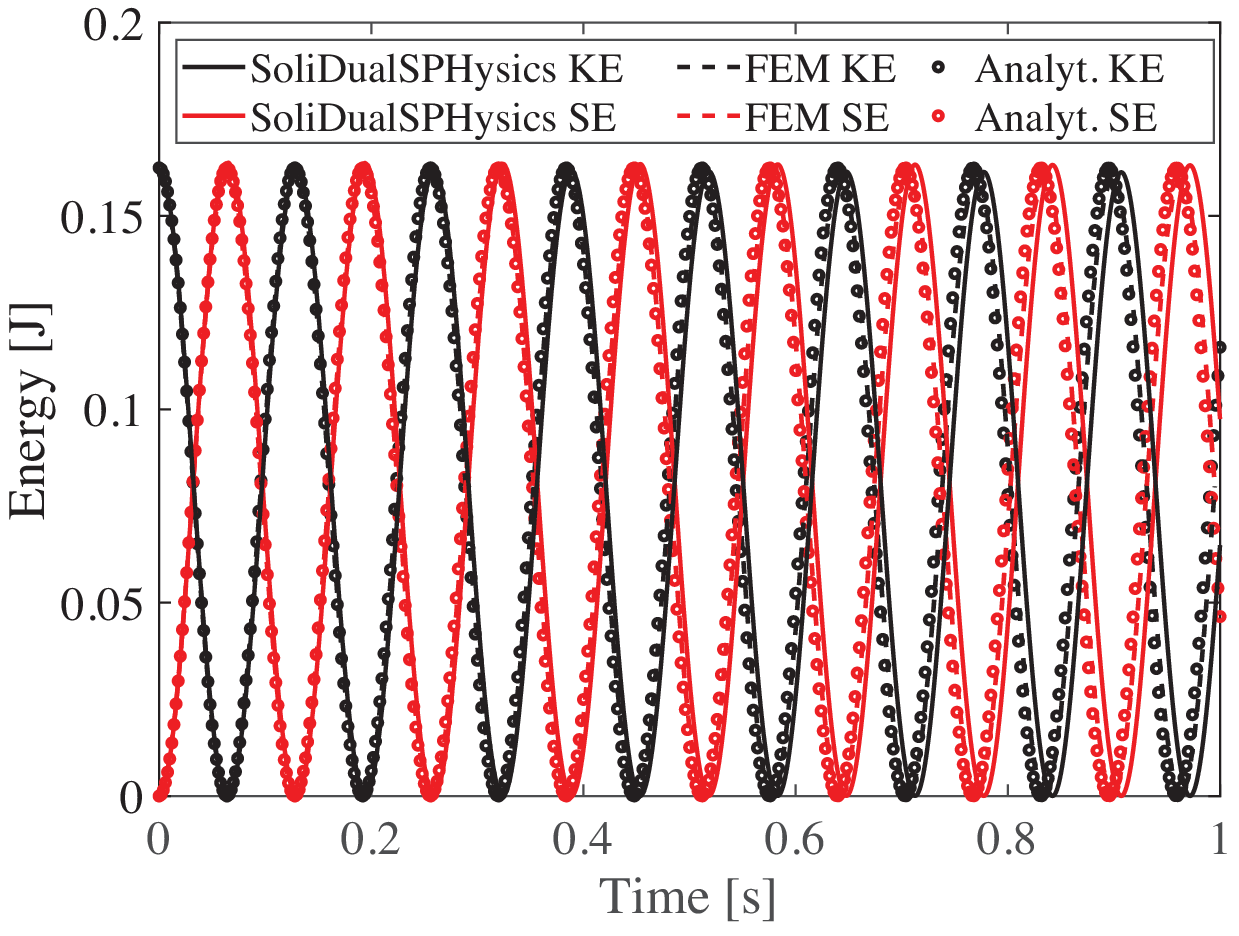}
         \caption{}
     \end{subfigure}
    \begin{subfigure}[b]{0.49\linewidth}
         \includegraphics[width=\linewidth]{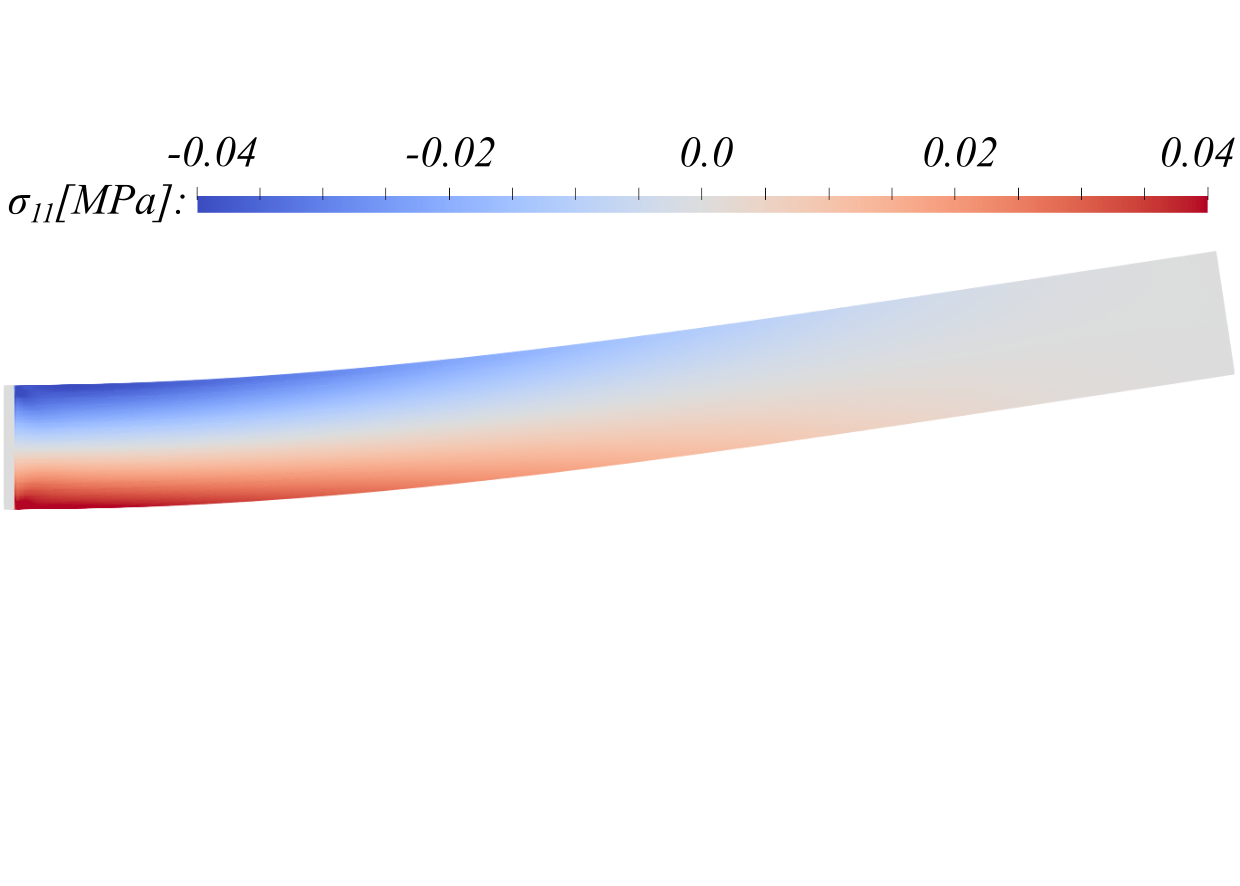}
         \caption{}
     \end{subfigure}
    \caption{Free oscillation of a cantilever beam. (a) Time history of the free-end deflection in the $X_3$-direction. (b) Time history of the free-end deflection in the $X_1$-direction. (c) Kinetic (KE) and strain (SE) energies for $\Delta x=0.125$~mm. (d) Cauchy stress $\sigma_{11}$ at $t=0.57$~s. Comparison is made with the analytical solution, finite element simulations, and reference results from the literature \cite{Rahimi2023AnSphbased,khayyer2018enhanced}. The analytical solution predicts zero $X_1$-deflection under the small-strain assumption.}
    \label{fig:2dcatilevergraphs}
\end{figure}

\autoref{fig:2dcatilevergraphs}a--b presents the time history of the free-end deflection in the $X_3$- and $X_1$-directions, respectively. Excellent agreement is observed with the analytical solution and reference FEM results. The energy evolution shown in \autoref{fig:2dcatilevergraphs}c demonstrates near-conservative exchange between kinetic (KE) and strain (SE) energies, indicating minimal numerical dissipation. The Cauchy stress $\sigma_{11}$ contours at $t=0.57$~s are displayed in \autoref{fig:2dcatilevergraphs}d and show close agreement with the results reported in \cite{khayyer2018enhanced,Rahimi2023AnSphbased}. 
Reproducibility is facilitated through \solidname' XML-based input structure (see \autoref{app:freeoscbeam}), which enables concise specification of complex initial fields and boundary conditions.

\subsubsection{Free Oscillation of a Cantilever Plate}
\begin{figure}[!htbp]
    \centering
    \includegraphics[width=0.6\linewidth]{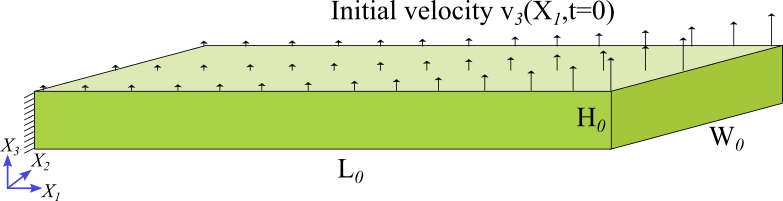}
    \caption{Free oscillation of a cantilever plate. Problem setup.}
    \label{fig:3dcatilevergeo}
\end{figure}

We next consider the three-dimensional counterpart of the cantilever beam example presented in \autoref{sec:freeoscibeam}. The cantilever plate geometry (\autoref{fig:3dcatilevergeo}) has dimensions $L_0=0.2$~m, $H_0=0.02$~m, and $W_0=0.06$~m. The plate is clamped at $X_1=0$ and initialized with the same velocity field defined in \Eqref{eq:freeoscivel}. The material parameters are identical to those of the 2D case, and the St.~Venant--Kirchhoff constitutive model is employed. The discretization uses a particle spacing of $\Delta x=0.5$~mm, resulting in approximately two million particles. The complete XML input file is provided in \autoref{app:freeoscplate}.
\begin{figure}[!htbp]
    \centering
    \begin{subfigure}[b]{0.49\linewidth}
         \includegraphics[width=\linewidth]{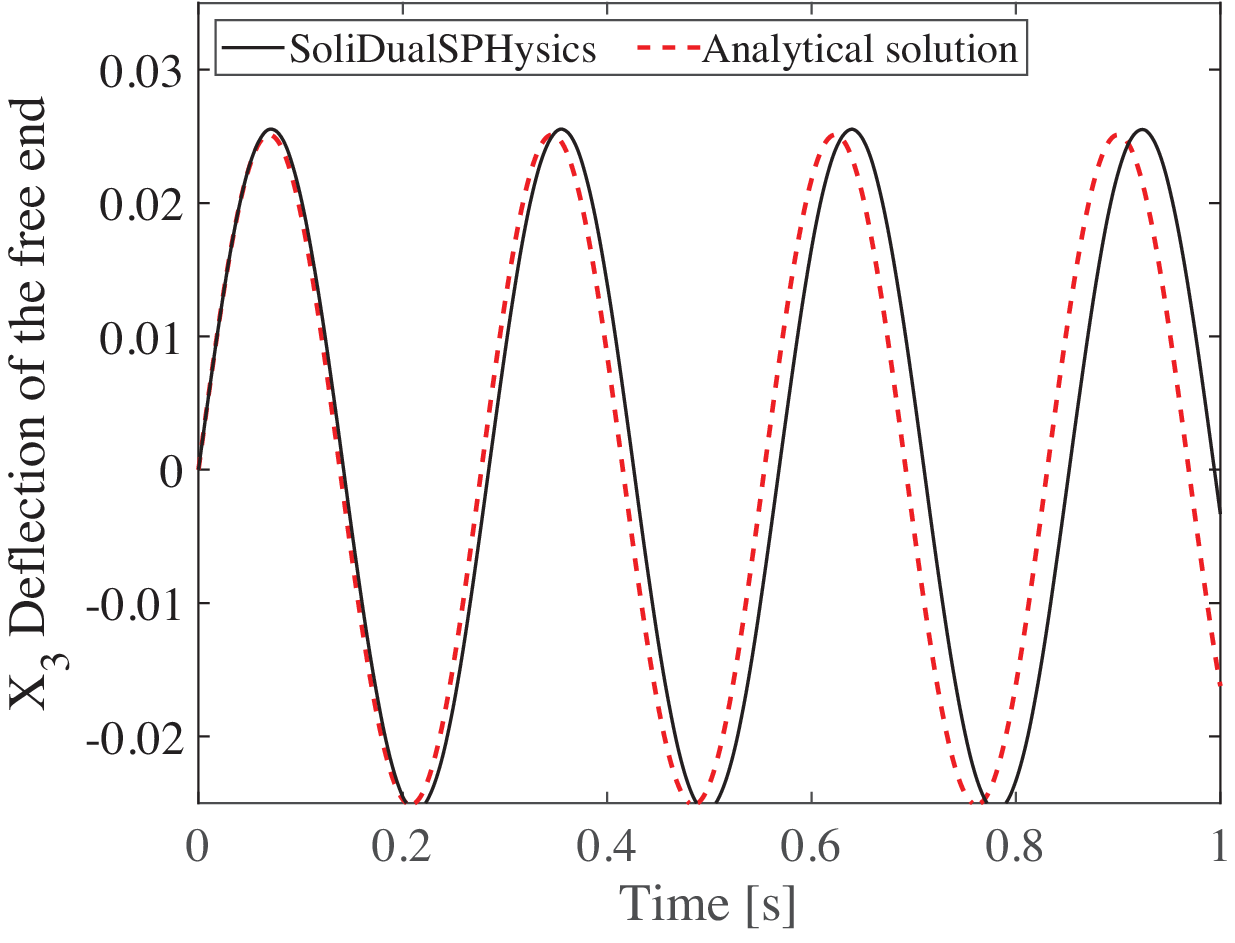}
         \caption{}
     \end{subfigure}
     \begin{subfigure}[b]{0.49\linewidth}
         \includegraphics[width=\linewidth]{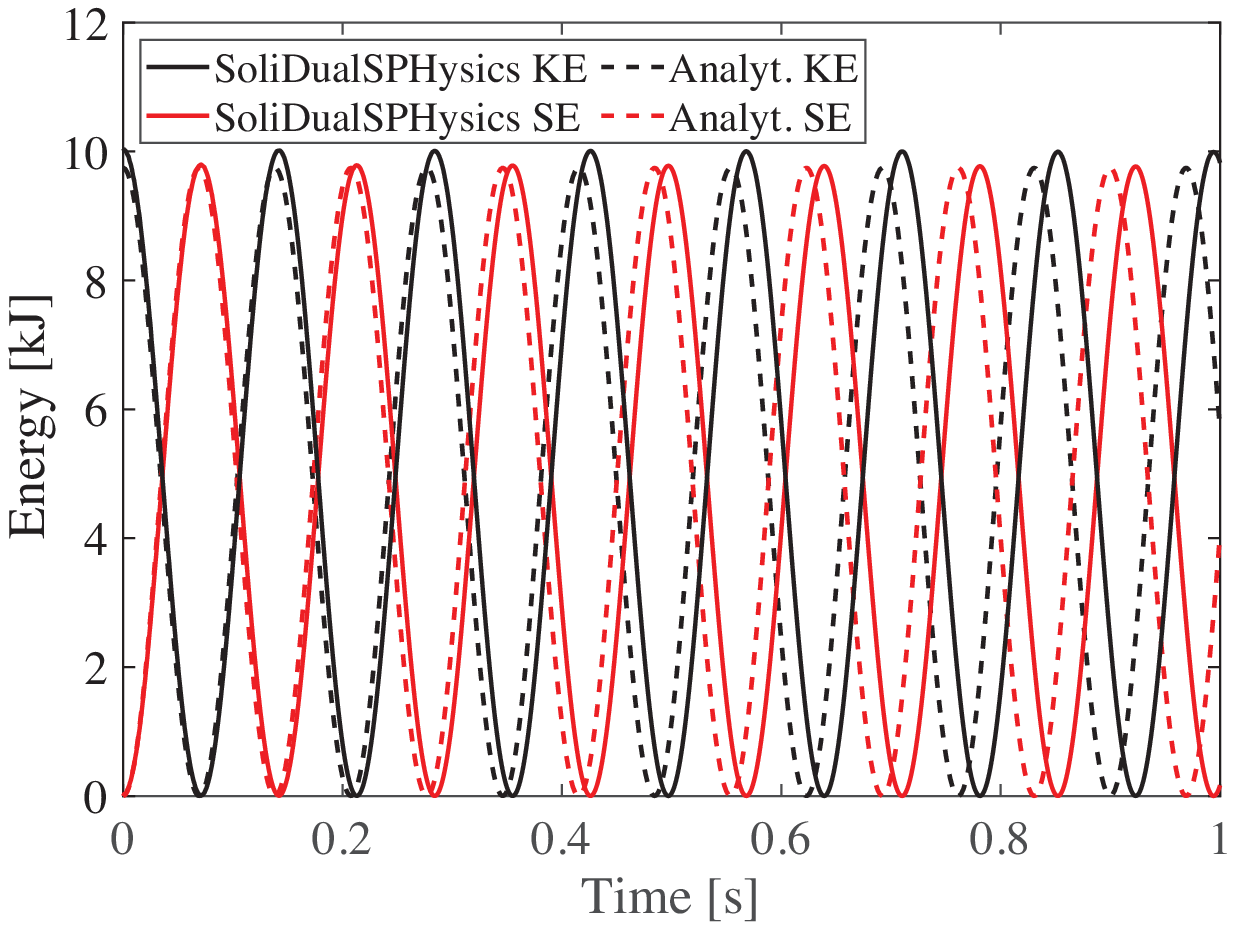}
         \caption{}
     \end{subfigure}
    \caption{Free oscillation of a cantilever plate. (a) Time history of the free-end deflection in the $X_3$-direction. (b) Kinetic (KE) and strain (SE) energy of the system. Comparison is made with the analytical small-strain reference solution. Since the analytical benchmark is based on linear Euler--Bernoulli theory, small deviations from the finite-strain numerical results are expected in both the oscillation frequency and the instantaneous energy evolution.}
    \label{fig:3dcatilevergraphs}
\end{figure}

\begin{figure}[!htbp]
    \centering
    \includegraphics[width=1.0\linewidth]{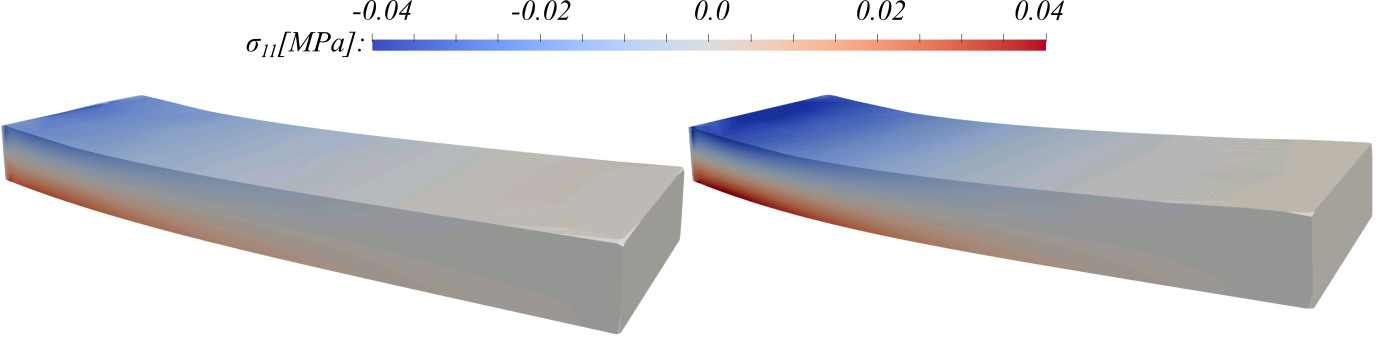}
    \caption{Free oscillation of a cantilever plate. Contours of Cauchy stress $\sigma_{11}$ at $t=35$~ms and $t=70$~ms.}
    \label{fig:freeosciplategeo}
\end{figure}

The analytical reference solution for the fundamental bending mode is identical in form to that of the beam example and is derived from linear Euler--Bernoulli theory. As in the 2D case, this solution serves as a small-strain benchmark for the oscillation response, whereas the numerical model is based on a three-dimensional finite-strain hyperelastic formulation. Accordingly, small deviations in the predicted oscillation frequency and energy evolution with respect to the analytical reference are expected. \autoref{fig:3dcatilevergraphs} compares the predicted tip deflection and energy evolution with the analytical small-strain reference. The dominant oscillation frequency is reproduced well, and the expected exchange between kinetic (KE) and strain (SE) energies is observed. In addition, the total numerical energy remains nearly constant over time, indicating stable integration and the absence of unphysical energy growth during the simulation. The Cauchy stress $\sigma_{11}$ contours shown in \autoref{fig:freeosciplategeo} further confirm physically consistent bending behavior. The peak stress occurs near the clamped region and evolves smoothly over time without spurious oscillations, consistent with an elastic, bending-dominated response.

\subsubsection{Large Deformation of a 3D Cantilever Beam}
\begin{figure}[!htbp]
    \centering
    \includegraphics[width=0.6\linewidth]{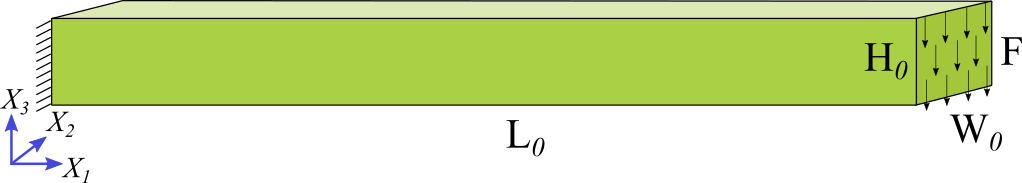}
    \caption{Large deformation of a 3D cantilever beam. Problem setup.}
    \label{fig:3dcatilevercolumngeo}
\end{figure}

We consider a three-dimensional cantilever beam subjected to a ramped, distributed surface traction to induce large deformation. The beam geometry (\autoref{fig:3dcatilevercolumngeo}) has length $L_0=0.1$~m, height $H_0=0.01$~m, and width $W_0=0.01$~m, and is clamped at $X_1=0$. The material is modeled using a compressible neo--Hookean constitutive law with density $\rho_0=7800~\mathrm{kg/m^3}$, Young’s modulus $E=210$~GPa, and Poisson's ratio $\nu=0.3$. The discretization uses a particle spacing of $\Delta x=1$~mm with $\mathrm{mapfac}=1$. A distributed surface traction with peak magnitude $F_{\max}=1.75\times10^{9}~\mathrm{N/m^2}$ is applied through a linear ramp over $T_{\max}=1$~s:
\begin{equation}
F(t)=
\begin{cases}
F_{\max}\,t/T_{\max} & t\le T_{\max},\\
F_{\max} & \text{otherwise}.
\end{cases}
\end{equation}
The traction is prescribed using user-defined expressions, as shown in \autoref{fig:defstrucdefcantcolumn}. The complete XML input file is provided in \autoref{app:deformcolumn3d}.
\begin{figure}[!htbp]
    \centering
    \begin{lstlisting}[style=xmlstyle]
<special>
  <mathexpressions>
    <userexpression id="1" comment="Math expression">
      <locals value="Fmax=-1.75e9; Tmax=1.0; xtip=0.099;"/>
       <expression value="if(x0>xtip,if(t<=Tmax,t/Tmax,1.0)*Fmax,skip)"/>
    </userexpression>
    <userexpression id="2" comment="Math expression">
      <expression value="if(x0<=0.0,0.0,skip)"/>
    </userexpression>
  </mathexpressions>
  <deformstrucs>
    <deformstrucbody mkbound="1">
      <bcforce type="2" ze="1" comment="Distributed load" />
      <bcvel xe="2" ye="2" ze="2" comment="Velocity constraint in x,y,z" /> 
      <density value="7800.0" comment="Mass density" units_comment="kg/m^3" />
      <youngmod value="210.0e9" comment="Young's Modulus" units_comment="Pa"/>
      <poissratio value="0.3" comment="Poisson's ratio" />
      <constitmodel value="2" comment="Constitutive model 2:neo--Hookean" />
      <artvisc factor1="0.1"  factor2="0.0" comment="Art. Visc. factor" />
      <mapfac value="1" comment="x1 refinement" />
      <measureplane comment="Measure tip displacement">
        <p1 x="100.4e-3" y="#Ly*0.455" z="#Lz*0.455" />
        <p2 x="100.4e-3" y="#Ly*0.555" z="#Lz*0.455" />
        <p3 x="100.4e-3" y="#Ly*0.555" z="#Lz*0.555" />
        <p4 x="100.4e-3" y="#Ly*0.455" z="#Lz*0.555" />
      </measureplane>
    </deformstrucbody>
  </deformstrucs>
</special>
    \end{lstlisting}
    \caption{Large deformation of a 3D cantilever beam. XML definition of the deformable structure body, boundary conditions, and measurement plane.}
    \label{fig:defstrucdefcantcolumn}
\end{figure}
\begin{figure}[!htbp]
    \centering
    \includegraphics[width=\linewidth]{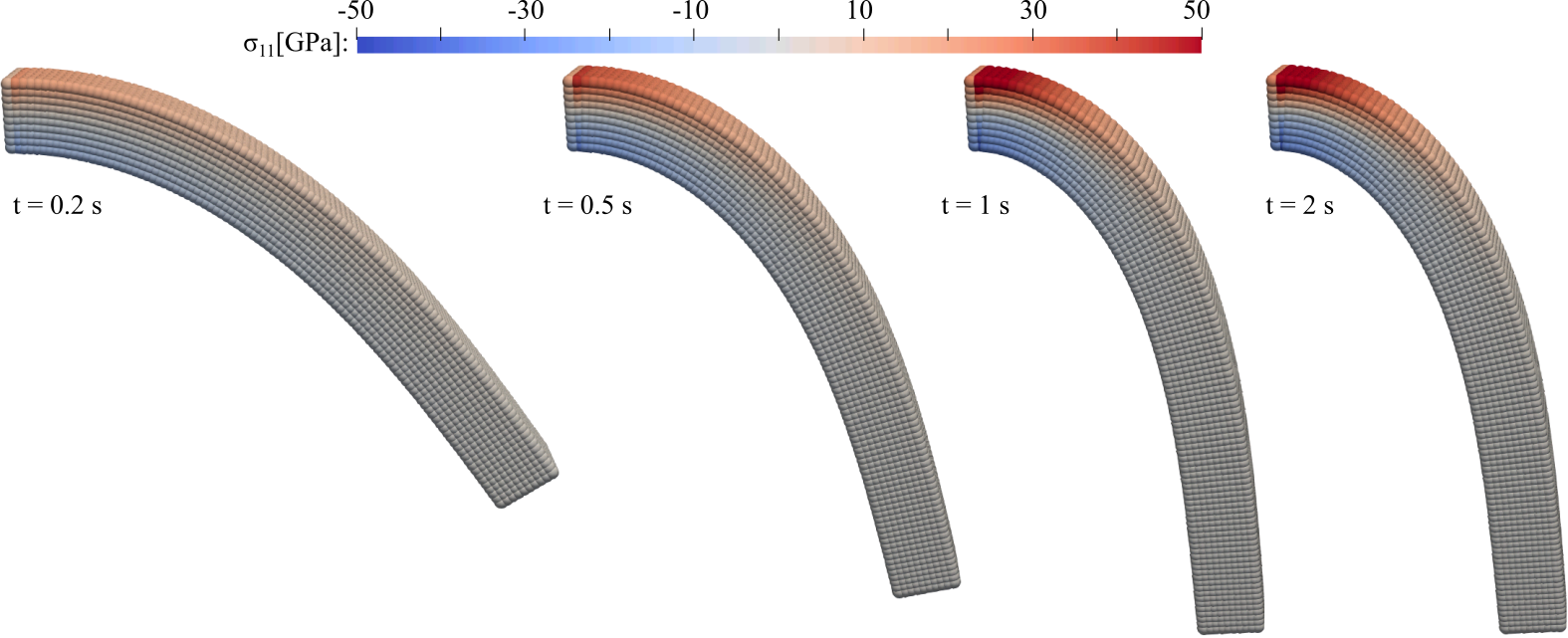}
    \caption{Large deformation of a 3D cantilever beam. Contour plots of $\sigma_{11}$ at various time instances.}
    \label{fig:3dcatilevercolumnstress}
\end{figure}
\begin{figure}[!htbp]
    \centering
    \includegraphics[width=0.5\linewidth]{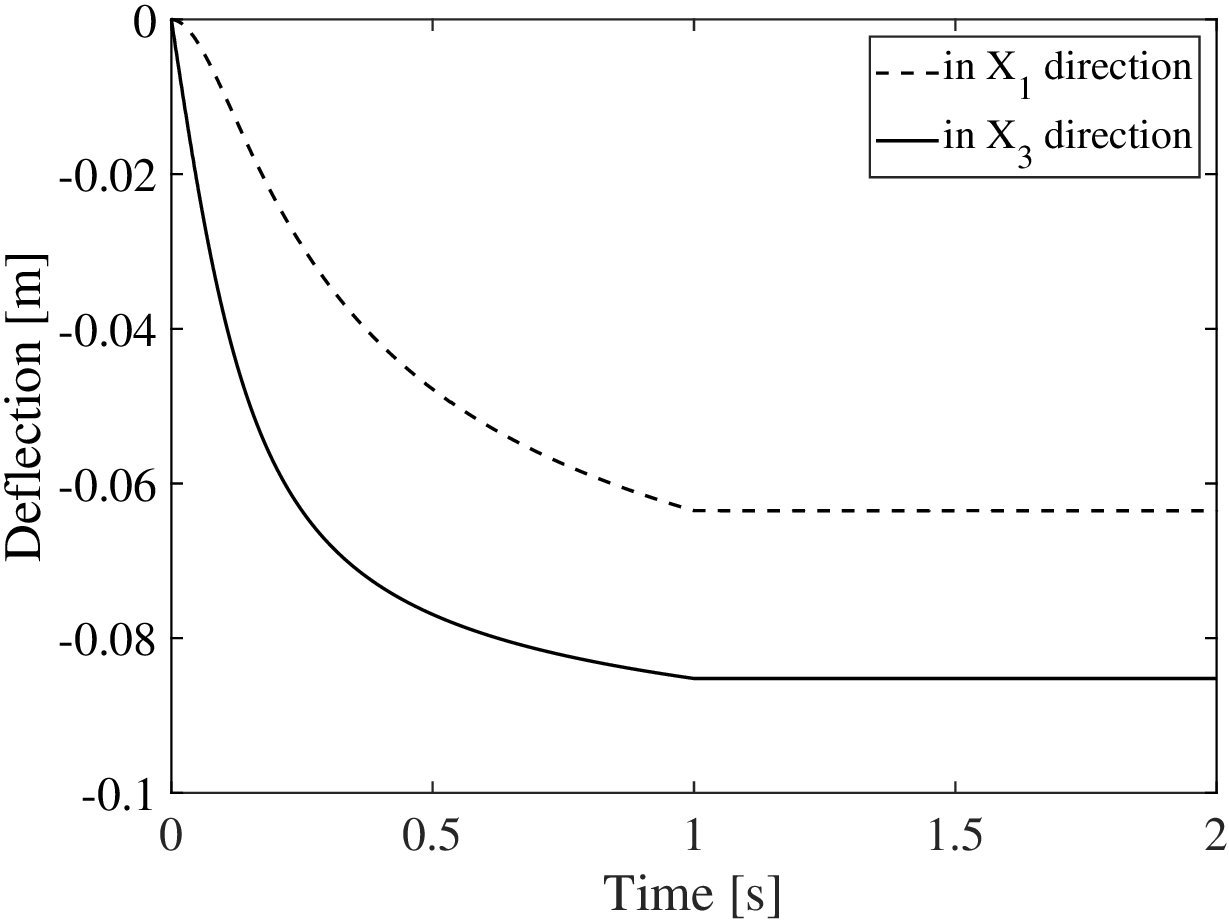}
    \caption{Large deformation of a 3D cantilever beam. Free end deflection in the $X_1$- and $X_3$-directions as a function of time.}
    \label{fig:3dcatilevercolumngraph}
\end{figure}
This example is included to demonstrate the application of force boundary conditions in \solidname via a ramped distributed surface traction. 

The stress contours in \autoref{fig:3dcatilevercolumnstress} exhibit the expected bending-dominated pattern, with peak $\sigma_{11}$ concentrated near the clamped end. The stress distribution evolves smoothly during loading and subsequent dynamic response, without spurious oscillations. The tip-deflection history shown in \autoref{fig:3dcatilevercolumngraph} demonstrates a smooth transition from the loading ramp to the dynamic response phase, with no evidence of numerical instability. The overall displacement and stress trends are in qualitative agreement with reference results reported in \cite{Rahimi2022AGeneralized,he2017coupled}.

\subsubsection{Twisting 3D Column}
\begin{figure}[!htbp]
    \centering
    \includegraphics[width=0.25\linewidth]{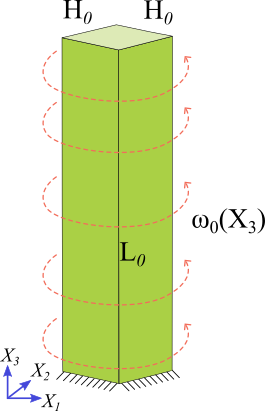}
    \caption{Twisting 3D column. Problem setup.}
    \label{fig:twistcolumngeo}
\end{figure}

This case assesses the framework's ability to resolve torsional dynamics and the associated propagation of stress waves. We consider a three-dimensional elastic column (\autoref{fig:twistcolumngeo}) of height $L_0=6$~m aligned with the $X_3$-axis and square cross-section of side length $H_0=1$~m. The column is assigned a non-uniform initial angular velocity about the $X_3$-axis and is subsequently clamped at its base ($X_3=0$) at $t=0$, allowing the system to evolve freely thereafter.
The prescribed angular velocity profile is
\begin{equation}
\omega_0(X_3)=105\,\sin\!\left(\frac{\pi X_3}{2L_0}\right),
\end{equation}
implemented through Cartesian velocity components using user-defined expressions, as shown in \autoref{fig:twistcolumncasedef}. The velocity field is applied only at $t=0$ to initialize the torsional motion, while the base of the column remains constrained throughout the simulation. The material is modeled as a compressible neo--Hookean solid with density $\rho_0=1100~\mathrm{kg/m^3}$, Young’s modulus $E=1.7\times10^{7}$~Pa, and Poisson's ratio $\nu=0.45$. The column is discretized using three different particle spacings, $\Delta x=0.025$~m, $0.05$~m, $0.1$~m. The complete case definition is provided in \autoref{app:twistingcolumn}.
\begin{figure}[!htbp]
    \centering
    \begin{lstlisting}[style=xmlstyle]
<special>
  <mathexpressions>
    <userexpression id="1" comment="User expression">
      <locals value="xcent=0.5; ycent=0.5; omega=105.0"/>
      <expression value="if(z0<=0.0,0.0,if(t<=0,omega*sin(0.2619047*z0)*(ycent-y0),skip))"/>
    </userexpression>
    <userexpression id="2" comment="User expression">
      <locals value="xcent=0.5; ycent=0.5; omega=105.0"/>
      <expression value="if(z0<=0.0,0.0,if(t<=0,omega*sin(0.2619047*z0)*(x0-xcent),skip))"/>
    </userexpression>
    <userexpression id="3" comment="User expression">
      <expression value="if(z0<=0.0,0.0,skip)"/>
    </userexpression>
  </mathexpressions>
  <deformstrucs>
    <deformstrucbody mkbound="1">
      <bcvel xe="1" ye="2" ze="3" comment="Velocity BC" />
      <density value="1100.0" comment="Mass density"/>
      <youngmod value="170.0e5" comment="Young's Modulus"/>
      <poissratio value="0.45" comment="Poisson's ratio" />
      <artvisc factor1="0.5" factor2="0.0" comment="Art. Visc." />
      <constitmodel value="2" comment="Const. model 2:neo--Hookean" />
      <mapfac value="3" />
    </deformstrucbody>
  </deformstrucs>
</special>
    \end{lstlisting}
    \caption{Twisting 3D column. Definition of deformable structure body and boundary conditions.}
    \label{fig:twistcolumncasedef}
\end{figure}
\begin{figure}[!htbp]
    \centering
    \includegraphics[width=\linewidth]{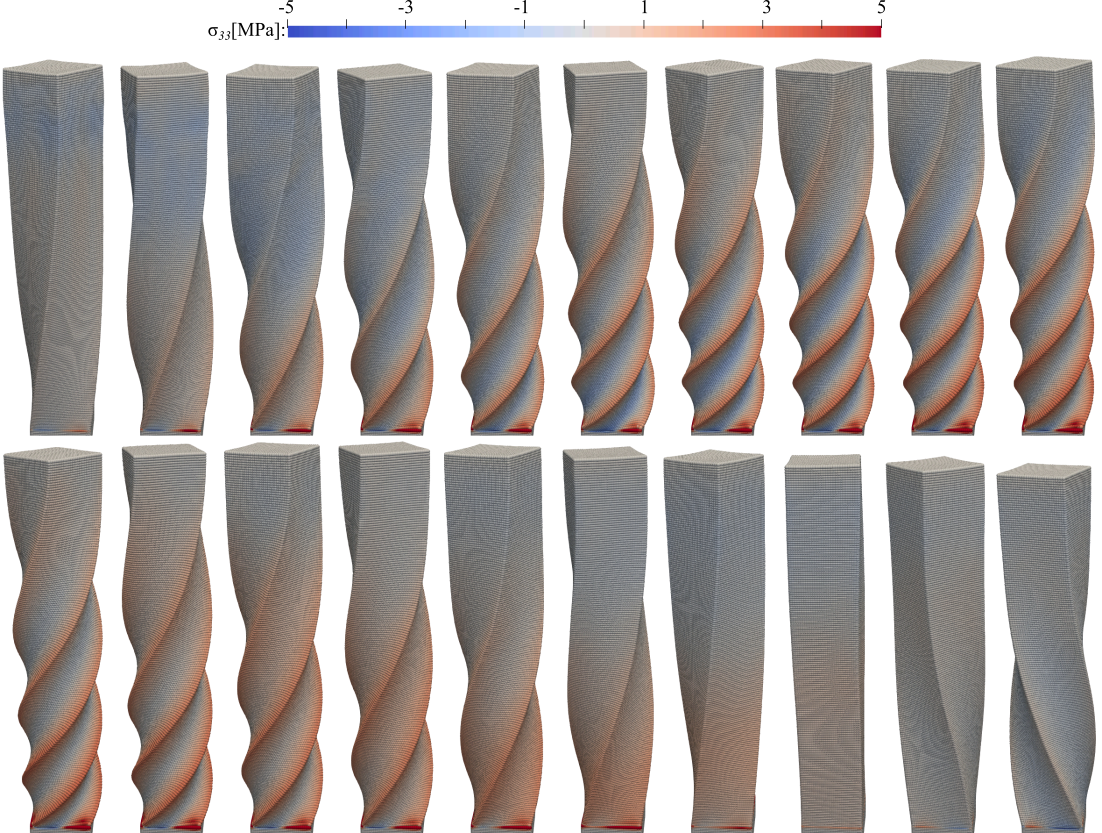}
    \caption{Twisting 3D column. Contours of Cauchy stress $\sigma_{33}$ during the first 200~ms of the simulation for $\Delta x=0.025$~m, shown at 10~ms intervals from left to right.}
    \label{fig:twistcolumnstress}
\end{figure}

\begin{figure}[!htbp]
    \centering
    \begin{subfigure}[b]{0.49\linewidth}
         \includegraphics[width=\linewidth]{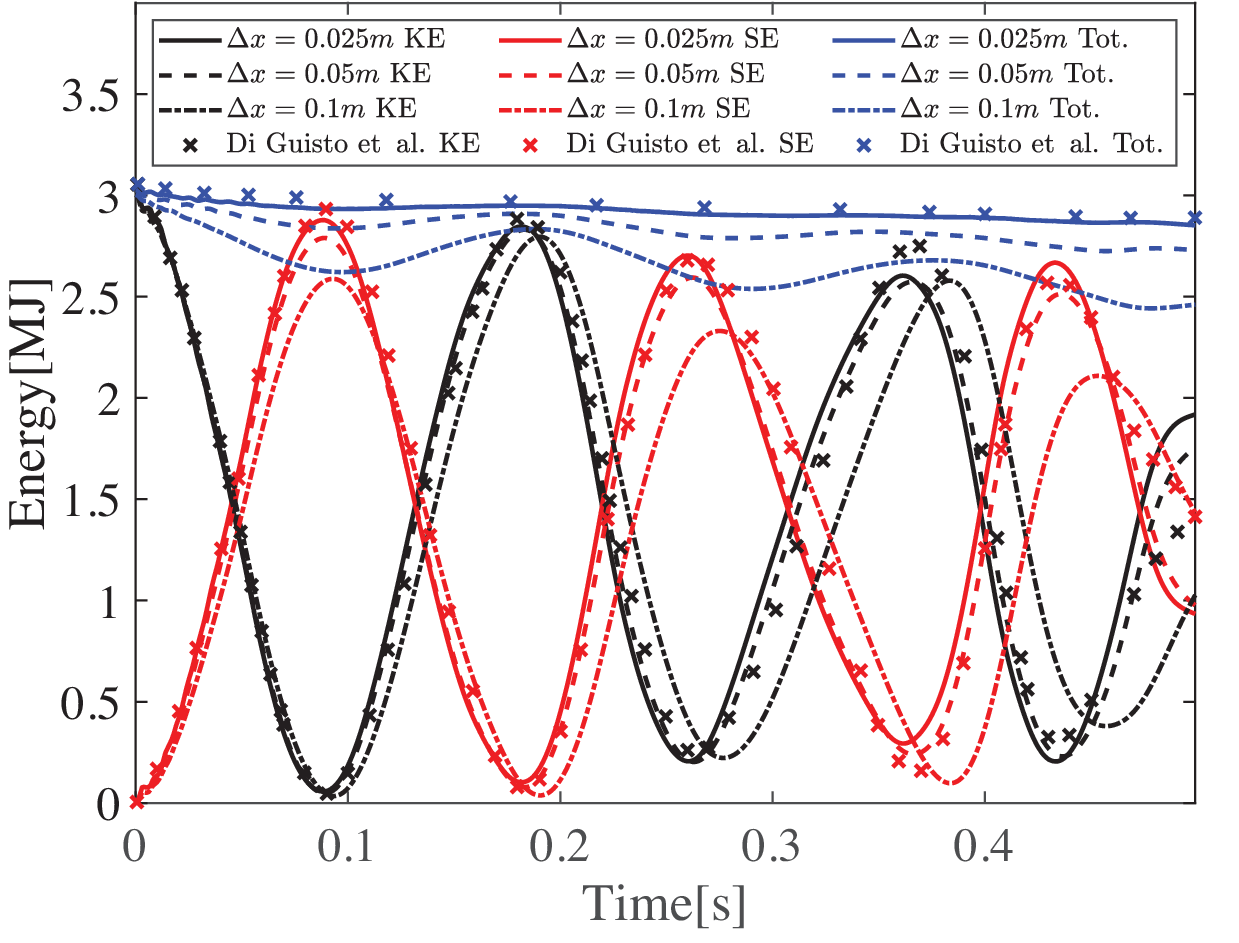}
         \caption{}
     \end{subfigure}
     \begin{subfigure}[b]{0.49\linewidth}
         \includegraphics[width=\linewidth]{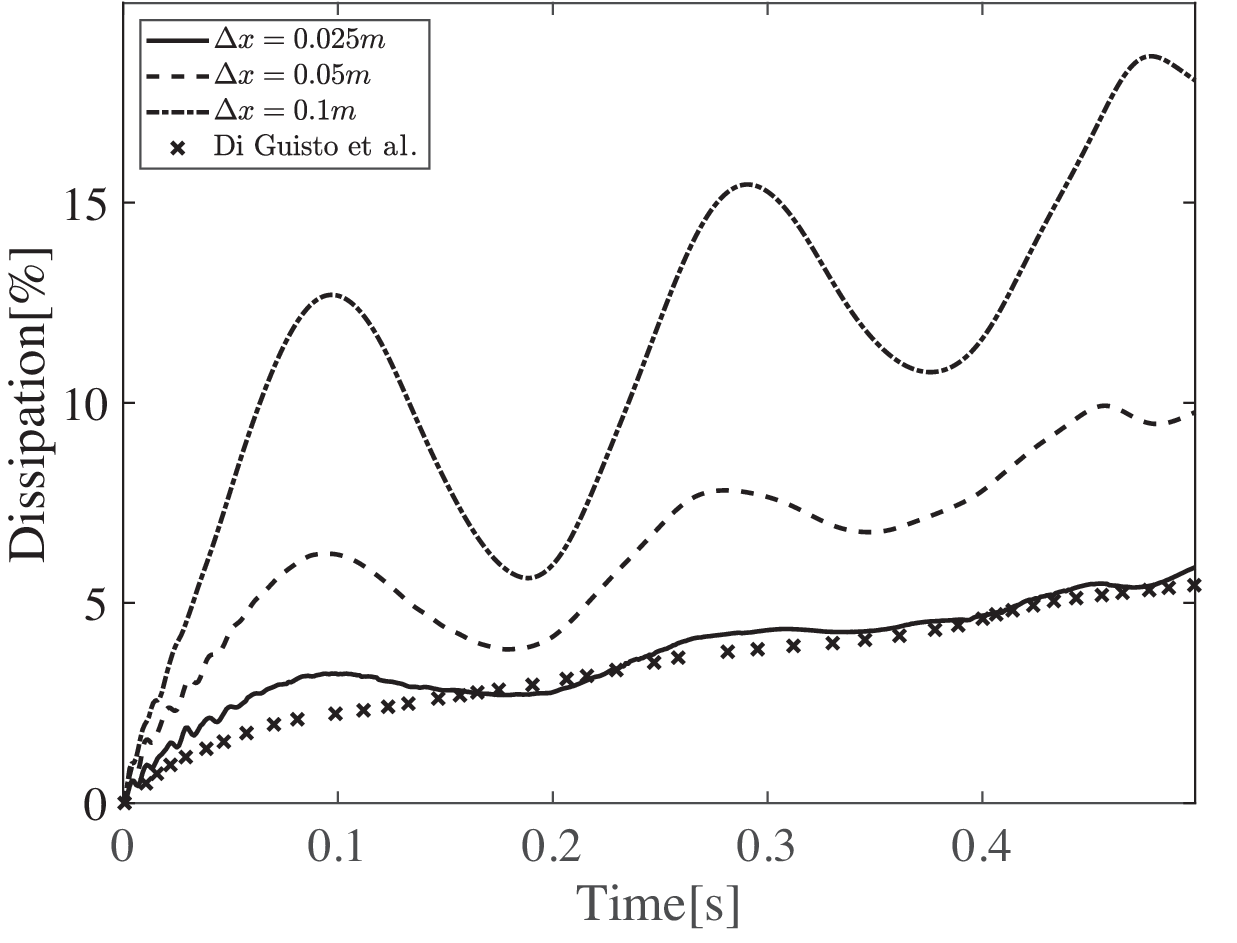}
         \caption{}
     \end{subfigure}
    \caption{Twisting 3D column. (a) Time history of total kinetic energy (KE), total strain energy (SE), and total energy of the column. (b) Total energy dissipation in the system. Results are plotted for different values of $\Delta x$ and compared with \cite{di2024first}.}
    \label{fig:twistcolumnenergies}
\end{figure}

\autoref{fig:twistcolumnstress} shows the evolution of $\sigma_{33}$ during the first 200~ms. Immediately after the base constraint is enforced, a clean torsional wave propagates along the column. The structure subsequently undergoes elastic recovery toward its undeformed configuration, completing approximately half a torsional cycle by $t\approx 180$~ms. The absence of spurious high-frequency oscillations in the stress field indicates stable wave propagation and adequate numerical regularization. The predicted dynamics are consistent with the trends reported in \cite{He2019AGpu,di2024first}. The corresponding energy histories and dissipation trends are reported in \autoref{fig:twistcolumnenergies}, showing the expected exchange between kinetic and strain energy, limited resolution-dependent total energy dissipation, and good agreement with the reference results of \cite{di2024first}.

\subsection{Fracture Cases}

\subsubsection{Dynamic Crack Branching}
\begin{figure}[!htbp]
    \centering
    \includegraphics[width=0.5\linewidth]{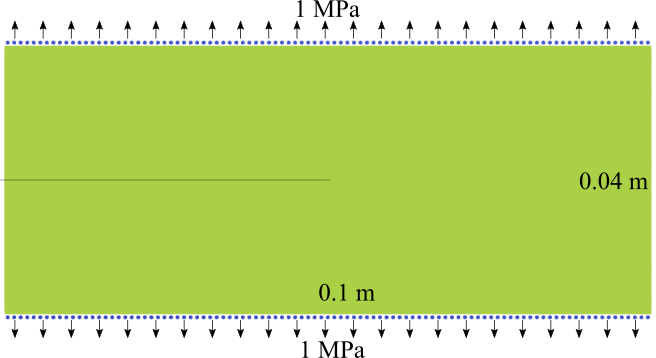}
    \caption{Dynamic crack branching. Problem setup.}
    \label{fig:dynamiccrackgeo}
\end{figure}

To demonstrate the framework's capability to model dynamic brittle fracture, we simulate the benchmark problem of crack branching in a pre-notched plate subjected to tensile loading \cite{Borden2012,rahimi2022Asmoothed}. The plate geometry ($L_0=100$~mm, $H_0=40$~mm) with a pre-existing notch of length 50~mm is shown in \autoref{fig:dynamiccrackgeo}. A constant traction boundary condition ($\sigma_{33}=1$~MPa) is applied on the top and bottom surfaces for the duration of the simulation. The material parameters are Young's modulus $E=32$~GPa, Poisson's ratio $\nu=0.2$, density $\rho_0=2450~\mathrm{kg/m^3}$, and fracture energy $G_c=3~\mathrm{J/m^2}$. The St. Venant--Kirchhoff constitutive model is employed in combination with the phase-field formulation, and plane-strain is assumed. The phase-field length scale is set to $\epsilon_0=0.125$~mm, and the particle spacing is $\Delta x=0.0625$~mm. The complete XML definition of the case is provided in \autoref{app:dyncrackbranch}.

\begin{figure}[!htbp]
    \centering
    \includegraphics[width=1.0\linewidth]{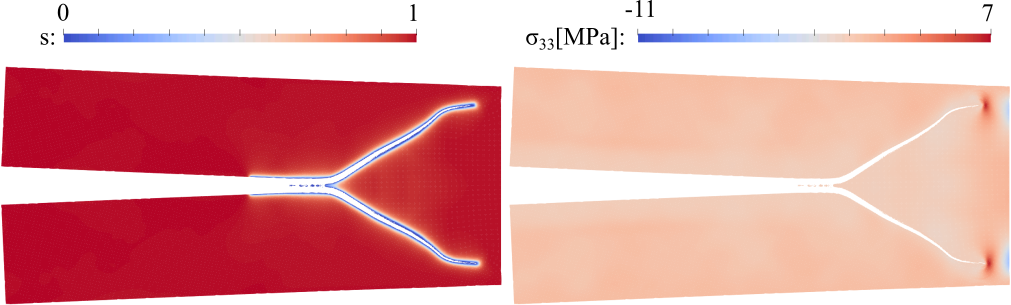}
    \caption{Dynamic crack branching. Contours of the phase-field variable (left) and the Cauchy stress component $\sigma_{33}$ (right) at $t=80~\mu$s. The deformed configuration is magnified by a factor of 50 for visualization.}
    \label{fig:dynamiccrackstress}
\end{figure}

\autoref{fig:dynamiccrackstress} presents contours of the phase-field variable (representing the crack path) and the $\sigma_{33}$ component of the Cauchy stress at approximately \SI{80}{\micro\second}. Crack initiation occurs at the notch tip at approximately \SI{14}{\micro\second}, followed by branching at approximately \SI{35}{\micro\second} with a divergence angle of about $60^\circ$. The crack propagates symmetrically, in agreement with previously reported results \cite{rahimi2022Asmoothed,Borden2012,Kamensky2018}. The stress concentration near the crack tip is consistent with theoretical expectations for dynamic brittle fracture.
\begin{figure}[!htbp]
    \centering
    \includegraphics[width=0.5\linewidth]{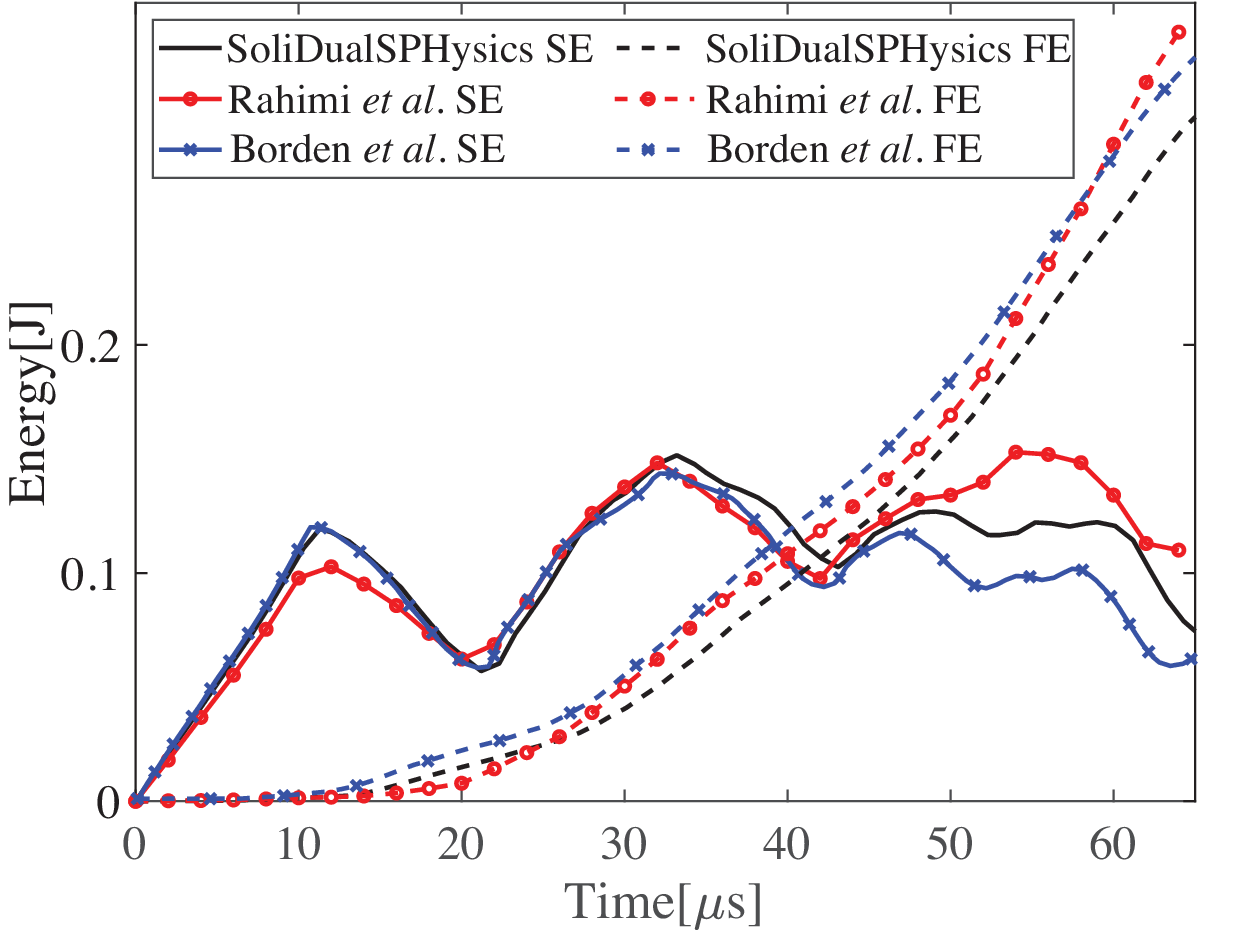}
    \caption{Dynamic crack branching. Evolution of strain energy (SE) and fracture energy (FE), compared with results from Rahimi \etal \cite{rahimi2022Asmoothed} and Borden \etal \cite{Borden2012}.}
    \label{fig:dynamiccrackgraphs}
\end{figure}
The evolution of strain energy (SE) and fracture energy (FE) is shown in \autoref{fig:dynamiccrackgraphs}. As crack propagation progresses, the strain energy decreases while the fracture energy increases, indicating proper energy transfer from stored elastic energy to crack surface formation. The predicted energy histories closely match the results reported in \cite{rahimi2022Asmoothed,Borden2012}.

\subsubsection{Kalthoff--Winkler Experiment}
\begin{figure}[!htbp]
    \centering
    \includegraphics[width=0.65\linewidth]{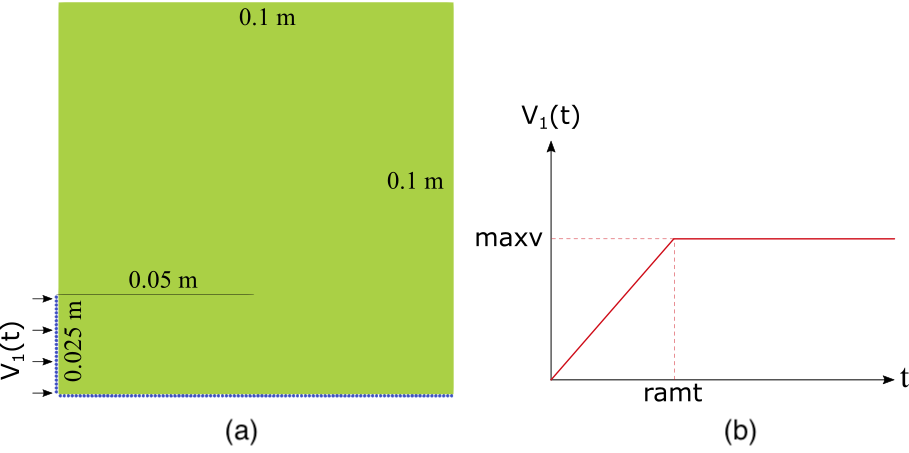}
    \caption{Kalthoff--Winkler experiment. (a) Problem setup. (b) Ramped velocity boundary condition applied to the impact region.}
    \label{fig:kaltwinkgeo}
\end{figure}

Here we simulate the classical Kalthoff--Winkler impact experiment \cite{kalthoff1988failure}, in which a pre-notched plate subjected to dynamic shear loading exhibits brittle crack initiation at a characteristic angle. This benchmark is widely used to assess the ability of numerical methods to capture dynamic fracture under mixed-mode loading. The plate geometry ($L_0=100$~mm, $H_0=100$~mm) with a pre-existing notch of length \SI{50}{\mm} is shown in \autoref{fig:kaltwinkgeo}. The material response is modeled using the St. Venant--Kirchhoff constitutive model coupled with the hyperbolic phase-field formulation for brittle fracture, and plane-strain is assumed. The material parameters are Young's modulus $E=\SI{190}{GPa}$, Poisson's ratio $\nu=0.3$, density $\rho_0=\SI{8000}{kg/m^3}$, and fracture energy $G_c=\SI{22.13}{kJ/m^2}$. The phase-field length scale and particle spacing are set to $\epsilon_0=0.15$~mm and $\Delta x=0.125$~mm, respectively.
Impact loading is applied as a ramped velocity boundary condition in the $X_1$-direction to the region labeled \texttt{mk="2"} (auxiliary geometry). The imposed velocity profile is
\begin{equation}
v_x(t) =
\begin{cases}
\frac{t}{t_0} \cdot \SI{16.5}{\m/\s} & t \leq t_0, \\
\SI{16.5}{\m/\s} & \text{otherwise},
\end{cases}
\end{equation}
where $t_0=\SI{1}{\micro\s}$. The time-dependent profile is implemented via nested \texttt{if} clauses in the XML input file using \texttt{<mathexpressions>}. The complete case definition is provided in \autoref{app:kaltwink}.

\begin{figure}[!htbp]
    \centering
    \includegraphics[width=0.8\linewidth]{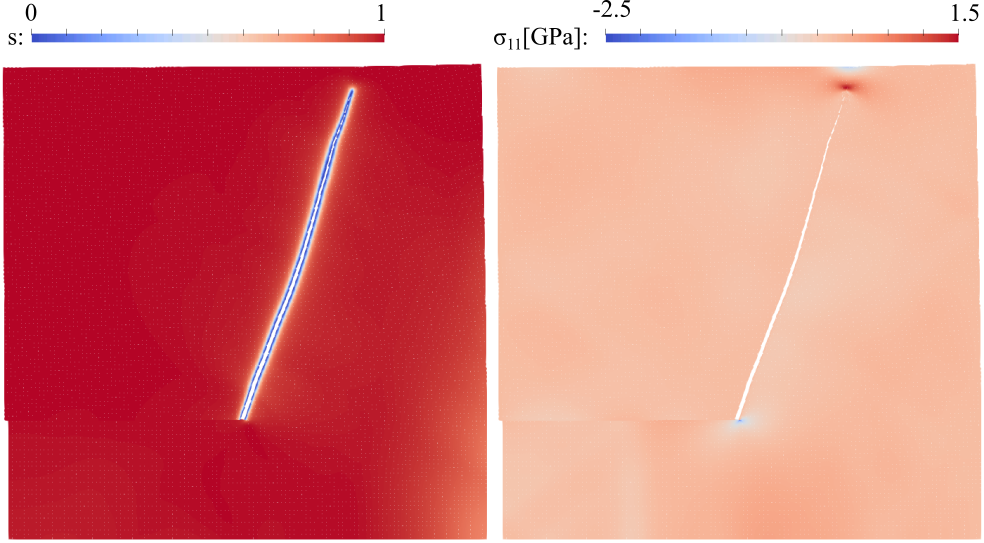}
    \caption{Kalthoff--Winkler experiment. Contours of the phase-field variable (crack path) and the Cauchy stress component $\sigma_{11}$ at $t=\SI{80}{\micro\second}$.}
    \label{fig:kalthoffstress}
\end{figure}

\begin{figure}[!htbp]
    \centering
    \includegraphics[width=0.5\linewidth]{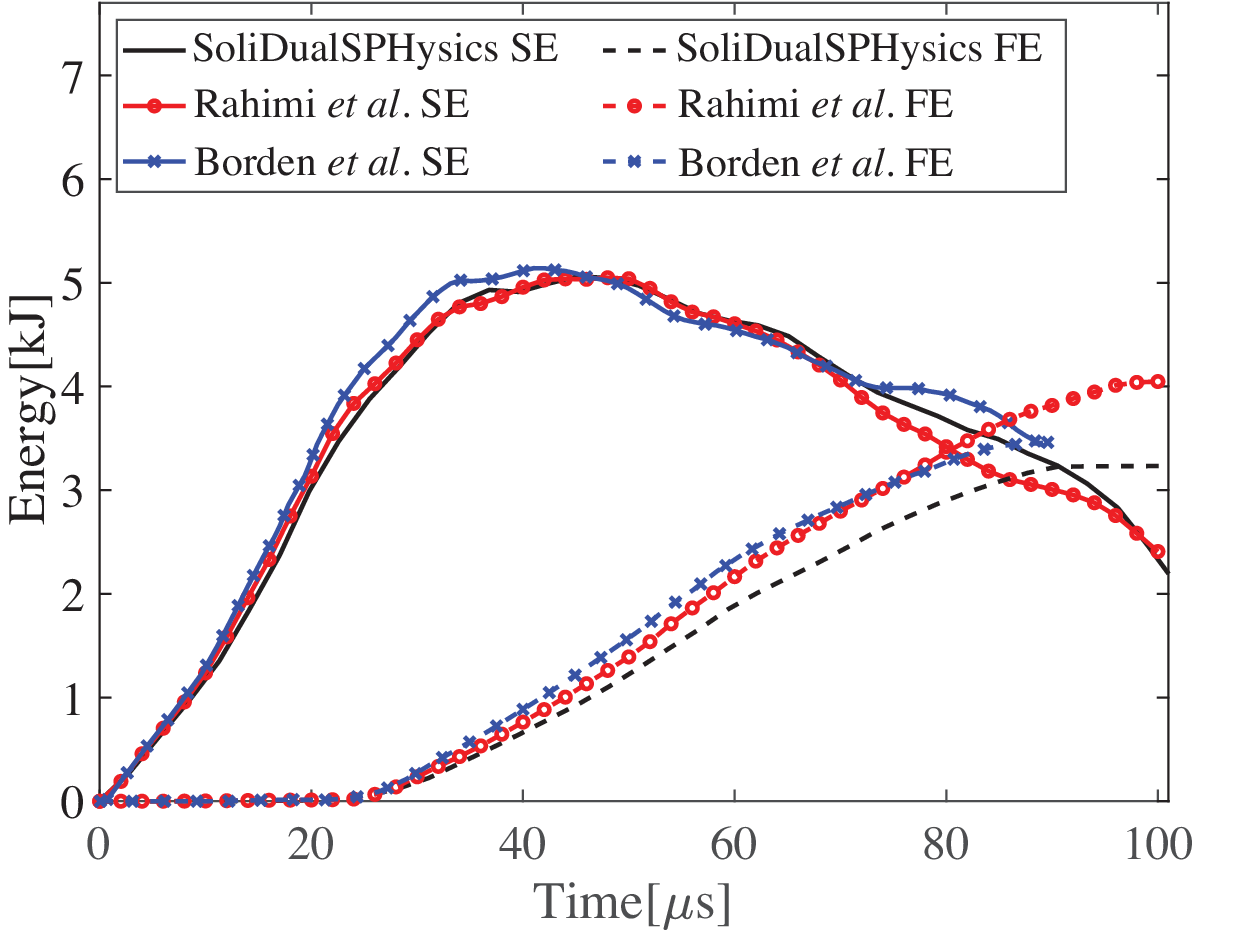}
    \caption{Kalthoff--Winkler experiment. Evolution of strain energy (SE) and fracture energy (FE), compared with results from Rahimi \etal \cite{rahimi2022Asmoothed} and Borden \etal \cite{Borden2012}.}
    \label{fig:kalthoffgraphs}
\end{figure}

\autoref{fig:kalthoffstress} shows the phase-field contours (crack path) and the $\sigma_{11}$ component of the Cauchy stress at $t=\SI{80}{\micro\second}$. Crack initiation occurs at approximately $t\approx\SI{23}{\micro\second}$, followed by diagonal crack propagation forming an angle of approximately $70^\circ$ with respect to the notch plane, consistent with experimental observations and prior numerical studies \cite{Borden2012,kalthoff1988failure,rahimi2022Asmoothed}. The stress field exhibits the expected concentration near the crack tip during propagation, without spurious oscillations.
The energy evolution is shown in \autoref{fig:kalthoffgraphs}. During crack propagation, strain energy (SE) decreases while fracture energy (FE) increases, indicating proper conversion of stored elastic energy into fracture surface energy. The predicted energy histories are in close agreement with the results reported in \cite{Borden2012,rahimi2022Asmoothed}.

\subsubsection{Four-Point Bending}\label{sec:four_point_bending}
\begin{figure}[!htbp]
    \centering
    \includegraphics[width=0.5\linewidth]{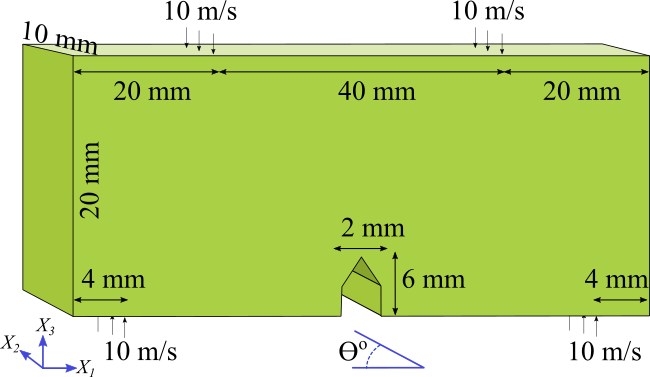}
    \caption{Four-point bending. Problem setup. $\theta_0$ denotes the initial tilt angle of the pre-existing crack in the $X_2$-direction.}
    \label{fig:fourpbendinggeo}
\end{figure}

As a final fracture benchmark, we simulate dynamic four-point bending of a pre-notched beam and investigate the influence of the initial crack inclination on subsequent crack growth. The specimen contains a pre-existing slit crack of width \SI{2}{mm} and height \SI{6}{mm}, introduced at three tilt angles $\theta_0\in\{90^\circ,60^\circ,45^\circ\}$ (tilted in the $X_2$-direction), following the benchmark of Hug \etal~\cite{hug20203d}. The beam geometry ($L_0=\SI{80}{mm}$, $H_0=\SI{20}{mm}$, $W_0=\SI{10}{mm}$) is shown in \autoref{fig:fourpbendinggeo}. The material properties are density $\rho_0=50~\mathrm{kg/m^3}$, Young’s modulus $E=12.44$~GPa, and Poisson's ratio $\nu=0.3$, and the St. Venant--Kirchhoff
constitutive model is employed. Brittle fracture is described using the hyperbolic phase-field formulation with fracture energy $G_c=11.8\times10^{3}~\mathrm{J/m^2}$ and phase-field length scale $\epsilon_0=0.25$~mm. The particle spacing is set to $\Delta x=0.2$~mm. Four-point bending is imposed through prescribed velocities applied over four localized loading/support regions on the top and bottom surfaces, with a maximum magnitude of $V_{\max}=\SI{10}{m/s}$ maintained for the duration of the simulation. The corresponding XML implementation of the velocity boundary conditions, phase-field, material parameters, and notch definition is shown in \autoref{fig:4pbendingcasedef}, whereas the complete case definition is provided in \autoref{app:4pbending}.

Strongly localized kinematic boundary conditions may induce non-physical damage nucleation in particle-based simulations due to steep gradients near constrained particles \cite{rahimi2021improved,rahimi2020ordinary,haghighat2023efficient}. To prevent spurious crack initiation at the loading points, we enforce a lower bound on the phase-field variable in small neighborhoods around the four boundary-condition regions using the \icard{restrictphi} keyword. Specifically, particles satisfying the user-defined expression with \texttt{id=1} are constrained to remain nearly intact ($s\ge 0.9999$), ensuring that fracture initiates only from the pre-existing notch while leaving the bulk response unaffected.
\begin{figure}[!htbp]
    \centering
    \begin{lstlisting}[style=xmlstyle]
<special>
  <mathexpressions>
    <userexpression id="1" comment="phi constrain">
       <expression value=" if(z0<0.00090 and x0>=0.002525 and x0<=0.005475, 0.9999, if(z0<0.00090 and x0>=0.074525 and x0<=0.077475, 0.9999, if(z0>0.01910 and x0>=0.018450 and x0<=0.021400, 0.9999, if(z0>0.01910 and x0>=0.058600 and x0<=0.061550, 0.9999,skip))))"/>
    </userexpression>
    <userexpression id="2" comment="z bc">
       <locals value="Velmax=10.0"/>
       <expression value="if(z0<0.00010 and x0>=0.003825 and x0<=0.004225, Velmax, if(z0<0.00010 and x0>=0.075625 and x0<=0.076175, Velmax, if(z0>0.01990 and x0>=0.01970 and x0<=0.020100, -Velmax, if(z0>0.01990 and x0>=0.059900 and x0<=0.060250, -Velmax, skip))))"/>
    </userexpression>
  </mathexpressions>
  <deformstrucs>
    <deformstrucbody mkbound="1">
       <bcvel ze="2" />
       <restrictphi value="1"/>
       <density value="50.0" />
       <youngmod value="12.44e9" />
       <poissratio value="0.3" />
       <fracture value="1" />
       <Gc value="11.8e3" />
       <pflenscale value="0.25e-3" />
       <notch>
         <p1 x="0.04" y="0.0e-3" z="-1.0e-3" />
         <p2 x="0.04" y="0.0e-3" z="0.0056" />
         <p3 x="#xc0" y="10.0e-3" z="0.0056" />
         <p4 x="#xc0" y="10.0e-3" z="-1.0e-3" />
       </notch>
    </deformstrucbody>
  </deformstrucs>
</special>
    \end{lstlisting}
    \caption{Four-point bending. Definition of the deformable body, velocity boundary conditions, and phase-field restriction near the loading regions.}
    \label{fig:4pbendingcasedef}
\end{figure}

\begin{figure}[!htbp]
    \centering
    \includegraphics[width=\linewidth]{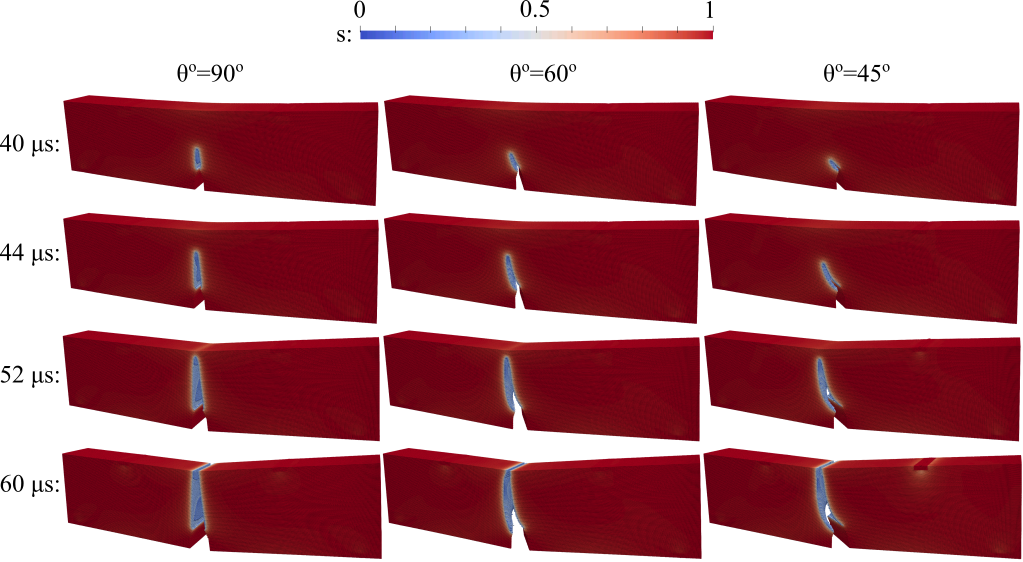}
    \caption{Four-point bending. Crack evolution visualized through contours of the phase-field variable $s$ for $\theta_0=90^\circ,60^\circ,45^\circ$ at selected times. Particles with $s<0.1$ are omitted for clarity.}
    \label{fig:fourpbendingcrackevo}
\end{figure}

\begin{figure}[!htbp]
    \centering
    \includegraphics[width=0.5\linewidth]{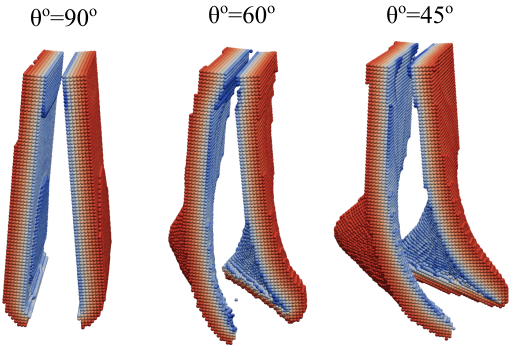}
    \caption{Four-point bending. Final fracture surfaces for $\theta_0=90^\circ,60^\circ,45^\circ$. Increased tortuosity and out-of-plane deformation are observed as the initial crack tilt enhances mixed-mode effects. Particles with $s<0.1$ are omitted for clarity.}
    \label{fig:fourpbendingcracksurface}
\end{figure}

\autoref{fig:fourpbendingcrackevo} presents the evolution of the phase-field variable for the three initial crack inclinations. In all cases, damage initiates at the pre-notched tip and evolves into a well-defined crack that propagates through the beam, demonstrating robust handling of mixed-mode crack growth under dynamic bending. For $\theta_0=90^\circ$, crack propagation is predominantly planar and mode-I dominated, rapidly localizing into a nearly straight fracture band. For inclined notches ($\theta_0=60^\circ$ and $45^\circ$), early crack growth initially follows the slit orientation and subsequently kinks as bending-induced tensile stresses dominate. This reorientation produces a curved damage band and stronger mixed-mode characteristics. At later stages, the crack reaches the tensile surface and extends along the top region, reflecting the transition from notch-controlled initiation to global bending-driven propagation.
The final fracture surfaces extracted from the damaged region are shown in \autoref{fig:fourpbendingcracksurface}. The $\theta_0=90^\circ$ case produces an essentially planar surface, whereas the inclined-notch cases develop more tortuous and non-planar fracture surfaces, indicating increased shear contribution and mixed-mode kinematics. Overall, the predicted crack paths and qualitative surface morphologies agree with the numerical and experimental trends reported in~\cite{hug20203d}, supporting the capability of \solidname to reproduce angle-dependent crack growth in three-dimensional bending.

\subsection{Plasticity Cases}

\subsubsection{Flyer Plate Impact}
\begin{figure}[!htbp]
    \centering
    \includegraphics[width=0.75\linewidth]{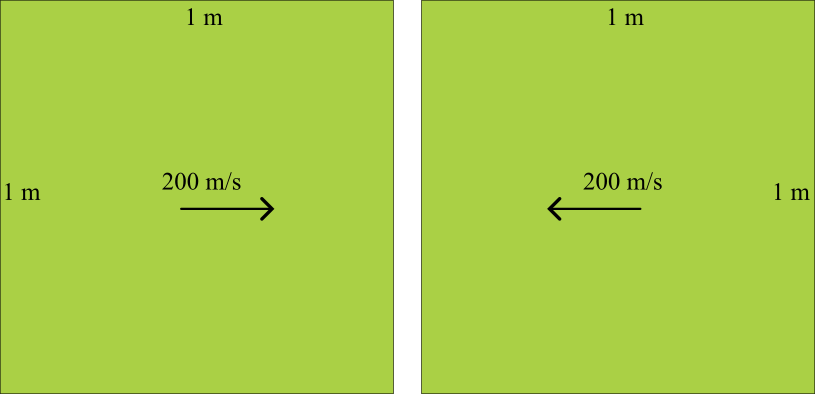}
    \caption{Flyer plate impact. Problem setup.}
    \label{fig:2dtaylorbargeo}
\end{figure}

This problem examines large-deformation plasticity and deformable--deformable contact by two identical plates impacting at a relative speed of $\SI{400}{\m/\s}$. The configuration is shown in \autoref{fig:2dtaylorbargeo}. Each plate is assigned an initial velocity of $\pm\SI{200}{\m/\s}$ (applied at $t=0$), resulting in a symmetric head-on collision about the mid-plane. The plates are modeled as square blocks of dimensions $1~\mathrm{m} \times 1~\mathrm{m}$ in a plane-strain configuration. The material response is described using the finite-strain $J_2$ elastoplastic model with isotropic hardening. Both plates are assigned density $\rho_0=7870~\mathrm{kg/m^3}$, Young's modulus $E=200$~GPa, Poisson's ratio $\nu=0.29$, initial yield stress $\sigma_{y0}=400$~MPa, and hardening modulus $H=100$~MPa. The particle spacing is set to $\Delta x=0.01$~m. Contact between the bodies is handled using the DEM-based interaction model with restitution coefficient $0.95$ and contact scaling coefficient $\texttt{contcoeff}=5$. The definition of the geometry and boundary conditions is shown in \autoref{fig:2dtaylbarcasedef}, whereas the complete XML setup can be found in \autoref{app:2dtaylorbar}.

\begin{figure}[!htbp]
    \centering
    \begin{lstlisting}[style=xmlstyle]
<special>
  <deformstrucs>
    <contcoeff value="5" />
    <deformstrucbody mkbound="1">
      <bcvel z="-200.0" tend="0.0"/>
      <density value="7870.0" />
      <youngmod value="200.0e9" />
      <poissratio value="0.29" />
      <artvisc factor1="0.05" factor2="0.0" />
      <constitmodel value="3" />
      <restcoef value="0.95" />
      <yieldstress value="4.0e8" />
      <hardening value="1.0e8" />
    </deformstrucbody>
    <deformstrucbody mkbound="2">
      <bcvel z="200.0" tend="0.0"/>
      <density value="7870.0" />
      <youngmod value="200.0e9" />
      <poissratio value="0.29" />
      <artvisc factor1="0.05" factor2="0.0" />
      <constitmodel value="3" />
      <restcoef value="0.95" />
      <yieldstress value="4.0e8" />
      <hardening value="1.0e8" />
    </deformstrucbody>
  </deformstrucs>
</special>
    \end{lstlisting}
    \caption{Flyer plate impact. Definition of deformable structure bodies and boundary conditions.}
    \label{fig:2dtaylbarcasedef}
\end{figure}

\begin{figure}[!htbp]
    \centering
    \includegraphics[width=1.0\linewidth]{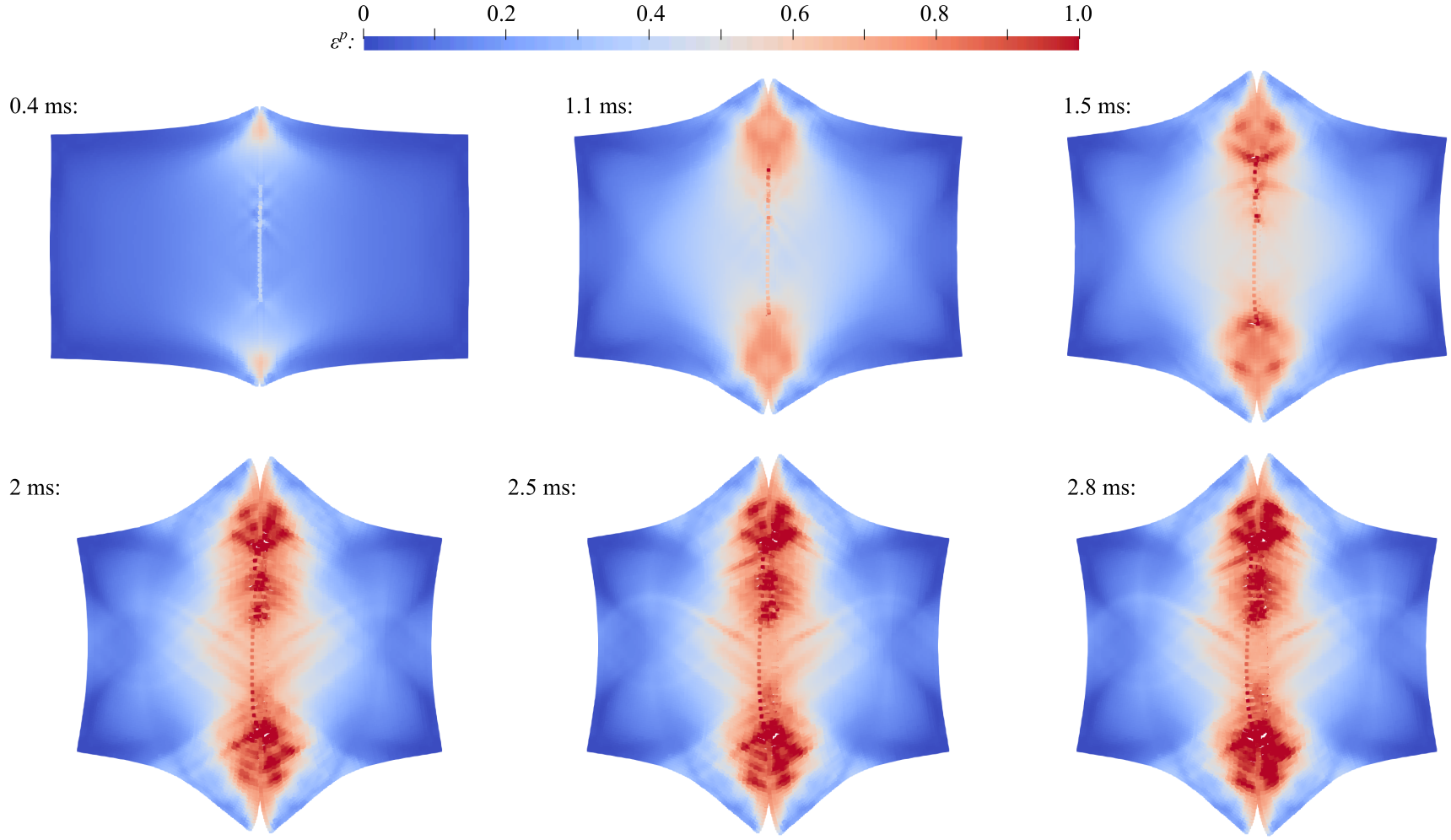}
    \caption{Flyer plate impact. Evolution of equivalent plastic strain at selected time instants following impact.}
    \label{fig:2dtaylorbarstrain}
\end{figure}

\autoref{fig:2dtaylorbarstrain} shows the evolution of equivalent plastic strain following impact. At early times, plasticity initiates at the contact interface and near the impacted corners, consistent with the high compressive stresses and sharp strain gradients generated by the collision. As time advances, the plastic zone expands away from the interface and the bodies exhibit the expected lateral spreading (barreling/mushrooming) under intense compressive loading.
At later times, as particle distributions become increasingly distorted, minor contact irregularities may appear due to the limitations of pairwise DEM-type interaction under extreme deformation. Such behavior is typical in particle-based impact simulations. A more advanced particle-to-surface contact formulation based on interface reconstruction could further improve robustness in highly distorted regimes; this extension is left for future work.

\subsubsection{3D Taylor Bar Impact}
\begin{figure}[!htbp]
    \centering
    \includegraphics[width=0.25\linewidth]{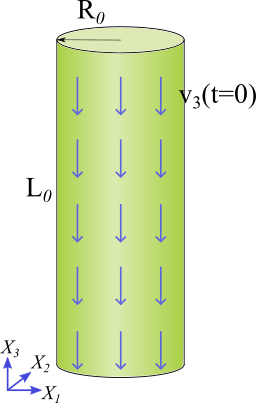}
    \caption{3D Taylor bar impact. Problem setup.}
    \label{fig:3dtaylorbargeo}
\end{figure}

Finally, we simulate the plastic deformation of a Taylor bar subjected to impact, a standard benchmark for assessing large-strain $J_2$ plasticity algorithms under high-rate loading \cite{chen2023simplified,montans2005computational}. The geometry is shown in \autoref{fig:3dtaylorbargeo} and consists of a cylindrical bar of initial length $L_0=32.4$~mm and radius $R_0=3.2$~mm. The material response is governed by the finite-strain $J_2$ plasticity model with isotropic hardening. The material parameters are density $\rho_0=\SI{8930}{kg/m^3}$, Young's modulus $E=\SI{117}{GPa}$, Poisson's ratio $\nu=0.35$, initial yield stress $\sigma_{y0}=\SI{400}{MPa}$, and hardening modulus $H=\SI{100}{MPa}$. The bar is discretized using three particle spacings of $\Delta x=0.1$~mm, $\Delta x=0.2$~mm, and $\Delta x=0.4$~mm and is assigned an initial axial velocity $v_3=-227~\mathrm{m/s}$. Impact against a rigid surface is modeled by enforcing zero axial velocity on particles located at the base ($X_3 \approx 0$), while the remainder of the bar retains the prescribed initial velocity at $t=0$. This condition is implemented via a user-defined expression (see \autoref{fig:3dtaylbarcasedef}), which sets $v_3=0$ for particles at the base and $v_3=-227~\mathrm{m/s}$ elsewhere at the initial time. The full case setup is provided in \autoref{app:taylorbar}.

\begin{figure}[!htbp]
    \centering
    \begin{lstlisting}[style=xmlstyle]
<special>
  <mathexpressions>
    <userexpression id="1">
      <locals value="Vinit=-227;"/>
      <expression value="if(z<1.0e-12,0.0,if(t<=0.0,Vinit,skip))"/>
    </userexpression>
  </mathexpressions>
  <deformstrucs>
    <deformstrucbody mkbound="1">
    <bcvel ze="1" />
    <density value="8930.0"/>
    <youngmod value="1.17e11" />
    <poissratio value="0.35" />
    <artvisc factor1="0.05" factor2="0.0" />
    <constitmodel value="3"/>
    <yieldstress value="400.0e6" />
    <hardening value="100.0e6" />
    </deformstrucbody>
  </deformstrucs>
</special>
    \end{lstlisting}
    \caption{3D Taylor bar impact. Definition of deformable structure body and boundary conditions.}
    \label{fig:3dtaylbarcasedef}
\end{figure}

\begin{figure}[!htbp]
    \centering
    \begin{subfigure}[b]{0.49\linewidth}
         \includegraphics[width=\linewidth]{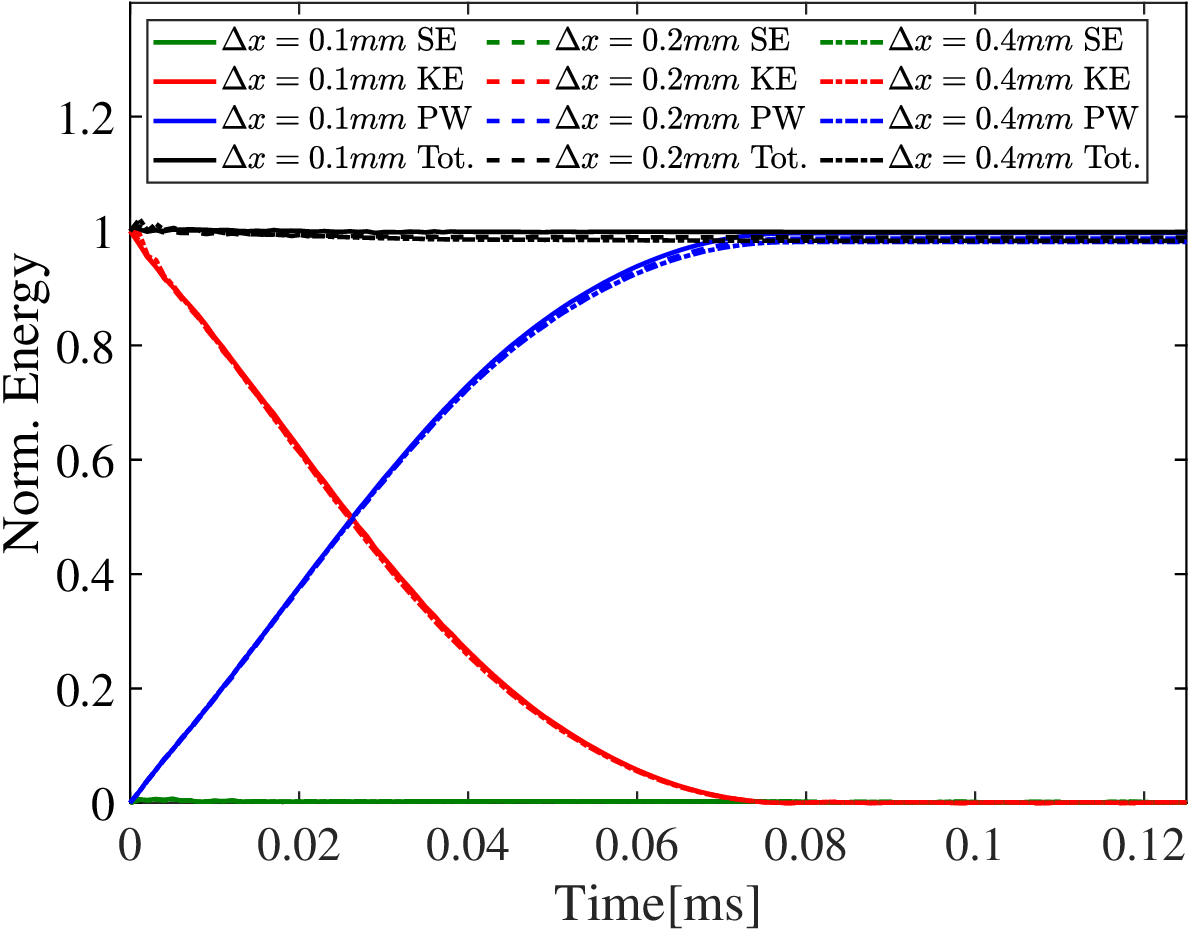}
         \caption{}
    \end{subfigure}
    \begin{subfigure}[b]{0.49\linewidth}
         \includegraphics[width=\linewidth]{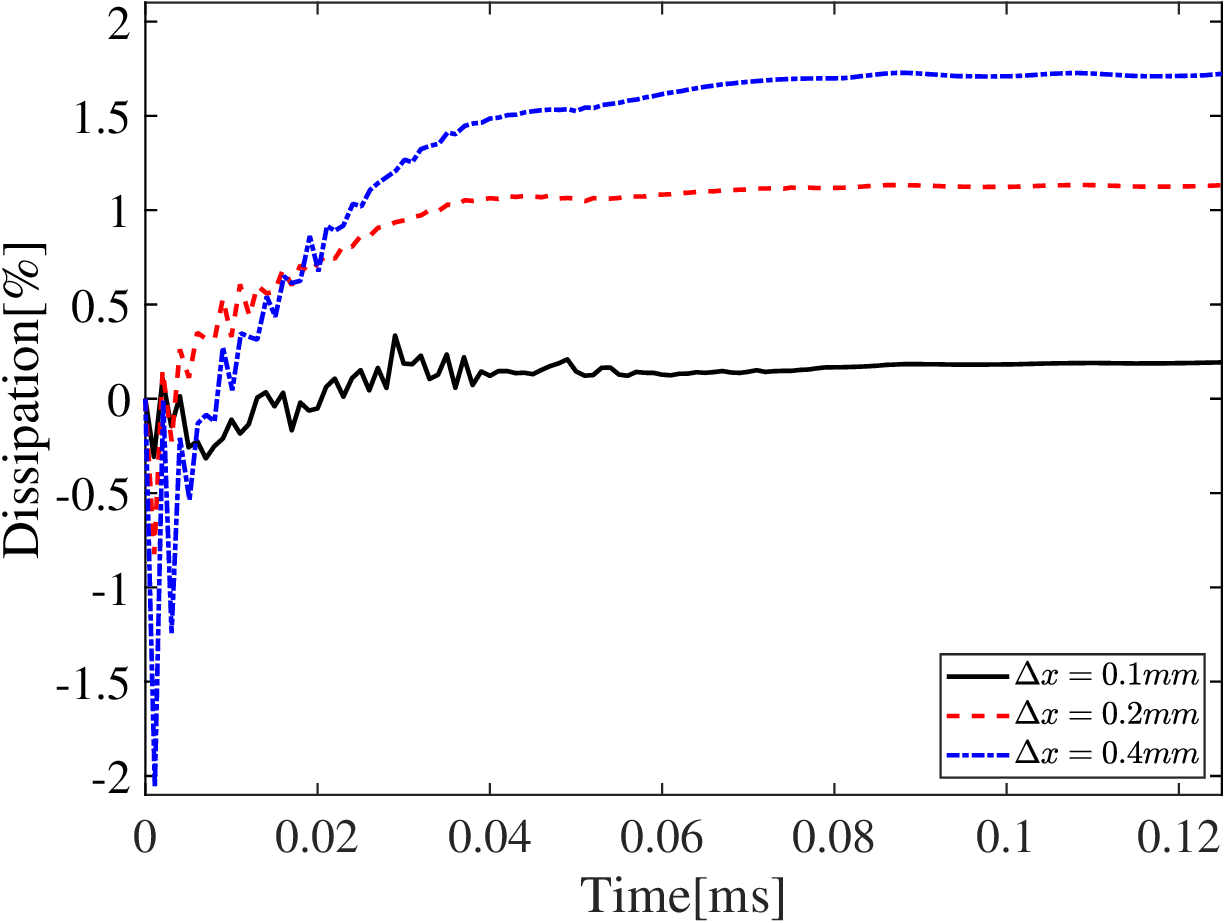}
         \caption{}
    \end{subfigure}
    \begin{subfigure}[b]{0.485\linewidth}
         \includegraphics[width=\linewidth]{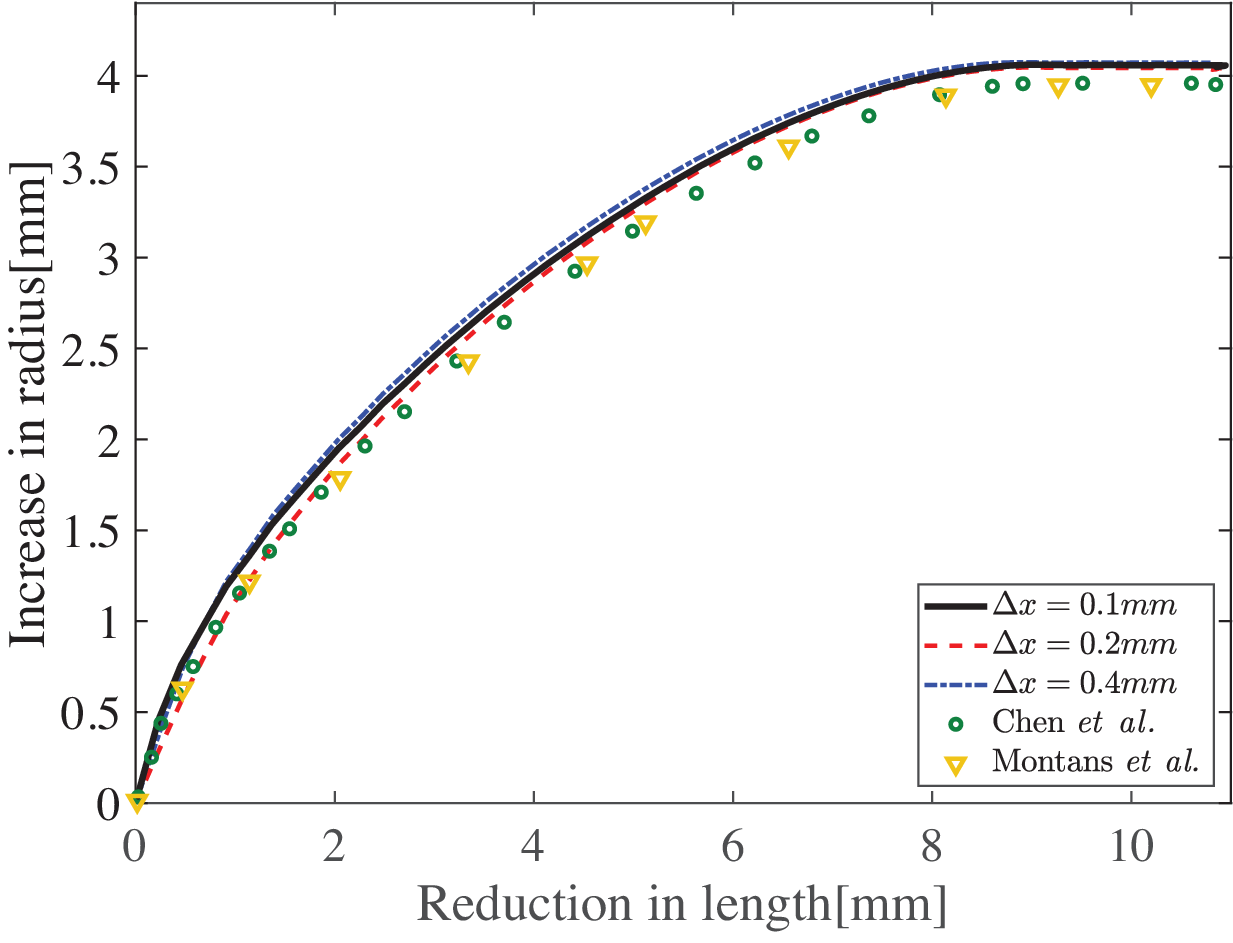}
         \caption{}
    \end{subfigure}
    \caption{3D Taylor bar impact. (a) Time evolution of energies normalized by their initial values of 247.45 J, 257.68 J, and 275.53 J for $\Delta x = 0.1$~mm, $\Delta x = 0.2$~mm, and $\Delta x = 0.4$~mm, respectively. The energies are normalized by their respective initial values because the particle-based approximation of the cylindrical geometry yields slightly different represented initial volumes, and therefore slightly different initial total energies, for different resolutions. (b) Total energy dissipation in the system. (c) Radial expansion versus axial shortening of the bar, compared with the reference solutions of Chen \etal \cite{chen2023simplified} and Montans \etal \cite{montans2005computational}.}

    \label{fig:3dtaylorbargraph}
\end{figure}

\autoref{fig:3dtaylorbargraph} presents the energy evolution, total energy dissipation, and the radial expansion versus axial shortening response for the three particle resolutions. In \autoref{fig:3dtaylorbargraph}a the kinetic energy (KE) decreases rapidly after impact, while the plastic/internal energy (PW) increases as the bar undergoes progressive inelastic deformation. After the initial transient, the energy curves approach a plateau, indicating that most of the impact-driven plastic work has already been accumulated. Since the cylindrical geometry is represented by a particle discretization, the total initial volume of the bar is not exactly identical for all particle spacings. Thus, the energies are normalized by their respective initial values to enable a meaningful comparison across resolutions. \autoref{fig:3dtaylorbargraph}b shows that the total energy dissipation remains small in all cases, confirming that the explicit integration and constitutive update remain numerically stable throughout the simulation.

The radius-shortening curves in \autoref{fig:3dtaylorbargraph}c show that all three discretizations reproduce the expected upset trend of the Taylor bar and remain in reasonable agreement with the reference solutions \cite{chen2023simplified,montans2005computational}. As expected, the coarsest discretization exhibits the largest deviation. The finer discretizations, $\Delta x=0.2$~mm and $\Delta x=0.1$~mm, both provide improved agreement, although the convergence is not strictly monotonic. This behavior is attributed to the geometric approximation of the cylindrical bar by particles, which leads to small resolution-dependent differences in the represented initial geometry, contact onset, and early plastic-flow evolution in this challenging large-deformation impact problem. Overall, the results indicate that the present formulation captures the correct global plastic flow and final upset behavior with satisfactory accuracy.
\begin{figure}[!htbp]
    \centering
    \includegraphics[width=1.0\linewidth]{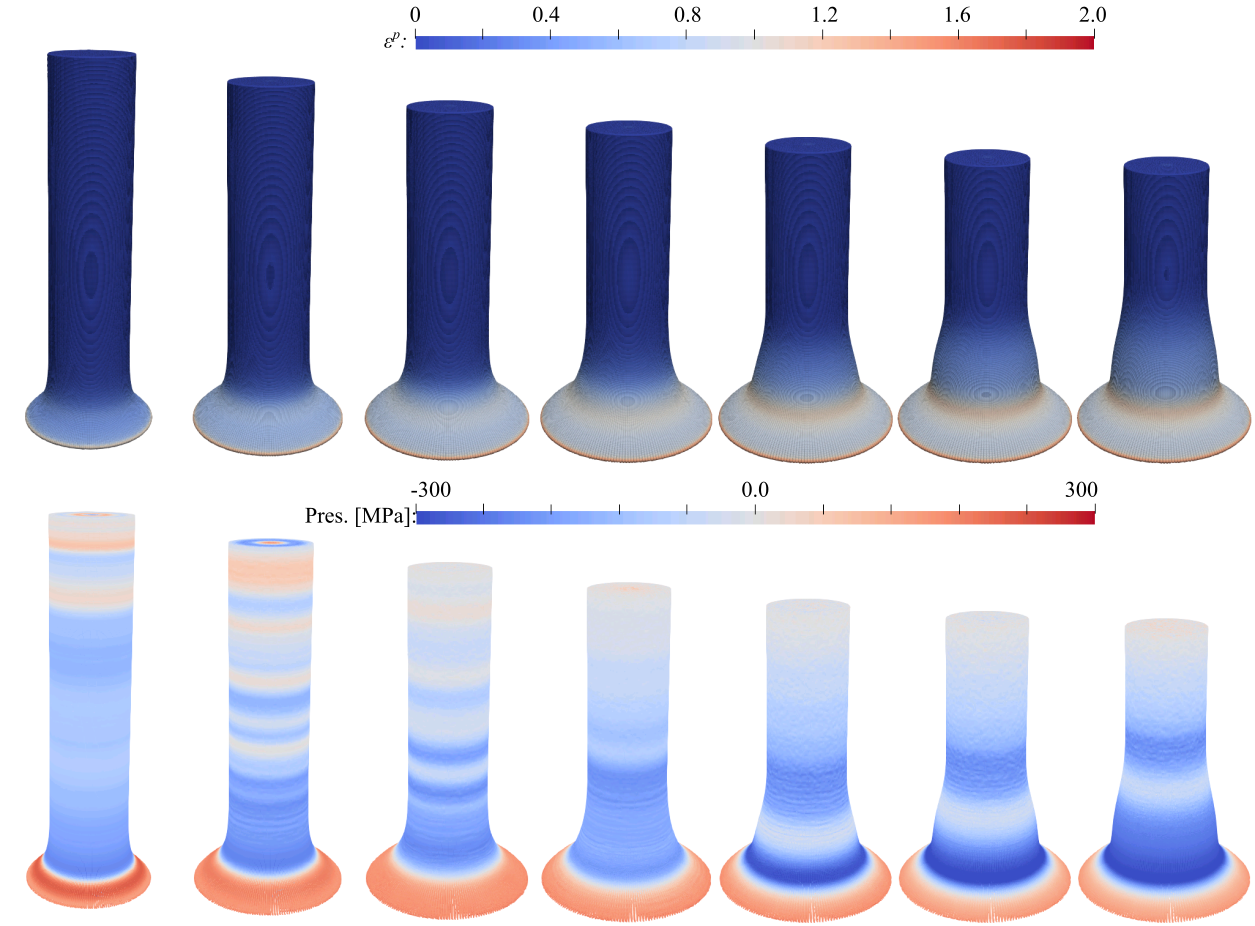}
    \caption{3D Taylor bar impact. Contours of equivalent plastic strain (top) and pressure (bottom) during the first 70~$\mu$s after impact, shown at 10~$\mu$s intervals from left to right for $\Delta x=0.1~$mm. Plastic deformation initiates near the impacted end and progressively spreads as mushrooming develops, while the pressure field highlights the strong compressive state generated at impact and its subsequent redistribution through the bar. The predicted patterns are consistent with published benchmark results \cite{chen2023simplified,montans2005computational}.}
    \label{fig:3dtaylorbarstrain}
\end{figure}

The equivalent plastic strain and pressure fields in \autoref{fig:3dtaylorbarstrain} provide further insight into the deformation mechanics. Immediately after impact, strong compressive pressure develops near the impacted end, accompanied by intense localization of plastic strain in the same region. As time progresses, the plastic zone spreads both radially and axially, and the bar develops the characteristic mushrooming shape associated with the Taylor impact benchmark. The pressure contours further confirm the strong compressive state near the contact surface and its gradual redistribution into the bulk material during the upsetting process. The predicted deformation, together with the strain and pressure distributions, is consistent with the expected physical response and with published results \cite{chen2023simplified,montans2005computational}.

\subsection{Performance Analysis}
\label{sec:performance}
This section evaluates the computational performance of the present software for the four-point bending benchmark under multiple constitutive and damage settings, and quantifies the benefits of CPU parallelism and GPU acceleration. The benchmark variants are:
\begin{itemize}
    \item CaseA1: St. Venant--Kirchhoff hyperelasticity,
    \item CaseA2: neo--Hookean hyperelasticity,
    \item CaseB1: fracture-enabled St. Venant--Kirchhoff hyperelasticity,
    \item CaseB2: fracture-enabled neo--Hookean hyperelasticity, and
    \item CaseC: finite-strain $J_2$ plasticity with isotropic hardening.
\end{itemize}
For each case, we keep numerical settings identical across hardware backends (particle resolution, time-integration settings, and solver parameters); across cases, only constitutive/fracture options differ.

\begin{figure}[!htbp]
    \centering
    \begin{subfigure}[b]{0.485\linewidth}
         \includegraphics[width=\linewidth]{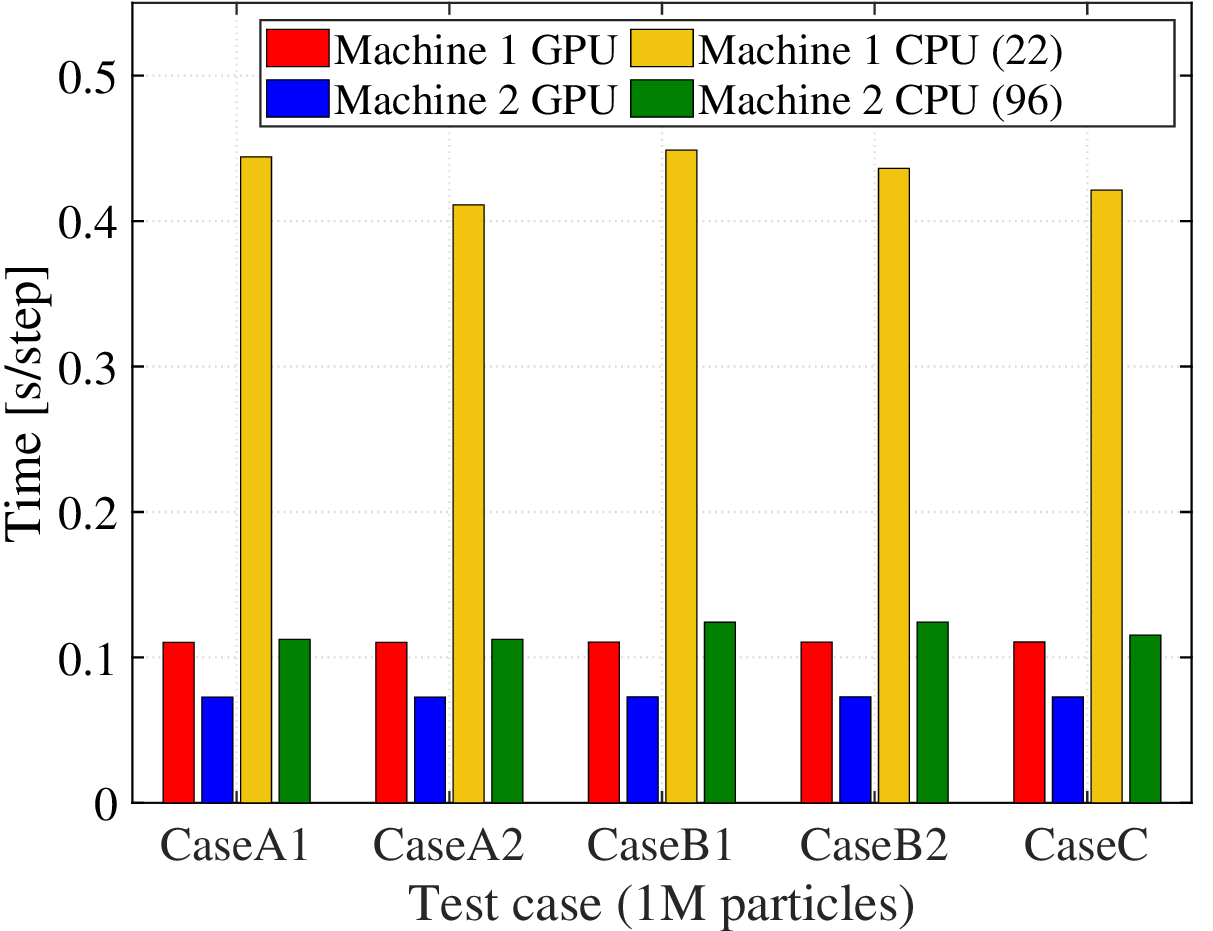}
         \caption{}
         \label{fig:perf_cases}
    \end{subfigure}
    \begin{subfigure}[b]{0.49\linewidth}
         \includegraphics[width=\linewidth]{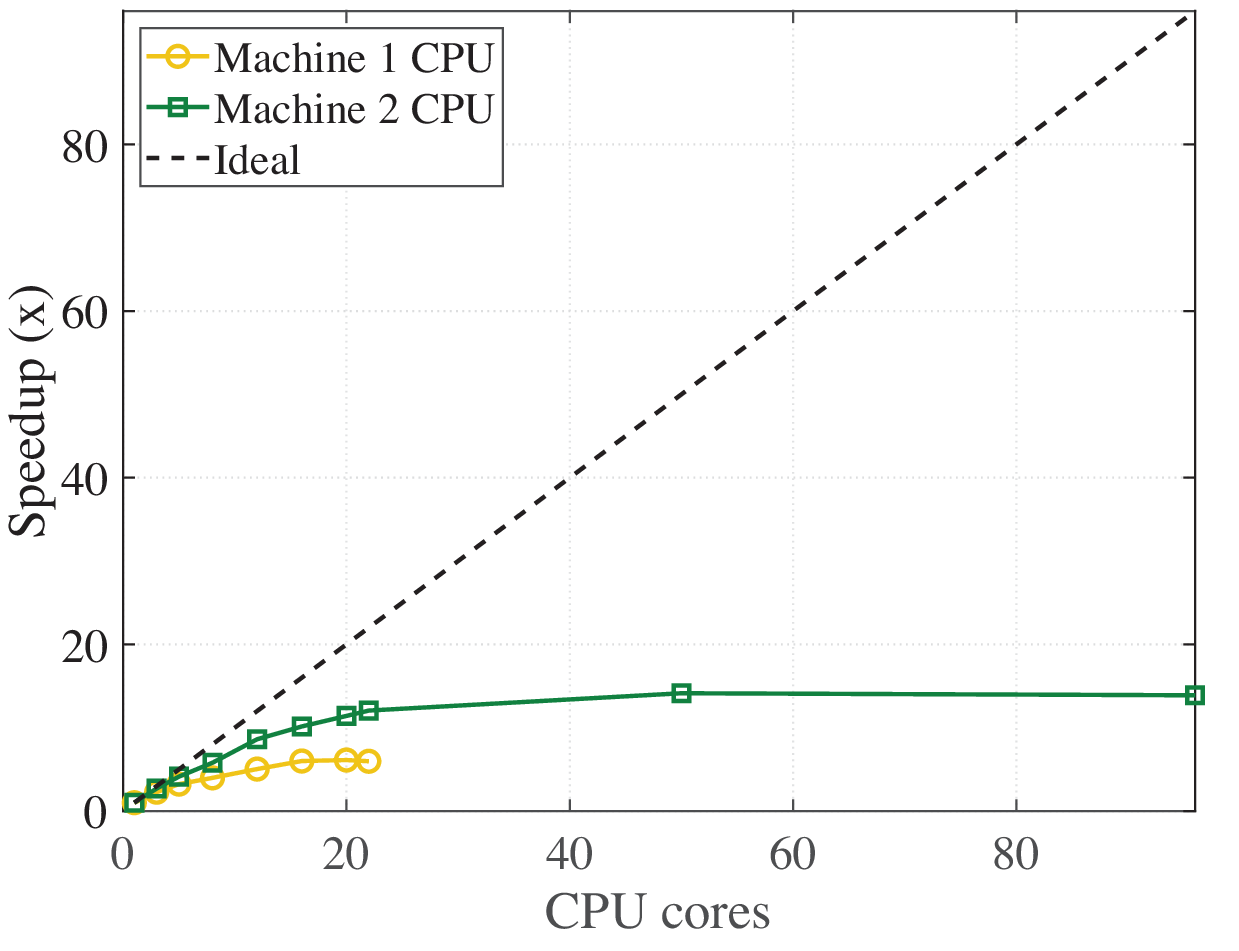}
         \caption{}
         \label{fig:perf_cpu_scaling}
    \end{subfigure}
    \begin{subfigure}[b]{0.49\linewidth}
         \includegraphics[width=\linewidth]{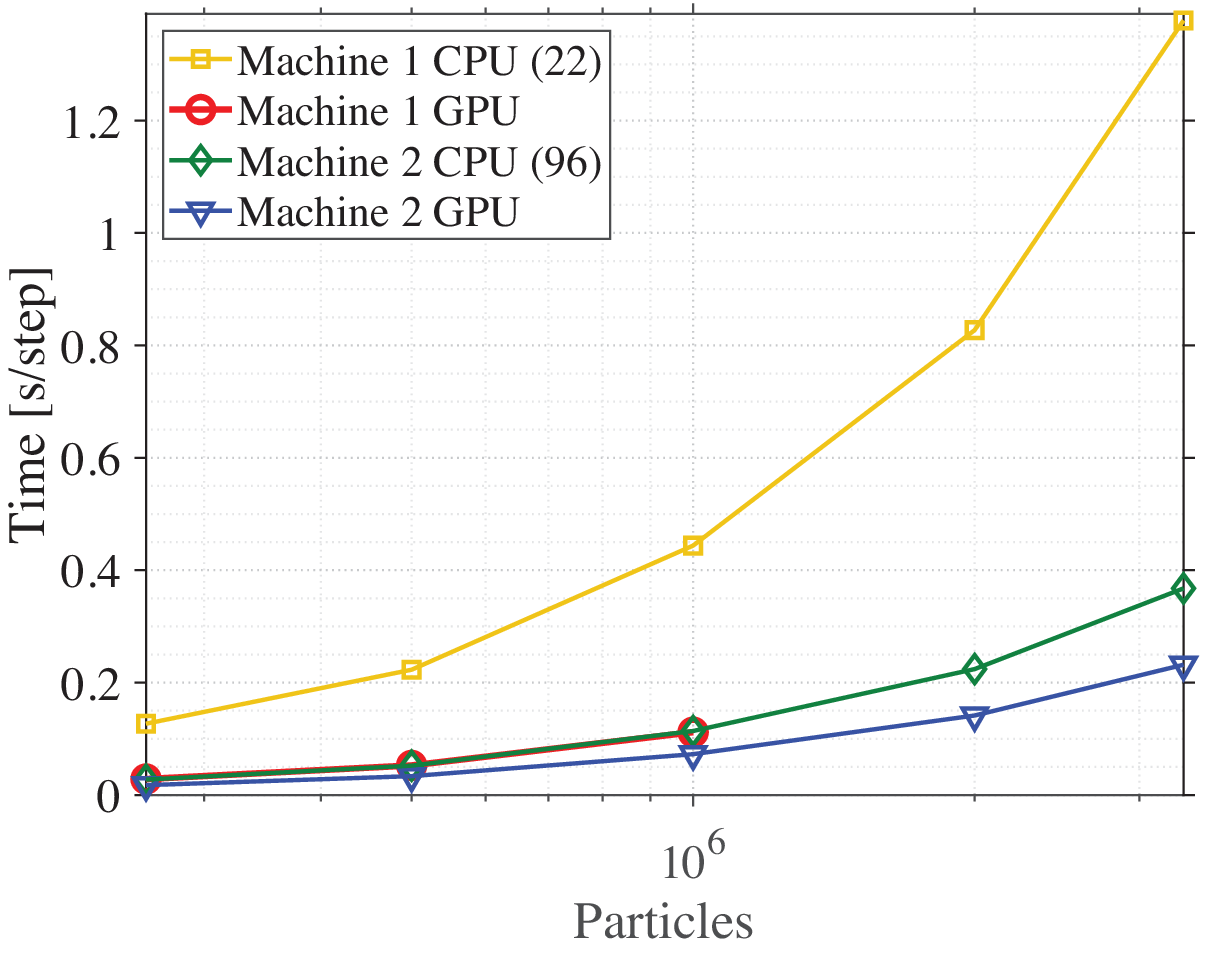}
         \caption{}
         \label{fig:perf_particles}
    \end{subfigure}
    \caption{(a) Time per step for the five benchmark variants at 1M particles on Machine~1 and Machine~2, comparing CPU execution (maximum available thread count) and GPU execution. (b) CPU strong-scaling for CaseA1 at 1M particles on Machine~1 and Machine~2; speedup is reported relative to the single-core baseline. (c) Problem-size scaling for CaseA1: time per step as a function of particle count on Machine~1 and Machine~2 for CPU (maximum available thread count) and GPU runs (particle count is shown on a logarithmic scale).}

    \label{fig:computtime}
\end{figure}

\autoref{fig:perf_cases} compares the five benchmark variants at a fixed particle count (1M particles) on both machines. Across all cases, the GPU time per step is highly stable: on Machine~1, GPU timings vary by less than 0.25\% across constitutive choices, and on Machine~2 the variation is below 0.15\%. This indicates that, at this scale, the per-step cost is dominated by kernels common to all configurations (e.g., neighbor interaction and particle update stages), while the incremental cost of switching between plasticity, SVK or NH constitutive models, or enabling fracture, remains comparatively small along the GPU path. On the CPU, the spread across cases is larger (approximately 9--11\%), consistent with the higher relative impact of constitutive and damage updates in a CPU execution path. Overall, \autoref{fig:perf_cases} confirms a consistent reduction in time per step with GPU execution on both machines.

On Machine~1, the GPU is consistently $\sim$3.7--4.1$\times$ faster than the 22-thread CPU configuration across the five cases (equivalently, $\sim$3.7--4.1$\times$ higher throughput).
On Machine~2, the GPU remains faster, but the gap is smaller because the many-core CPU is substantially stronger: the GPU provides approximately $\sim$1.55--1.71$\times$ speedup over the 96-thread CPU configuration, depending on the case.
Overall, these results show that the GPU implementation delivers robust acceleration, while the relative gain depends on the strength of the CPU subsystem and the parallel efficiency attained by the CPU implementation.

\autoref{fig:perf_cpu_scaling} reports CPU strong-scaling for CaseA1 at 1M particles. Both platforms exhibit the expected decrease in time per step as the number of CPU threads increases, followed by diminishing returns once memory bandwidth, synchronization overheads, and parallel-region costs become significant. On Machine~1, scaling improves up to the full 22-thread configuration, reaching an overall speedup of $\sim$6$\times$ relative to the single-core baseline, after which benefits saturate. On Machine~2, the stronger CPU subsystem yields better scaling, reaching $\sim$12$\times$ speedup at 22 threads and continuing to improve up to around 50 threads (speedup $\sim$14$\times$), after which scaling plateaus and may slightly regress at the largest thread count shown. This behavior is consistent with strong-scaling limits in bandwidth-bound particle interaction kernels.

\autoref{fig:perf_particles} evaluates CaseA1 across multiple particle counts using the maximum machine capacity on each backend. The results exhibit near-linear scaling of time per step with particle count (equivalently, throughput decreases approximately inversely with particle count), consistent with the expected $O(N)$ cost of particle updates when the average neighbor count is controlled.
For example, increasing the problem size from 260k to 3.345M particles reduces throughput by roughly an order of magnitude while the particle count increases by about 12.9$\times$, indicating close-to-linear scaling in the dominant kernels.
The Machine~2 GPU consistently delivers the best performance over the tested range and maintains a stable advantage over the Machine~2 CPU configuration. At the largest particle counts, some deviations from ideal linear scaling are observed, especially for Machine~1 CPU, consistent with increasing memory-bandwidth pressure in the particle interaction kernels.
 In summary, \autoref{fig:computtime} demonstrates:
\begin{itemize}
    \item stable per-step cost across a range of constitutive models and fracture settings (especially on GPU),
    \item significant acceleration from GPU execution on both platforms, and
    \item strong-scaling benefits on CPU up to moderate thread counts.
\end{itemize}

These trends indicate that further CPU performance gains would likely require improved memory locality and reduced synchronization overhead, while additional GPU improvements would primarily stem from kernel-level optimizations and reduced memory traffic.

\section{Conclusions}
\label{sec:conclusions}
We introduced \solidname, an open-source, GPU-accelerated extension of DualSPHysics for the simulation of standalone solid mechanics problems involving hyperelasticity, finite-strain plasticity, and brittle fracture within a unified smoothed particle hydrodynamics (SPH) framework. The governing equations are discretized using a total Lagrangian SPH formulation, while brittle fracture is incorporated through a hyperbolic phase-field approach that enables crack initiation, propagation, branching, and coalescence without explicit crack tracking. Finite-strain $J_2$ plasticity is formulated using a multiplicative decomposition, ensuring objectivity and thermodynamic consistency under large deformations. The framework further supports user-defined mathematical expressions for prescribing complex time- and space-dependent boundary conditions, and leverages DualSPHysics' native CPU/GPU parallel architecture to achieve high computational efficiency.

The numerical examples show very good agreement with published computational results and experimental observations, and demonstrate that the software reproduces the major qualitative and quantitative features of the considered applications. The performance study indicates near-linear problem-size scaling in the dominant kernels and substantial GPU acceleration relative to multi-core CPU execution, with expected strong-scaling saturation at high thread counts.

Overall, this work establishes \solidname as a robust and extensible solid mechanics and fracture module within the DualSPHysics ecosystem. We view the present framework as an open-source backbone for further developments by the computational mechanics community, and acknowledge that a number of additional modeling and stabilization improvements can be incorporated in future versions. In that spirit, we welcome contributions from the community to further expand and improve the capabilities of the code.

\section*{Code Availability}
The source code of SoliDualSPHysics is openly available at: \url{https://github.com/naqibr/SoliDualSPHysics}.

\section*{Acknowledgments}
This work was supported by the National Science Foundation under Grant No. 2545336. The authors are grateful to the developers of the open-source DualSPHysics code.
\clearpage
\appendix

% Title for the appendix part
\section*{Appendices: XML Input files for \solidname cases}
\addcontentsline{toc}{section}{Appendices: XML Input files for \solidname cases}

\setcounter{section}{0}
\renewcommand{\thesection}{\arabic{section}}
\renewcommand{\theHsection}{app.\arabic{section}}
\renewcommand{\sectionautorefname}{Appendix}

\section{Free oscillation of a cantilever beam}\label{app:freeoscbeam}
\begin{lstlisting}[style=xmlstyle]
<?xml version="1.0" encoding="UTF-8" ?>
<case>
  <casedef>
    <constantsdef>
      <gravity x="0" y="0" z="0.0" comment="Gravitational acceleration" units_comment="m/s^2" />
      <rhop0 value="997" comment="Reference density of the fluid" units_comment="kg/m^3" />
      <hswl value="0" auto="true" comment="Maximum still water level to calculate speedofsound using coefsound" units_comment="metres (m)" />
      <gamma value="7" comment="Polytropic constant for water used in the state equation" />
      <speedsystem value="10" auto="false" comment="Maximum system speed (by default the dam-break propagation is used)" />
      <coefsound value="10" comment="Coefficient to multiply speedsystem" />
      <coefh value="1.0" comment="Coefficient to calculate the smoothing length (h=coefh*sqrt(3*dp^2) in 3D)" />
      <cflnumber value="0.2" comment="Coefficient to multiply dt" />
    </constantsdef>
    <mkconfig boundcount="240" fluidcount="9" />
    <geometry>
      <definition dp="1.0e-3" units_comment="metres (m)">
        <pointmin x="-20.5e-3" y="0.5e-3" z="-20.5e-3" />
        <pointmax x="220.5e-3" y="0.5e-3" z="40.5e-3" />
      </definition>
      <commands>
        <mainlist>
          <newvar Lxs="-1.5e-3" Lxf="200.5e-3"/>
          <newvar Lys="0.5e-3" Lyf="19.0e-3"/>
          <newvar Lzs="0.5e-3" Lzf="19.0e-3"/>
          <newvar LzMn="9.5e-3" LzMp="10.5e-3"/>
          <setdrawmode mode="full" />
          <setshapemode> actual | bound </setshapemode>
          <setmkbound mk="1" />
          <drawbox>
            <boxfill>solid</boxfill>
            <point x="#Lxs" y="#Lys" z="#Lzs" />
            <size x="#Lxf" y="#Lyf" z="#Lzf" />
          </drawbox>
        </mainlist>
      </commands>
    </geometry>
    <motion>
      <objreal ref="1">
        <begin mov="1" start="0" />
        <mvnull id="1" />
      </objreal>
    </motion>
  </casedef>
  <execution>
    <special>
      <mathexpressions>
        <userexpression id="1" comment="Math expression">
          <locals value="L0=0.2; kw=9.375; cs=57.0"/>
          <expression value="if(x0<=0.0,0.0,if(t<=0.0,0.01 * cs * ((cos(kw*L0)+cosh(kw*L0))*(cosh(kw*x0)-cos(kw*x0)) + (sin(kw*L0)-sinh(kw*L0))*(sinh(kw*x0)-sin(kw*x0)))/ ((cos(kw*L0)+cosh(kw*L0))*(cosh(kw*L0)-cos(kw*L0)) + (sin(kw*L0)-sinh(kw*L0))*(sinh(kw*L0)-sin(kw*L0))),skip))"/>
        </userexpression>
        <userexpression id="2" comment="Math expression">
          <expression value="if(x0<=0.0,0.0,skip)"/>
        </userexpression>
      </mathexpressions>
      <deformstrucs>
        <deformstrucbody mkbound="1">
          <bcvel ze="1" xe="2" ye="2" comment="Velocity BC" />          
          <density value="1000.0" comment="Mass density" units_comment="kg/m^3" />
          <u_mu value="0.715e6" comment="Shear Modulus" units_comment="Pa" />
          <u_bulk value="3.25e6" comment="Bulk Modulus" units_comment="Pa" />
          <constitmodel value="1" comment="Constitutive model 1:SVK" />
          <artvisc factor1="0.015" factor2="0.01" comment=" Art. Visc." />
          <mapfac value="4" comment="x4 refinement" />
          <measureplane comment="Measure tip displacement">
            <p1 x="199.999e-3" y="#Lys" z="#LzMn" />
            <p2 x="199.999e-3" y="#Lyf + 0.5e-3" z="#LzMn" />
            <p3 x="199.999e-3" y="#Lyf + 0.5e-3" z="#LzMp" />
            <p4 x="199.999e-3" y="#Lys" z="#LzMp" />
          </measureplane>
        </deformstrucbody>
      </deformstrucs>
    </special>
    <parameters>
      <parameter key="StepAlgorithm" value="1" comment="Step Algorithm 1:Verlet, 2:Symplectic" />
      <parameter key="Kernel" value="2" comment="Interaction Kernel 1:Cubic Spline, 2:Wendland" />
      <parameter key="Visco" value="8.92678034e-7" comment="Viscosity value" />
      <parameter key="TimeMax" value="1.0" comment="Time of simulation" units_comment="seconds" />
      <parameter key="TimeOut" value="0.001" comment="Time out data" units_comment="seconds" />
      <simulationdomain comment="Defines domain of simulation">
        <posmin x="default-10%" y="default- 10%" z="default - 10%" comment="e.g.: x=0.5, y=default-1, z=default-10%" />
        <posmax x="default+10%" y="default+ 10%" z="default + 10%" />
      </simulationdomain>
    </parameters>
  </execution>
</case>
\end{lstlisting}

\section{Free oscillation of a cantilever plate}\label{app:freeoscplate}
\begin{lstlisting}[style=xmlstyle]
<?xml version="1.0" encoding="UTF-8" ?>
<case>
  <casedef>
    <constantsdef>
      <gravity x="0" y="0" z="0.0"/>
      <rhop0 value="997" />
      <hswl value="0" auto="true" />
      <gamma value="7" />
      <speedsystem value="10" auto="false" />
      <coefsound value="10" />
      <coefh value="1.0" />
      <cflnumber value="0.2" />
    </constantsdef>
    <mkconfig boundcount="240" fluidcount="9"/>
    <geometry>
      <definition dp="1.0e-3" units_comment="metres (m)">
        <pointmin x="-20.5e-3" y="-40.5e-3" z="-20.5e-3"/>
        <pointmax x="220.5e-3" y="100.5e-3" z="40.5e-3"/>
      </definition>
      <commands>
        <mainlist>
          <newvar Lxs="-1.5e-3" Lxf="200.5e-3"/>
          <newvar Lys="0.5e-3" Lyf="59.0e-3"/>
          <newvar Lzs="0.5e-3" Lzf="19.0e-3"/>
          <newvar LzMn="9.5e-3" LzMp="10.5e-3"/>
          <newvar LyMn="0.0" LyMp="60.0e-3"/>
          <setdrawmode mode="full"/>
          <setshapemode> actual | bound </setshapemode>
          <setmkbound mk="1"/>
          <drawbox>
            <boxfill>solid</boxfill>
            <point x="#Lxs" y="#Lys" z="#Lzs"/>
            <size x="#Lxf" y="#Lyf" z="#Lzf"/>
          </drawbox>
        </mainlist>
      </commands>
    </geometry>
    <motion>
      <objreal ref="1">
        <begin mov="1" start="0"/>
        <mvnull id="1"/>
      </objreal>
    </motion>
  </casedef>
  <execution>
    <special>
      <mathexpressions>
        <userexpression id="1" comment="Math expression">
          <locals value="L0=0.2; kw=9.375; cs=57.0"/>
          <expression value="if(x0<=0.0,0.0,if(t<1.0e-12,0.01 * cs * ((cos(kw*L0)+cosh(kw*L0))*(cosh(kw*x0)-cos(kw*x0)) + (sin(kw*L0)-sinh(kw*L0))*(sinh(kw*x0)-sin(kw*x0)))/ ((cos(kw*L0)+cosh(kw*L0))*(cosh(kw*L0)-cos(kw*L0)) + (sin(kw*L0)-sinh(kw*L0))*(sinh(kw*L0)-sin(kw*L0))),skip))"/>
        </userexpression>
        <userexpression id="2" comment="Math expression">
          <expression value="if(x0<=0.0,0.0,skip)"/>
        </userexpression>
      </mathexpressions>
      <deformstrucs>
        <deformstrucbody mkbound="1">
          <bcvel ze="1" xe="2" ye="2" comment="Velocity BC"/>          
          <density value="1000.0" />
          <u_mu value="0.715e6" />
          <u_bulk value="3.25e6" />
          <constitmodel value="1" comment="Const. model 1:SVK"/>
          <artvisc factor1="0.1" factor2="0.0" />
          <mapfac value="2" comment="x2 refinement"/>
          <measureplane comment="Measure tip displacement at the free end">
            <p1 x="199.999e-3" y="#LyMn" z="#LzMn"/>
            <p2 x="199.999e-3" y="#LyMp" z="#LzMn"/>
            <p3 x="199.999e-3" y="#LyMp" z="#LzMp"/>
            <p4 x="199.999e-3" y="#LyMn" z="#LzMp"/>
          </measureplane>
        </deformstrucbody>
      </deformstrucs>
    </special>
    <parameters>
      <parameter key="StepAlgorithm" value="1" />
      <parameter key="Kernel" value="2" />
      <parameter key="Visco" value="8.92678034e-7" />
      <parameter key="TimeMax" value="1.0" />
      <parameter key="TimeOut" value="0.001" />
      <simulationdomain >
        <posmin x="default-10%" y="default- 10%" z="default - 10%" />
        <posmax x="default+10%" y="default+ 10%" z="default + 10%"/>
      </simulationdomain>
    </parameters>
  </execution>
</case>
\end{lstlisting}

\section{Large deformation of a 3D cantilever beam}\label{app:deformcolumn3d}
\begin{lstlisting}[style=xmlstyle]
<?xml version="1.0" encoding="UTF-8" ?>
<case>
  <casedef>
    <constantsdef>
      <gravity x="0" y="0" z="0.0" />
      <rhop0 value="997" />
      <hswl value="0" auto="true" />
      <gamma value="7" />
      <speedsystem value="10" auto="false" />
      <coefsound value="10" />
      <coefh value="1.0" />
      <cflnumber value="0.2" />
    </constantsdef>
    <mkconfig boundcount="240" fluidcount="9"/>
    <geometry>
      <definition dp="1.0e-3" units_comment="metres (m)">
        <pointmin x="-5.5e-3" y="-1.5e-3" z="-1.5e-3"/>
        <pointmax x="110.5e-3" y="20.5e-3" z="20.5e-3"/>
      </definition>
      <commands>
        <mainlist>
          <newvar Lx="100.0e-3" Ly="9.0e-3" Lz="9.0e-3"/>
          <setdrawmode mode="full"/>
          <setmkbound mk="1"/>
          <drawbox>
            <boxfill>solid</boxfill>
            <point x="-1.5e-3" y="0.50e-3" z="0.50e-3"/>
            <size x="101.0e-3" y="#Ly" z="#Lz"/>
          </drawbox>
        </mainlist>
      </commands>
    </geometry>
    <motion>
      <objreal ref="1">
        <begin mov="1" start="0"/>
        <mvnull id="1"/>
      </objreal>
    </motion>
  </casedef>
  <execution>
    <special>
      <mathexpressions>
        <userexpression id="1" comment="Math expression">
          <locals value="Fmax=-1.75e9; Tmax=1.0; xtip=0.099;"/>
           <expression value="if(x0>xtip,if(t<=Tmax,t/Tmax,1.0)*Fmax,skip)"/>
        </userexpression>
        <userexpression id="2" comment="Math expression">
          <expression value="if(x0<=0.0,0.0,skip)"/>
        </userexpression>
      </mathexpressions>
      <deformstrucs>
        <deformstrucbody mkbound="1">
          <bcforce type="2" ze="1" comment="Distributed load"/>
          <bcvel xe="2" ye="2" ze="2" comment="Velocity constraint in x,y,z"/> 
          <density value="7800.0" />
          <youngmod value="210.0e9" />
          <poissratio value="0.3" />
          <constitmodel value="2" comment="Const. model 2:Neo-Hookean"/>
          <artvisc factor1="0.1" factor2="0.0" />
          <mapfac value="1" comment="x1 refinement"/>
          <measureplane comment="Measure tip displacement">
            <p1 x="100.4e-3" y="#Ly*0.455" z="#Lz*0.455"/>
            <p2 x="100.4e-3" y="#Ly*0.555" z="#Lz*0.455"/>
            <p3 x="100.4e-3" y="#Ly*0.555" z="#Lz*0.555"/>
            <p4 x="100.4e-3" y="#Ly*0.455" z="#Lz*0.555"/>
          </measureplane>
        </deformstrucbody>
      </deformstrucs>
    </special>
    <parameters>
      <parameter key="StepAlgorithm" value="1" />
      <parameter key="Kernel" value="2" />
      <parameter key="Visco" value="8.92678034e-7" />
      <parameter key="CoefDtMin" value="0.01" />
      <parameter key="TimeMax" value="2.0" />
      <parameter key="TimeOut" value="0.002" />
      <simulationdomain >
        <posmin x="default-10%" y="default- 10%" z="default - 10%"/>
        <posmax x="default+10%" y="default+ 10%" z="default + 10%"/>
      </simulationdomain>
    </parameters>
  </execution>
</case>
\end{lstlisting}

\section{Twisting 3D column}\label{app:twistingcolumn}
\begin{lstlisting}[style=xmlstyle]
<?xml version="1.0" encoding="UTF-8" ?>
<case>
  <casedef>
    <constantsdef>
      <gravity x="0" y="0" z="0.0"/>
      <rhop0 value="997" />
      <hswl value="0" auto="true" />
      <gamma value="7" />
      <speedsystem value="10" auto="false" />
      <coefsound value="10" />
      <coefh value="1.0" />
      <cflnumber value="0.01"/>
    </constantsdef>
    <mkconfig boundcount="240" fluidcount="9" />
    <geometry>
      <definition dp="0.1" units_comment="metres (m)">
        <pointmin x="-1.05" y="-1.05" z="-1.05" />
        <pointmax x="1.55" y="1.55" z="6.55" />
      </definition>
      <commands>
        <mainlist>
          <newvar Lx="0.95" Ly="0.95" Lz="5.95"/>
          <setdrawmode mode="full" />
          <setmkbound mk="1" />
          <drawbox>
            <boxfill>solid</boxfill>
            <point x="0.05" y="0.05" z="-0.4" />
            <size x="#Lx" y="#Ly" z="6.35" />
          </drawbox>
        </mainlist>
      </commands>
    </geometry>
    <motion>
      <objreal ref="1">
        <begin mov="1" start="0" />
        <mvnull id="1" />
      </objreal>
    </motion>
  </casedef>
  <execution>
    <special>
      <mathexpressions>
        <userexpression id="1" comment="User expression">
          <locals value="xcent=0.5; ycent=0.5; omega=105.0"/>
          <expression value="if(z0<=0.0,0.0,if(t<=0,omega*sin(0.2619047*z0)*(ycent-y0),skip))"/>
        </userexpression>
        <userexpression id="2" comment="User expression">
          <locals value="xcent=0.5; ycent=0.5; omega=105.0"/>
          <expression value="if(z0<=0.0,0.0,if(t<=0,omega*sin(0.2619047*z0)*(x0-xcent),skip))"/>
        </userexpression>
        <userexpression id="3" comment="User expression">
          <expression value="if(z0<=0.0,0.0,skip)"/>
        </userexpression>
      </mathexpressions>
      <deformstrucs>
        <deformstrucbody mkbound="1">
          <bcvel xe="1" ye="2" ze="3" comment="Velocity BC" />
          <density value="1100.0" comment="Mass density"/>
          <youngmod value="170.0e5" comment="Young Modulus"/>
          <poissratio value="0.45" comment="Poisson ratio" />
          <artvisc factor1="0.5" factor2="0.0" comment="Art. Visc." />
          <constitmodel value="2" comment="Const. model 2:Neo-Hookean" />
          <mapfac value="4" />
        </deformstrucbody>
      </deformstrucs>
    </special>
    <parameters>
      <parameter key="StepAlgorithm" value="1" />
      <parameter key="Kernel" value="2" />
      <parameter key="Visco" value="8.92678034e-7" />
      <parameter key="CoefDtMin" value="0.001" />
      <parameter key="TimeMax" value="1.5"/>
      <parameter key="TimeOut" value="0.001"/>
      <simulationdomain>
        <posmin x="default-10%" y="default- 10%" z="default - 10%" />
        <posmax x="default+10%" y="default+ 10%" z="default + 10%" />
      </simulationdomain>
    </parameters>
  </execution>
</case>
\end{lstlisting}

\section{Dynamic crack branching}\label{app:dyncrackbranch}
\begin{lstlisting}[style=xmlstyle]
<?xml version="1.0" encoding="UTF-8" ?>
<case>
  <casedef>
    <constantsdef>
      <gravity x="0" y="0" z="0.0" />
      <rhop0 value="997" />
      <hswl value="0" auto="true" />
      <gamma value="7" />
      <speedsystem value="10" auto="false" />
      <coefsound value="10" />
      <coefh value="1.0" />
      <cflnumber value="0.2" />
    </constantsdef>
    <mkconfig boundcount="240" fluidcount="9" />
    <geometry>
      <definition dp="0.125e-3" units_comment="metres (m)">
        <pointmin x="-1.06125e-3" y="0.50e-3" z="-1.06125e-3" />
        <pointmax x="100.06125e-3" y="0.50e-3" z="50.06125e-3" />
      </definition>
      <commands>
        <mainlist>
          <setdrawmode mode="full" />
          <setmkbound mk="3" />
          <setshapemode> actual </setshapemode>
          <drawbox>
            <boxfill>solid</boxfill>
            <point x="0.06125e-3" y="0.06125e-3" z="39.9385e-3" />
            <size x="99.9385e-3" y="0.9385e-3" z="0.125e-3" />
          </drawbox>
          <setmkbound mk="2" />
          <setshapemode> actual </setshapemode>
          <drawbox>
            <boxfill>solid</boxfill>
            <point x="0.06125e-3" y="0.06125e-3" z="-0.06125e-3" />
            <size x="99.9385e-3" y="0.9385e-3" z="0.06125e-3" />
          </drawbox>
          <setmkbound mk="1" />
          <setshapemode> actual </setshapemode>
          <drawbox>
            <boxfill>solid</boxfill>
            <point x="0.06125e-3" y="0.0" z="0.06125e-3" />
            <size x="99.9385e-3" y="1.0e-3" z="39.9385e-3" />
          </drawbox>
        </mainlist>
      </commands>
    </geometry>
    <motion>
      <objreal ref="1">
        <begin mov="1" start="0" />
        <mvnull id="1" />
      </objreal>
    </motion>
  </casedef>
  <execution>
    <special>
      <deformstrucs>
        <deformstrucbody mkbound="1">
          <nbsrange value="1"/>
          <bcforce type="2" mkid="3" z="1.0e6"/>
          <bcforce type="2" mkid="2" z="-1.0e6"/>
          <notch>  
            <p1 x="-2.0e-3" y="-5.0e-3" z="0.02" />
            <p2 x="50.030625e-3" y="-5.0e-3" z="0.02" />
            <p3 x="50.030625e-3" y="25.0e-3" z="0.02" />
            <p4 x="-2.0e-3" y="25.0e-3" z="0.02" />
          </notch>
          <density value="2450.0"/>
          <youngmod value="32.0e9" />
          <poissratio value="0.2" />
          <constitmodel value="1" />
          <artvisc factor1="0.2" factor2="0.0"/>
          <fracture value="1" />
          <Gc value="3.0" />
          <pflim value="0.05" />
          <pflenscale value="0.12501e-3" />
          <mapfac value="2" />
        </deformstrucbody>
      </deformstrucs>
    </special>
    <parameters>
      <parameter key="StepAlgorithm" value="1" />
      <parameter key="Kernel" value="2" />
      <parameter key="Visco" value="8.92678034e-7" />
      <parameter key="CoefDtMin" value="0.001" />
      <parameter key="TimeMax" value="120.0e-6" />
      <parameter key="TimeOut" value="1.0e-6" />
      <simulationdomain>
        <posmin x="default-10%" y="default- 10%" z="default - 10%" />
        <posmax x="default+10%" y="default+ 10%" z="default + 10%" />
      </simulationdomain>
    </parameters>
  </execution>
</case>
\end{lstlisting}

\section{Kalthoff--Winkler experiment}\label{app:kaltwink}
\begin{lstlisting}[style=xmlstyle]
<?xml version="1.0" encoding="UTF-8" ?>
<case>
  <casedef>
    <constantsdef>
      <gravity x="0" y="0" z="0.0"/>
      <rhop0 value="997"/>
      <hswl value="0" auto="true" />
      <gamma value="7" />
      <speedsystem value="10" auto="false" />
      <coefsound value="10" />
      <coefh value="1.0" />
      <cflnumber value="0.2"/>
    </constantsdef>
    <mkconfig boundcount="240" fluidcount="9" />
    <geometry>
      <definition dp="1.0e-3" units_comment="metres (m)">
        <pointmin x="-2.0e-3" y="0.50e-3" z="-2.0e-3" />
        <pointmax x="101.0e-3" y="0.50e-3" z="100.0e-3" />
      </definition>
      <commands>
        <mainlist>
          <setdrawmode mode="full" />
          <setmkbound mk="1" />
          <drawbox>
            <boxfill>solid</boxfill>
            <point x="0.5e-3" y="0.0" z="0.5e-3" />
            <size x="99.5e-3" y="10.0e-3" z="99.5e-3" />
          </drawbox>
          <setmkbound mk="2" />
          <drawbox>
            <boxfill>solid</boxfill>
            <point x="-0.5e-3" y="0.0" z="0.5e-3" />
            <size x="0.5e-3" y="10.0e-3" z="24.5e-3" />
          </drawbox>
          <setmkbound mk="3" />
          <drawbox>
            <boxfill>solid</boxfill>
            <point x="0.5e-3" y="0.0" z="-0.5e-3" />
            <size x="99.5e-3" y="10.0e-3" z="0.5e-3" />
          </drawbox>
        </mainlist>
      </commands>
    </geometry>
    <motion>
      <objreal ref="1">
        <begin mov="1" start="0" />
        <mvnull id="1" />
      </objreal>
    </motion>
  </casedef>
  <execution>
    <special>
      <mathexpressions>
        <userexpression id="2">
          <locals value="maxv=16.5; ramt=1.0e-6"/>
          <expression value="if(t>ramt,maxv,t/ramt*maxv)"/>
        </userexpression>
      </mathexpressions>
      <deformstrucs>
        <deformstrucbody mkbound="1">
          <nbsrange value="1"/>
          <bcvel mkid="2" xe="2"/>
          <bcvel mkid="3" z="0.0"/>
          <density value="8000.0" />
          <youngmod value="190.0e9" />
          <poissratio value="0.3" />
          <constitmodel value="1" />
          <artvisc factor1="0.1" />
          <fracture value="1" />
          <Gc value="22.13e3" />
          <pflim value="0.05" />
          <pflenscale value="0.15e-3" />
          <mapfac value="8" />
          <notch>
            <p1 x="0.0e-3" y="-1.0e-3" z="25.6e-3" />
            <p2 x="50.0e-3" y="-1.0e-3" z="25.6e-3" />
            <p3 x="50.0e-3" y="1.0e-3" z="25.6e-3" />
            <p4 x="0.0e-3" y="1.0e-3" z="25.6e-3" />
          </notch>
        </deformstrucbody>
      </deformstrucs>
    </special>
    <parameters>
      <parameter key="StepAlgorithm" value="1" />
      <parameter key="Kernel" value="2" />
      <parameter key="Visco" value="8.92678034e-7" />
      <parameter key="CoefDtMin" value="0.001" />
      <parameter key="TimeMax" value="120.0e-6" />
      <parameter key="TimeOut" value="1.0e-6"/>
      <simulationdomain>
        <posmin x="default-10%" y="default- 10%" z="default - 10%" />
        <posmax x="default+10%" y="default+ 10%" z="default + 10%" />
      </simulationdomain>
    </parameters>
  </execution>
</case>
\end{lstlisting}
\section{Four-point bending}\label{app:4pbending}
\begin{lstlisting}[style=xmlstyle]
<?xml version="1.0" encoding="UTF-8" ?>
<case>
  <casedef>
    <constantsdef>
      <gravity x="0" y="0" z="0.0" />
      <rhop0 value="997" />
      <hswl value="0" auto="true" />
      <gamma value="7" />
      <speedsystem value="10" auto="false" />
      <coefsound value="10" />
      <coefh value="1.0" />
      <cflnumber value="0.1" />
    </constantsdef>
    <mkconfig boundcount="240" fluidcount="9" />
    <geometry>
      <definition dp="0.2e-3" units_comment="metres (m)">
        <pointmin x="-10.0e-3" y="-10.0e-3" z="-30.0e-3" />
        <pointmax x="90.0e-3" y="15.0e-3" z="30.0e-3" />
      </definition>
      <commands>
        <mainlist>
        <!-- <xc0="40.0e-3" for 90 degree crack /> -->
        <!-- <xc0="34.0e-3" for 60 degree crack /> -->
        <!-- <xc0="30.0e-3" for 45 degree crack /> -->
          <newvar xc0="40.0e-3"/>
          <setdrawmode mode="face" />
          <setshapemode>dp | bound </setshapemode>
          <setmkbound mk="1" />
          <drawfilevtk file="Shape_90Lambda.vtk" objname="Beam" autofill="true"/>
          <shapeout file="FourPBend" />	
        </mainlist>
      </commands>
    </geometry>
    <motion>
      <objreal ref="1">
      <begin mov="1" start="0" />
        <mvnull id="1" />
      </objreal>
    </motion>
  </casedef>
  <execution>
     <special>
       <mathexpressions>
         <userexpression id="1" comment="phi constrain">
            <expression value=" if(z0<0.00090 and x0>=0.002525 and x0<=0.005475, 0.9999, if(z0<0.00090 and x0>=0.074525 and x0<=0.077475, 0.9999, if(z0>0.01910 and x0>=0.018450 and x0<=0.021400, 0.9999, if(z0>0.01910 and x0>=0.058600 and x0<=0.061550, 0.9999,skip))))"/>
         </userexpression>
         <userexpression id="2" comment="z bc">
            <locals value="Velmax=10.0"/>
            <expression value="if(z0<0.00010 and x0>=0.003825 and x0<=0.004225, Velmax, if(z0<0.00010 and x0>=0.075625 and x0<=0.076175, Velmax, if(z0>0.01990 and x0>=0.01970 and x0<=0.020100, -Velmax, if(z0>0.01990 and x0>=0.059900 and x0<=0.060250, -Velmax, skip))))"/>
         </userexpression>
       </mathexpressions>
       <deformstrucs>
         <deformstrucbody mkbound="1">
            <bcvel ze="2" />
            <restrictphi value="1"/>
            <density value="50.0" />
            <youngmod value="12.44e9" />
            <poissratio value="0.3" />
            <fracture value="1" />
            <Gc value="11.8e3" />
            <pflenscale value="0.25e-3" />
            <notch>
              <p1 x="0.04" y="0.0e-3" z="-1.0e-3" />
              <p2 x="0.04" y="0.0e-3" z="0.0056" />
              <p3 x="#xc0" y="10.0e-3" z="0.0056" />
              <p4 x="#xc0" y="10.0e-3" z="-1.0e-3" />
            </notch>
         </deformstrucbody>
       </deformstrucs>
     </special>
     <parameters>
       <parameter key="StepAlgorithm" value="1" />
       <parameter key="Kernel" value="2" />
       <parameter key="Visco" value="8.92678034e-7" />
       <parameter key="CoefDtMin" value="0.0001" />
       <parameter key="TimeMax" value="250.0e-6" />
       <parameter key="TimeOut" value="2.0e-6" />
       <simulationdomain>
         <posmin x="default-10%" y="default- 10%" z="default - 10%" />
         <posmax x="default+10%" y="default+ 10%" z="default + 10%" />
       </simulationdomain>
     </parameters>
  </execution>
</case>
\end{lstlisting}
\section{Flyer plate impact}\label{app:2dtaylorbar}
\begin{lstlisting}[style=xmlstyle]
<?xml version="1.0" encoding="UTF-8" ?>
<case>
  <casedef>
    <constantsdef>
      <gravity x="0" y="0" z="0.0" />
      <rhop0 value="997" />
      <hswl value="0" auto="true" />
      <gamma value="7" />
      <speedsystem value="10" auto="false" />
      <coefsound value="10" />
      <coefh value="1.3" />
      <cflnumber value="0.01" />
    </constantsdef>
    <mkconfig boundcount="240" fluidcount="9" />
    <geometry>
      <definition dp="0.01" units_comment="metres (m)">
        <pointmin x="-2.0" y="0" z="-1.0" />
        <pointmax x="2.0" y="0" z="5.0" />
      </definition>
      <commands>
        <mainlist>
          <newvar Lx="1.0" Ly="1.0" Lz="1.0"/>
          <setdrawmode mode="full" />
          <setmkbound mk="1" />
          <drawbox>
            <boxfill>solid</boxfill>
            <point x="0.0" y="0.0" z="1.1" />
            <size x="#Lx" y="#Ly" z="#Lz" />
          </drawbox>
          <setmkbound mk="2" />
          <drawbox>
            <boxfill>solid</boxfill>
            <point x="0.0" y="0.0" z="0.05" />
            <size x="#Lx" y="#Ly" z="#Lz" />
          </drawbox>
        </mainlist>
      </commands>
    </geometry>
    <motion>
      <objreal ref="1">
        <begin mov="1" start="0" />
        <mvnull id="1" />
      </objreal>
      <objreal ref="2">
        <begin mov="2" start="0" />
        <mvnull id="2" />
      </objreal>
    </motion>
  </casedef>
  <execution>
    <special>
      <deformstrucs>
        <contcoeff value="5" />
        <deformstrucbody mkbound="1">
          <bcvel z="-200.0" tend="0.0"/>
          <density value="7870.0" />
          <youngmod value="200.0e9" />
          <poissratio value="0.29" />
          <artvisc factor1="0.05" factor2="0.0" />
          <constitmodel value="3" />
          <restcoef value="0.95" />
          <yieldstress value="4.0e8" />
          <hardening value="1.0e8" />
        </deformstrucbody>
        <deformstrucbody mkbound="2">
          <bcvel z="200.0" tend="0.0"/>
          <density value="7870.0" />
          <youngmod value="200.0e9" />
          <poissratio value="0.29" />
          <artvisc factor1="0.05" factor2="0.0" />
          <constitmodel value="3" />
          <restcoef value="0.95" />
          <yieldstress value="4.0e8" />
          <hardening value="1.0e8" />
        </deformstrucbody>
      </deformstrucs>
    </special>
    <parameters>
      <parameter key="StepAlgorithm" value="1" />
      <parameter key="Kernel" value="2" />
      <parameter key="Visco" value="8.92678034e-7" />
      <parameter key="CoefDtMin" value="0.001" />
      <parameter key="TimeMax" value="100.0e-4" />
      <parameter key="TimeOut" value="0.1e-4" comment="Time out data" units_comment="seconds" />
      <simulationdomain >
        <posmin x="default-10%" y="default- 10%" z="default - 10%" />
        <posmax x="default+10%" y="default+ 10%" z="default + 10%" />
      </simulationdomain>
    </parameters>
  </execution>
</case>
\end{lstlisting}
\section{3D Taylor bar impact}\label{app:taylorbar}
\begin{lstlisting}[style=xmlstyle]
<?xml version="1.0" encoding="UTF-8" ?>
<case>
  <casedef>
    <constantsdef>
      <gravity x="0" y="0" z="0.0"/>
      <rhop0 value="997" />
      <hswl value="0" auto="true"  />
      <gamma value="7" />
      <speedsystem value="10" auto="false" />
      <coefsound value="10" />
      <coefh value="1.0" />
      <cflnumber value="0.02" />
    </constantsdef>
    <mkconfig boundcount="240" fluidcount="9" />
    <geometry>
      <definition dp="0.2e-3" units_comment="metres (m)">
        <pointmin x="-10.0e-3" y="-10.0e-3" z="-20.0e-3" />
        <pointmax x="10.0e-3" y="10.0e-3" z="50.0e-3" />
      </definition>
      <commands>
        <mainlist>
          <newvar Rd="3.2e-3" Lz="32.4e-3"/>
          <setdrawmode mode="full" />
          <setmkbound mk="1" />
          <setfrdrawmode auto="true"/>
          <drawcylinder radius="#Rd">
            <point x="0.0" y="0.0" z="0.0"/>
            <point x="0.0" y="0.0" z="#Lz"/>
          </drawcylinder>
        </mainlist>
      </commands>
    </geometry>
    <motion>
      <objreal ref="1">
        <begin mov="1" start="0" />
        <mvnull id="1" />
      </objreal>
    </motion>
  </casedef>
  <execution>
    <special>
      <mathexpressions>
        <userexpression id="1">
          <locals value="Vinit=-227;"/>
          <expression value="if(z<1.0e-12,0.0,if(t<=0.0,Vinit,skip))"/>
        </userexpression>
      </mathexpressions>
      <deformstrucs>
        <deformstrucbody mkbound="1">
          <bcvel ze="1" />
          <density value="8930.0"/>
          <youngmod value="1.17e11" />
          <poissratio value="0.35" />
          <artvisc factor1="0.05" factor2="0.0" />
          <constitmodel value="3"/>
          <yieldstress value="400.0e6" />
          <hardening value="100.0e6" />
        </deformstrucbody>
      </deformstrucs>
    </special>
    <parameters>
      <parameter key="StepAlgorithm" value="1" />
      <parameter key="Kernel" value="2" />
      <parameter key="Visco" value="8.92678034e-7" />
      <parameter key="CoefDtMin" value="0.001" />
      <parameter key="TimeMax" value="2.5e-4" />
      <parameter key="TimeOut" value="0.01e-4" />
      <simulationdomain >
        <posmin x="default-10%" y="default- 10%" z="default - 10%" />
        <posmax x="default+10%" y="default+ 10%" z="default + 10%" />
      </simulationdomain>
    </parameters>
  </execution>
</case>=
\end{lstlisting}

\bibliographystyle{unsrt}
\bibliography{main}

\end{document}